\documentclass{article}
\usepackage{graphicx}
\usepackage{amsmath}
\begin{document}

\begin{center}
{\Large{\bf The Physics of $2 \neq 1 + 1$}} \\ \vspace*{8mm} 
Yanhua Shih \\
Department of Physics \\
University of Maryland, Baltimore County, \\ Baltimore, MD 21250 
\end{center}

\vspace{1mm}

\begin{center}
\parbox{13.75cm}{{\textbf{Abstract}}: One of the most surprising consequences of 
quantum mechanics is the entanglement of two or more distant particles.  In an 
entangled EPR two-particle system, the value of the momentum (position) for 
neither single subsystem is determined.  However, if one of the subsystems is 
measured to have a certain momentum (position), the other subsystem is 
determined to have a unique corresponding value, despite the distance 
between them.  This peculiar behavior of an entangled quantum system has 
surprisingly been observed experimentally in two-photon temporal 
and spatial correlation measurements, 
such as ``ghost" interference and ``ghost" imaging. This article  addresses 
the fundamental concerns behind these experimental observations and to 
explore the nonclassical nature of two-photon superposition by emphasizing 
the physics of $2 \neq 1 + 1$.}
\end{center}

\vspace{2mm}

\section{Introduction}

\hspace{5.5mm} In quantum theory, \emph{a particle} is allowed to exist  in a 
set of orthogonal states simultaneously.  A vivid picture of this concept might be
Schr\"odinger's cat, where his cat  is in a state of both alive and dead simultaneously.
In mathematics, the concepts of ``alive" and ``dead" are expressed through the idea 
of orthogonality.  In quantum mechanics, the superpositions of these orthogonal 
states are used to describe the physical reality of a quantum object.  In this respect 
the superposition principle is indeed a mystery when compared with our everyday 
experience.

In this article, we discuss another surprising consequence of quantum mechanics, 
namely that of quantum entanglement.  Quantum entanglement involves a multi-particle 
system in a coherent superposition of orthogonal states.  Here again Schr\"odinger's 
cat is a nice way of cartooning the strangeness of quantum entanglement.  Now 
imagine two Schr\"odinger's cats ÒpropagatingÓ to separate distant locations.  
The two cats are nonclassical by means of the following two criteria: 
(1) each of the cats is in 
a state of alive and dead simultaneously; (2) the two must be observed 
to be both alive or both dead whenever we observe them, despite their separation. 
There would probably be no concern if our observations were based on a large 
number of alive-alive or dead-dead twin cats, pair by pair, with say a 50\% chance 
to observe a dead-dead or alive-alive pair.  However, we are talking about a single 
pair of cats with this single pair being in the state of alive-alive and dead-dead 
simultaneously, and, in addition each of the cats in the pair must be alive and dead 
simultaneously.  The superposition of multi-particle states with these entangled properties 
represents a troubling concept to classical theory.  These concerns 
derive not only from the fact that the superposition of multi-particle states has no 
classical counterpart, but also because it represents a nonlocal behavior which 
may never be understood classically.

The concept of quantum entanglement started in 1935 \cite{EPR}.  Einstein, Podolsky and Rosen, 
suggested a {\em gedankenexperiment} and introduced an entangled two-particle system 
based on the superposition of two-particle wavefunctions.  The EPR system is composed of two 
distant interaction-free particles which are characterized by the following wavefunction: 
\begin{eqnarray}\label{eprst}  
\Psi(x_1, \, x_2) \ = \ \frac{1}{2\pi \hbar}\int dp_1 dp_2 \, \delta(p_1+p_2) \,  
e^{i p_1 (x_1-x_{0})/\hbar}  e^{i p_2 x_2/\hbar} \ = \ \delta(x_1-x_2- x_0)  
\end{eqnarray}
where $e^{i p_1 (x_1-x_{0})/\hbar} $ and $e^{i p_2 x_2/\hbar}$ are the eigenfunctions
with eigenvalues $p_1=p$ and $p_2=-p$ of the momentum operators $\hat{p}_1$ 
and $\hat{p}_2$ associated with particles 1 and 2, respectively. $x_1$ and $x_2$ are the 
coordinate variables to describe the positions of particles 1 and 2, respectively; and $x_{0}$
is a constant.   The EPR state is very peculiar.  Although there is no interaction between the 
two distant particles, the two-particle superposition cannot be factorized into a 
product of two individual superpositions of two particles.  Remarkably, quantum theory allow 
for such states.  

What can we learn from the EPR state of Eq.~(\ref{eprst})?   

(1) In coordinate representation, the wavefunction is a delta function $\delta(x_1-x_2-x_0)$.    
The two particles are separated in space with a constant value of $x_{1}-x_{2}=x_{0}$,  
although the coordinates $x_{1}$ and $x_{2}$ of the two particles are both unspecified.   

(2) The delta wavefunction $\delta(x_1-x_2-x_0)$
is the result of the superposition of plane wavefunctions for free particle one, 
$e^{i p_1 (x_1-x_0)/\hbar}$, and free particle two, $e^{i p_2 x_2 /\hbar}$, with a particular
distribution $\delta(p_1+p_2)$.  It is $\delta(p_1+p_2)$ that made the superposition special.
Although the momentum of particle one and particle two may take on any values, the delta 
function restricts the superposition to only those terms in which the total momentum 
of the system takes a constant value of zero.   

Now, we transfer the wavefunction from coordinate representation to momentum representation:  
\begin{eqnarray}\label{eprst-1} 
\Psi(p_1, \, p_2) \ = \ \frac{1}{2\pi \hbar}\int dx_1  dx_2 \, \delta(x_1-x_2- x_0) \, 
e^{-i p_1 (x_1-x_0)/\hbar}  e^{-i p_2 x_2 /\hbar} \ = \ \delta(p_1+p_2).
\end{eqnarray} 

What can we learn from the EPR state of Eq.~(\ref{eprst-1})?   

(1) In momentum representation, the wavefunction is a delta function $\delta(p_1+p_2)$.    
The total momentum of the two-particle system takes a constant value of $p_{1}+p_{2}=0$, 
although the momenta $p_{1}$ and $p_{2}$ are both unspecified.   

(2) The delta wavefunction $\delta(p_1+p_2)$
is the result of the superposition of plane wavefunctions for free particle one, 
$e^{-i p_1 (x_1-x_0)/\hbar}$, and free particle two, $e^{-i p_2 x_2 /\hbar}$, with a particular
distribution $\delta(x_1-x_2- x_0)$.  It is $\delta(x_1-x_2- x_0) $ that made the superposition 
special.  Although the coordinates of particle one and particle two may take on any values, 
the delta function restricts the superposition to only those terms in which $x_{1} - x_{2}$ 
is a constant value of $x_{0}$.   
    
In an EPR system, the value of the momentum (position) for neither single subsystem is 
determined.  However, if one of the subsystems is measured to be at a certain momentum 
(position), the other one is determined with a unique corresponding value, despite the 
distance between them.    An idealized EPR state of a two-particle system is therefore 
characterized by $\Delta(p_1+p_2)=0$ and $\Delta (x_1- x_2)=0$ simultaneously, even if 
the momentum and position of each individual free particle are completely undefined, i.e., 
$\Delta p_j \sim \infty$ and $\Delta x_j \sim \infty$, $j=1,2$.    In other words, each of the 
subsystems may have completely random values or all possible values of momentum and 
position in the course of their motion, but the correlations of the two subsystems are 
determined with certainty whenever a joint measurement is performed.  

The EPR states of Eq.~(\ref{eprst}) and Eq.~(\ref{eprst-1}) are simply the results of the 
quantum mechanical \textit{superposition of two-particle states}.   
The physics behind EPR states is far beyond the acceptable limit of Einstein.  

Does a free particle have a defined momentum and position in the state of Eq.~(\ref{eprst}) 
and Eq.~(\ref{eprst-1}), regardless of whether we measure it or not?  On one hand, the 
momentum and position of neither independent particle is specified and the superposition 
is taken over all possible values of the momentum and position.  We may have to believe 
that the particles do not have any defined momentum and position, or have all possible 
values of momentum and position within the superposition, during the course of their 
motion.  On the other hand, if the measured momentum (position) of one particle 
uniquely determines the momentum (position) of the other distant particle, 
it would be hard for anyone who believes no action-at-a-distance to imagine that the 
momenta (position) of the two particles are not predetermined with defined values 
before the measurement.   EPR thus put us into a paradoxical situation. 
It seems reasonable for us to ask the same question that EPR had asked in 1935:
``Can quantum-mechanical description of physical reality be considered complete?'' \cite{EPR}

In their 1935 article, Einstein, Podolsky and Rosen argued that the existence
 of the entangled two-particle state of Eq.~(\ref{eprst}) and  Eq.~(\ref{eprst-1}), 
a straightforward quantum mechanical superposition of two-particle states, 
led to the violation of the uncertainty principle of quantum theory.  
To draw their conclusion, EPR started from the following criteria.

{\em Locality}: there is no action-at-a-distance;

{\em Reality}: ``if, without in any way disturbing a system, we can predict
with certainty the value of a physical quantity, then there exist an element
of physical reality corresponding to this quantity.'' According to the delta 
wavefunctions, we can predict with certainty the result of measuring the
momentum (position) of particle 1 by measuring the momentum (position) of 
particle 2, and the measurement of particle 2 cannot cause any disturbance to 
particle 1, if the measurements are space-like separated events. Thus, both the 
momentum and position of particle 1 must be elements of physical reality 
regardless of whether we measure it or not.  This, however, is not allowed by 
quantum theory.  Now consider:

{\em Completeness}: ``every element of the physical reality must have a counterpart 
in the complete theory.'' This led to the question as the title of their 1935 article: 
``Can Quantum-Mechanical Description of Physical Reality Be Considered Complete?'' 

EPR's arguments were never appreciated by Copenhagen.  Bohr criticized
EPR's criterion of physical reality \cite{Bohr}: ``it is too narrow". 
However, it is perhaps not easy to find a wider criterion. A memorable quote from Wheeler, 
``No elementary quantum phenomenon is a phenomenon until it is a recorded 
phenomenon'', summarizes what Copenhagen has been trying to teach us \cite{Wheeler}. 
By 1927, most physicists accepted the Copenhagen
interpretation as the standard view of quantum formalism.  Einstein, however,
refused to compromise.  As Pais recalled in his book, during a walk around 1950,
Einstein suddenly stopped and ``asked me if I really believed that the moon (pion)
exists only if I look at it.''  \cite{Pais}

\vspace{6mm}

There has been arguments considering $\Delta(p_1+p_2)\Delta (x_1- x_2)=0$
a violation of the uncertainty principle.   This argument is false.  
It is easy to find that $p_1+p_2$ and $x_1-x_2$ are not conjugate variables.   
As we know, non-conjugate variables correspond to commuting operators in
quantum mechanics, if the corresponding operators exist.\footnote{It is possible 
that no quantum mechanical operator is associated 
with a measurable variable, such as time $t$.  From this perspective, an uncertainty 
relation based on variables rather than operators is more general.}   To have
$\Delta(p_1+p_2)=0$ and $\Delta (x_1- x_2)=0$ simultaneously, or to have 
$\Delta(p_1+p_2)\Delta (x_1- x_2)=0$, is not a violation of the uncertainty principle.
This point can easily be seen from the following two dimensional Fourier transforms:
\begin{eqnarray*}
&& \Psi(x_1, \, x_2) 
\ = \ \frac{1}{2\pi \hbar}\int dp_1 \, dp_2 \, \delta(p_1+p_2) \, e^{i p_1
(x_1-x_0) / \hbar} \, e^{i p_2 x_2/\hbar}  \\ 
&=& \frac{1}{2\pi \hbar} \int  d(p_1+p_2) \, \delta(p_1+p_2) \, e^{i (p_1+p_2)(x'_1+x_2)/ 2\hbar}  
\int d(p_1-p_2)/2 \, e^{i (p_1 -p_2)(x'_1-x_2)/ 2 \hbar}  
\\ \nonumber &=& 1\times \delta(x_1 - x_2 - x_0) 
\end{eqnarray*}
where $x' = x_1-x_0$;
\begin{eqnarray*}
&& \Psi(p_1, \, p_2) \ = \ \frac{1}{2\pi \hbar}\int dx_1 \, dx_2  \, \delta(x_1-x_2-x_0) \, e^{-i p_1
(x_1-x_0) /\hbar} \, e^{-i p_2 x_2 /\hbar}  \\ 
&=&\frac{1}{2\pi \hbar} \int d(x'_1+x_2) \,  e^{-i (p_1+p_2)(x'_1+x_2)/ 2 \hbar}   
\int d(x'_1-x_2)/2  \, \delta(x'_1- x_2) \, 
e^{-i (p_1 - p_2)(x'_1-x_2)/ 2 \hbar}  \\ \nonumber
&=& \delta(p_1+p_2) \times 1. 
\end{eqnarray*}
The Fourier conjugate variables are $(x_1+ x_2) \Leftrightarrow (p_1+p_2)$ 
and $(x_1- x_2) \Leftrightarrow (p_1-p_2)$.
Although it is possible to have $\Delta(x_1- x_2)\sim 0$ and 
$ \Delta (p_1+p_2) \sim 0$ simultaneously, the uncertainty 
relations must hold for the Fourier conjugates $\Delta(x_1+ x_2) \Delta(p_1+p_2) \geq \hbar$, 
and $\Delta(x_1- x_2) \Delta (p_1-p_2) \geq \hbar$; with $\Delta(p_1-p_2) \sim \infty$ 
and $\Delta(x_1+ x_2) \sim \infty$.    

As a matter of fact, in their 1935 paper, Einstein-Podolsky-Rosen never questioned 
$\Delta(x_1- x_2) \, \Delta (p_1+p_2) = 0$ as a violation of the uncertainty principle.  
The violation of the uncertainty principle was probably not Einstein's concern at all, although
their 1935 paradox was based on the argument of the uncertainty principle.  
What really bothered Einstein so much?   For all of his life, Einstein, a true believer of realism, 
never accepted that a particle does not have a defined momentum and position during its
motion, but rather is specified by a probability amplitude of certain a momentum and position.  
``God does not play dice" was the most vivid criticism from Einstein to refuse the 
Schr\"odinger's cat.   The entangled two-particle system was used as an example
to clarify and to reinforce Einstein's realistic opinion.   To Einstein, the acceptance of 
Schr\"odinger's cat perhaps means action-at-a-distance or an inconsistency 
between quantum mechanics and the theory of relativity, when dealing with 
the entangled EPR two-particle system.    
Let us follow Copenhagen to consider that \textit{each particle} in an EPR 
pair has no defined momentum and position, or has all possible momentum 
and position within the superposition state, i.e., imagine
$\Delta p_j \neq 0$, $\Delta x_j  \neq 0$, $j=1,2$, for 
\textit{each single-particle} until the measurement.  Assume the measurement 
devices are particle counting devices able to identify the 
position of each particle among an ensemble of particles. 
For each registration of a particle the measurement device 
records a value of its position.  No one can predict what value is registered for each 
measurement; the best knowledge we may have is the probability to register that value.  
If we further assume no physical interaction 
between the two distant particles and believe no action-at-a-distance exist in nature, 
we would also believe that no matter 
how the two particles are created, the two registered values must be independent
of each other.   Thus, the value of $x_1 - x_2$ is unpredictable within the uncertainties
of $\Delta x_1$ and $\Delta x_2$.   The above statement is also valid for the momentum 
measurement.
Therefore, after a set of measurements on a large number of particle pairs,  the statistical 
uncertainty of 
the measurement on $p_1+p_2$ and $x_1-x_2$ must obey the following inequalities:
\begin{eqnarray}\label{classicalineq}
& & \Delta  (p_1+ p_2) = \sqrt{(\Delta p_{1})^{2}+(\Delta p_{2})^{2}} > Max(\Delta p_{1},\Delta p_{2})
\\  \nonumber 
& & \Delta (x_1- x_2)  = \sqrt{(\Delta x_{1})^{2}+(\Delta x_{2})^{2}} > Max(\Delta x_{1},\Delta x_{2}). 
\end{eqnarray}
Eq.~(\ref{classicalineq}) is obviously true in statistics, especially when we are sure 
that no disturbance is possible between the two independent-local measurements.   
This condition can be easily realized by making the two measurement events 
space-like separated events.    The classical inequality of Eq.~(\ref{classicalineq}) 
would not allow $\Delta (p_1 + p_2) = 0$ and $\Delta (x_1 - x_2) = 0$ as required 
in the EPR state, unless $\Delta p_1 = 0$, $\Delta p_2 = 0$, $\Delta x_1 = 0$ and 
$\Delta x_2 = 0$, simultaneously.  Unfortunately, the assumption of $\Delta p_1 = 0$, 
$\Delta p_2 = 0$, $\Delta x_1 = 0$, $\Delta x_2 = 0$ cannot be true because it violates 
the uncertainty relations 
$\Delta p_1\Delta x_1 \geq \hbar $ and $\Delta p_2\Delta x_2 \geq \hbar$.

In a non-perfect entangled system, the uncertainties of  
$p_1+ p_2$ and $x_1- x_2$ may differ from zero. Nevertheless, the 
measurements may still satisfy the EPR inequalities \cite{Inequality}:
\begin{eqnarray}\label{QMineq}
& & \Delta (p_1+ p_2) < min(\Delta p_1,\Delta p_2) \\ \nonumber 
& & \Delta (x_1- x_2) < min(\Delta x_1,\Delta x_2).
\end{eqnarray}

The apparent contradiction between the classical inequality Eq.~(\ref{classicalineq}) 
and the EPR  inequality Eq.~(\ref{QMineq}) deeply troubled Einstein.  
While one sees the measurements of $p_1+ p_2$ and $x_1- x_2$ of the two 
distant individual free particles satisfying Eq.~(\ref{QMineq}), but believing 
Eq.~(\ref{classicalineq}), one might easily be trapped into concluding either there is a 
violation of the uncertainty principle or there exists action-at-a-distance.  

\vspace{6mm}

Is it possible to have a realistic theory which provides correct predictions of the 
behavior of a particle similar to quantum theory and, at the same time,
respects the description of physical reality by EPR as ``complete''?  
Bohm and his followers have attempted a ``hidden variable
theory'', which seemed to satisfy these requirements \cite{Bohm}. The
hidden variable theory was successfully applied to many different quantum
phenomena until 1964, when Bell proved a theorem to show that an inequality,
which is violated by certain quantum mechanical statistical predictions, can
be used to distinguish local hidden variable theory from quantum mechanics \cite{Bell}. 
Since then, the testing of Bell's inequalities became a standard
instrument for the study of fundamental problems of quantum theory \cite{Shimony}.  
The experimental testing of Bell's inequality started from the early 1970's. 
Most of the historical experiments concluded the violation of the Bell's inequalities and thus
disproved the local hidden variable theory \cite{Shimony}\cite{Aspect}\cite{Shih}.    

In the following, we examine a simple yet popular realistic model to 
simulate the behavior of the entangled EPR system.  This model concerns
an ensemble of classically correlated particles instead of the quantum mechanical
superposition of a particle.  In terms of ``cats", this model is based on 
the measurement of a large number of twin cats in which 50\% are alive-alive 
twins and 50\% are dead-dead twins.  This model refuses the concept of
 Schr\"odinger's cat which requires \emph{a cat} to be alive and dead simultaneously, 
and \emph{each pair} of cats involved in a joint detection event is in the state of 
alive-alive and dead-dead simultaneously.

\vspace{3mm}

In this model, we may have three different states:     

\vspace{3mm}

(1) State one,  each single pair of particles 
holds defined momenta $p_1=$ constant and $p_2=$ constant with $p_1+p_2=0$.   
From pair to pair, the values of $p_1$ and $p_2$ may vary significantly. The sum of $p_1$ 
and $p_2$, however, keeps a constant of zero.  Thus, each joint detection of the two 
distant particles measures precisely the constant values of $p_1$ and $p_2$ and 
measures $p_1+p_2=0$.   The uncertainties of  $\Delta p_1$ and $\Delta p_2$ 
only have statistical meaning in terms of the measurements of an ensemble.    
This model successfully simulated 
$\Delta (p_1+p_2)=0$ based on the measurement of a large number of classically 
correlated particle pairs.  This is, however, only half of the EPR story.  Can we have 
$\Delta (x_1 -x_2) = 0$ simultaneously in this model?   
We do have $\Delta x_1\sim \infty$ and $\Delta x_2\sim \infty$, 
otherwise the uncertainty principle will be violated. The position correlation, 
however, can never achieve $\Delta (x_1 -x_2) = 0$ by any means.    

\vspace{3mm}

(2) State two, each single pair of particles holds a well defined position $x_1=$ 
constant and $x_2=$ constant with $x_1-x_2=x_0$.   From pair to pair,
the values of $x_1$ and $x_2$ may vary significantly. The difference of $x_1$ 
and $x_2$, however, maintains a constant of $x_0$.  Thus, each joint detection of the two 
distant particles measures precisely the constant values of $x_1$ and $x_2$ and 
measures $x_1-x_2=x_0$.   The uncertainties of  $\Delta x_1$ and $\Delta x_2$ 
only have statistical meaning in terms of the measurements of an ensemble.    
This model successfully simulated  
$\Delta (x_1-x_2)=0$ based on the measurement of a large number of classically 
correlated particle pairs.  This is, however, only half of the EPR story.  Can we have 
$\Delta (p_1 + p_2) = 0$ simultaneously in this model?   
We do have $\Delta p_1\sim \infty$ and $\Delta p_2\sim \infty$, 
otherwise the uncertainty principle will be violated.  The momentum correlation, 
however, can never achieve 
$\Delta (p_1 + p_2) = 0$ by any means.

The above two models of classically correlated particle pairs can never achieve both
$\Delta(p_1+p_2)=0$ and $\Delta (x_1- x_2)=0$.  What would happen if we combine
the two parts together?  This leads to the third model of classical simulation.

\vspace{3mm}

(3) State three, among a large number of classically correlated particle pairs, we
assume 50\% to be in state one and the other 50\% state two.   The $p_1+p_2$ 
measurements would have 50\% chance with $p_1+p_2=0$ and 50\% chance with 
$p_1+p_2=$ random value.  On the other hand, the $x_1-x_2$ measurements 
would have 50\% chance with $x_1-x_2=x_0$ and 50\% chance with $x_1-x_2=$ 
random value.  What are the statistical uncertainties on the measurements of 
$(p_1+p_2)$ and $(x_1-x_2)$ in this case?  
If we focus on only these events of state one, the statistical uncertainty on the 
measurement of $(p_1+ p_2)$  is $\Delta(p_1+p_2)=0$, and if we focus on these 
events of state two, the statistical uncertainty on the measurement of $(x_1-x_2)$ 
is $\Delta(x_1-x_2)=0$; however, if we consider all the measurements together, 
the statistical uncertainties on the measurements of $(p_1+p_2)$ and $(x_1-x_2)$, 
are both infinity: $\Delta(p_1+p_2)=\infty$ and $\Delta(x_1-x_2)=\infty$.   

\vspace{3mm}

In conclusion, classically correlated particle pairs may partially simulate EPR 
correlation with three types of optimized observations: 
\begin{eqnarray*}
(1) & & \Delta(p_1+p_2)=0 \,\, (100\%) \,\, \& \,\, \Delta (x_1- x_2)= \infty \,\, (100\%);  \\
(2) & & \Delta(x_1-x_2)=0 \,\, (100\%) \,\, \& \,\, \Delta (p_1+ p_2)= \infty \,\, (100\%);  \\
(3) & & \Delta(p_1+p_2)=0 \,\, (50\%) \,\, \& \,\, \Delta (x_1- x_2)= 0 \,\, (50\%);  \\
\end{eqnarray*}

\vspace{1mm}

\hspace{-6.5mm}Within one setup of experimental measurements, only the entangled 
EPR states result in the simultaneous observation of 
\begin{eqnarray*}
& & \Delta(p_1+p_2)=0 \,\, (100\%)  \,\, \& \,\, 
\Delta (x_1- x_2)=0 \,\, (100\%) \\ \nonumber
& &  \Delta p_1 \sim \infty, \hspace{2mm} \Delta p_2 \sim \infty, \hspace{2mm}
\Delta x_1 \sim \infty, \hspace{2mm} \Delta x_2 \sim \infty.
\end{eqnarray*}
We thus have a tool, besides the testing of Bell's inequality, to distinguish quantum 
entangled states from classically correlated particle pairs.     

\section{Entangled state}

\hspace{5.5mm}The entangled state of a two-particle system was 
mathematically formulated by Schr\"{o}dinger \cite{Schrodinger}. 
Consider a pure state for a system 
composed of two distinguishable subsystems

\begin{equation}  \label{schst}
\left| \Psi \right\rangle =\sum_{a,b} c(a,b)\left| a\right\rangle
\left| b\right\rangle
\end{equation}
where \{$\mid a\rangle $\} and \{$\mid b\rangle $\} are two sets of
orthogonal vectors for subsystems 1 and 2, respectively.
If $c(a,b)$ does not factor into a product of the form 
$f(a)\times g(b)$, then it follows that the state does not factor into a
product state for subsystems 1 and 2:
\begin{equation}\label{schst-2}
\hat{\rho} \, = \, | \Psi \rangle \langle \Psi | \, = \sum_{a,b}  \, c(a,b)  | a \rangle | b \rangle  
\sum_{a',b'} \, c^*(a',b') 
\langle b' | \langle a' | \, \neq \, \hat{\rho} _{1} \times \hat{\rho} _2,
\end{equation}   
where $\hat{\rho}$ is the density operator, the state was defined by 
Schr\"{o}dinger as an entangled state.

Following this notation, the first classic entangled state of a two-particle system, 
the EPR state of Eq.~(\ref{eprst}) and Eq.~(\ref{eprst-1}), is thus written as:
\begin{eqnarray}\label{eprst-10}
| \Psi \rangle_{EPR} = \sum_{x_1,x_2} \delta ( x_1 - x_2 + x_0)\,\,  | \, x_1 \, \rangle |\,  x_2 \, \rangle
= \sum_{p_1,p_2} \delta ( p_1 + p_2 )\,\,  | \, p_1 \, \rangle |\,  p_2 \, \rangle,
\end{eqnarray}
where we have described the entangled two-particle system as the coherent superposition
of the momentum eigenstates as well as the coherent superposition of the position 
eigenstates.    The  two $\delta$-functions in Eq.~(\ref{eprst-10}) represent, 
respectively and simultaneously, the perfect position-position and 
momentum-momentum correlation.   Although the two distant particles are 
interaction-free, the superposition selects only the eigenstates which are
specified by the $\delta$-function.    We may use the following statement to summarize  
the surprising feature of the EPR state: {\em the values of the momentum and the position 
for neither interaction-free single subsystem is determinated. However, if one of the 
subsystems is measured to be at a certain value of momentum and/or position, the momentum 
and/or position of the other one is 100\% determined, despite the distance between them}.  

It should be emphasized again that Eq.~(\ref{eprst-10}) is true, simultaneously, in the 
conjugate space of momentum and position.  This is different from classically 
correlated states
\begin{equation}\label{classicalClrrelated-p}
\hat{\rho} =\sum_{p_1,p_2}\delta ( p_1 + p_2 )\,\,  | \, p_1 \, \rangle |\,  p_2 \, \rangle
 \langle \, p_2 \, | \langle \, p_1 \, |,
\end{equation}
or
\begin{equation}\label{classicalClrrelated-x}
\hat{\rho} =\sum_{x_1,x_2}\delta ( x_1 - x_2 + x_0)\,\,  | \, x_1 \, \rangle |\,  x_2 \, \rangle
 \langle \, x_2 \, | \langle \, x_1 \, |.
\end{equation}
Eq.~(\ref{classicalClrrelated-p}) and Eq.~(\ref{classicalClrrelated-x}) represent 
mixed states.  Eq.~(\ref{classicalClrrelated-p}) and Eq.~(\ref{classicalClrrelated-x}) 
cannot be true simultaneously as we have discussed earlier.  Thus, we can distinguish 
entangled states from classically correlated states 
through the measurements of the EPR inequalities of Eq.~(\ref{QMineq}).

\subsection*{Two-photon state of spontaneous parametric down-conversion}

\hspace{5.5mm}The state of a signal-idler photon pair created in spontaneous 
parametric down-conversion (SPDC) is a typical EPR state \cite{KlyshkoBook}\cite{IEEE03}. 
Roughly speaking, the process of SPDC involves sending a pump laser beam into a 
nonlinear material, such as a non-centrosymmetric crystal. Occasionally,  the nonlinear
interaction leads to the annihilation of a high
frequency pump photon and the simultaneous creation of a pair of lower
frequency signal-idler photons forming an entangled two-photon state:
\begin{eqnarray}\label{state1}
\left| \Psi \right\rangle = \Psi_0 \, \sum_{s,i}\delta \left( \omega
_{s}+\omega _{i}-\omega _{p}\right) \delta \left( {\bf k}_{s}+{\bf
k}_{i}-{\bf k} _{p}\right) a_{s}^{\dagger }({\bf
k}_{s})\, a_{i}^{\dagger }({\bf k}_{i})\mid 0\rangle 
\end{eqnarray}
where $\omega _{j}$, {\bf k$_{j}$ (}j = s, i, p) are the frequency
and wavevector of the signal (s), idler (i), and pump (p),
$a_{s}^{\dagger }$ and $a_{i}^{\dagger }$ are creation
operators for the signal and the idler photon, respectively, and 
$\Psi_0$ is the normalization constant.  We
have assumed a CW monochromatic laser pump, i.e., $\omega _{p}$
and {\bf k}$_{p}$ are considered as constants.  The two delta functions
in Eq.~(\ref{state1}) are technically named as the phase matching condition
\cite{KlyshkoBook}\cite{Yariv}:
\begin{eqnarray}
\omega_{p} =  \omega_{s} + \omega_{i}, \hspace{10mm}
{\bf k}_{p} = {\bf k}_{s} + {\bf k}_{i}. \label{eq:phsmtch}
\end{eqnarray}
The names {\em signal} and {\em idler} are
historical leftovers. The names perhaps came about due to the
fact that in the early days of SPDC, most of the experiments were
done with non-degenerate processes. One radiation was in the
visible range (and thus easily observable, the signal), while the
other was in the IR range (usually not measured, the idler).  We will
see in the following discussions that the role of the idler is no
any less important than that of the signal.  
The SPDC process is referred to as type-I if the signal and idler
photons have identical polarizations, and type-II if they have
orthogonal polarizations.  The process is said to be {\em
degenerate} if the SPDC photon pair has the same free space
wavelength (e.g. $\lambda_{i }=\lambda_{s} = 2\lambda_{p}$), and
{\em nondegenerate} otherwise. In general, the pair exit the
crystal {\em non-collinearly}, that is, propagate to different
directions defined by the second equation in Eq.~(\ref{eq:phsmtch})
and Snell's law.  In addition, the pair
may also exit {\em collinearly}, in the same direction, together
with the pump.

The state of the signal-idler pair can be derived, quantum mechanically, 
by the first order perturbation theory with the help of the nonlinear interaction
Hamiltonian.  The SPDC interaction arises in a nonlinear crystal driven by a
pump laser beam.  The polarization, i.e., the dipole moment per
unit volume, is given by
\begin{equation}
P_{i}=\chi^{(1)}_{i,j}E_{j}+\chi^{(2)}_{i,j,k}E_{j}E_{k}+
\chi^{(3)}_{i,j,k,l}E_{j}E_{k}E_{l}+...
\end{equation}
where $\chi^{(m)}$ is the $mth$ order electrical susceptibility
tensor.  In SPDC, it is the second order nonlinear susceptibility
$\chi^{(2)}$ that plays the role.  The second order nonlinear
interaction Hamiltonian can be written as
\begin{equation}
H=\epsilon_{0}\int_{V}d\mathbf{r}\ \chi^{(2)}_{ijk}\ E_{i}E_{j}E_{k}
    \label{H}
\end{equation}
where the integral is taken over the interaction volume $V$.

It is convenient to use the Fourier representation for the
electrical fields in Eq.~(\ref{H}):
\begin{eqnarray}
{\bf E}({\bf r},\,t)=\int d{\bf k}\ [\ {\bf E}^{(-)}({\bf k})
e^{-i(\omega({\bf k})t-{\bf k}\cdot{\bf r})} +{\bf
E}^{(+)}({\bf k}) e^{i(\omega({\bf k})t-{\bf k}\cdot{\bf r})}\ ].
    \label{fourier}
\end{eqnarray}
Substituting Eq.~(\ref{fourier}) into Eq.~(\ref{H}) and keeping only the
terms of interest, we obtain the SPDC Hamiltonian in the
interaction representation:
\begin{eqnarray}\label{HH}
&& H_{int}(t) \\ \nonumber
&=&\epsilon_{0}\int_{V}d \mathbf{r}\int d {\bf
k}_{s}\,d {\bf k}_{i} \, \chi_{lmn}^{(2)}
E_{p\,l}^{(+)}e^{i(\omega_{p}t-{\bf k}_{p}\cdot{\bf r})} 
E_{s\,m}^{(-)}e^{-i(\omega_{s}({\bf k}_{s})t-{\bf k}_{s}\cdot{\bf r})}
E_{i\,n}^{(-)}e^{-i(\omega_{i}({\bf k}_{i})t-{\bf k}_{i}\cdot{\bf
r})}+h.c.,
\end{eqnarray}
where $h.c.$ stands for Hermitian conjugate.  To simplify the
calculation, we have also assumed the pump field to be a monochromatic 
plane wave  with wave vector ${\bf k}_{p}$ and frequency
$\omega_{p}$.

It is easily noticeable that in Eq.~(\ref{HH}), the volume
integration can be done for some simplified cases.  At this point,
we assume that $V$ is infinitely large.  Later, we will see that
the finite size of $V$ in longitudinal and/or transversal
directions may have to be taken into account. For an infinite
volume $V$, the interaction Hamiltonian Eq.~(\ref{HH}) is written
as
\begin{eqnarray}
H_{int}(t)=\epsilon_{0}\int d {\bf k}_{s}\,d {\bf k}_{i}\,
\chi^{(2)}_{lmn}\, E_{p\,l}^{(+)} E_{s\,m}^{(-)}E_{i\,n}^{(-)}
\, \delta({\bf k}_{p}-{\bf k}_{s}-{\bf k}_{i})
e^{i(\omega_{p}-\omega_{s}({\bf k}_{s})-\omega_{i}({\bf
k}_{i}))t}+h.c.  \label{Hi1}
\end{eqnarray}
It is reasonable to consider the pump field to be classical, which is
usually a laser beam, and quantize the signal and idler fields,
which are both at the single-photon level:
\begin{eqnarray}
E^{(-)}({\bf k}) = i\sqrt{\frac{2\pi\hbar\omega}{V}}a^{\dagger}({\bf k}),
\ \ \ E^{(+)}({\bf k}) = i\sqrt{\frac{2\pi\hbar\omega}{V}}a({\bf k}),
\label{qvfield}
\end{eqnarray}
where $a^{\dagger}({\bf k})$ and $a({\bf k})$ are photon creation
and annihilation operators, respectively. The state of the emitted
photon pair can be calculated by applying the first order perturbation
\begin{equation}
|\Psi\rangle=-\,\frac{i}{\hbar}\int dt\, H_{int}(t)\ |0\rangle.
    \label{state}
\end{equation}
By using vacuum $|0\rangle$ for the initial state in
Eq.~(\ref{state}), we assume that there is no input radiation in any
signal and idler modes, that is, we have a spontaneous parametric
down conversion (SPDC) process.

Further assuming an infinite interaction time, evaluating the time
integral in Eq.~(\ref{state}) and omitting altogether the
constants and slow (square root) functions of $\omega$, we obtain
the \emph{entangled} two-photon state of Eq.~(\ref{state1}) in the
form of an integral \cite{IEEE03}:
\begin{eqnarray}\label{state10}
|\Psi\rangle=\Psi_{0} \int d {\bf k}_{s}d {\bf k}_{i}\,
\delta[\omega_{p}-\omega_{s}({\bf k}_{s})-\omega_{i}({\bf
k}_{i})] \delta({\bf k}_{p}-{\bf k}_{s}-{\bf k}_{i})
a^{\dagger}_{s}({\bf k}_{s})a^{\dagger}_{i}({\bf k}_{i})
|0\rangle  
\end{eqnarray}
where $\Psi_{0}$ is a normalization constant which has absorbed all
omitted constants.  

\vspace{5mm}
\begin{figure}[htb]
\centering
    \includegraphics[width=110mm]{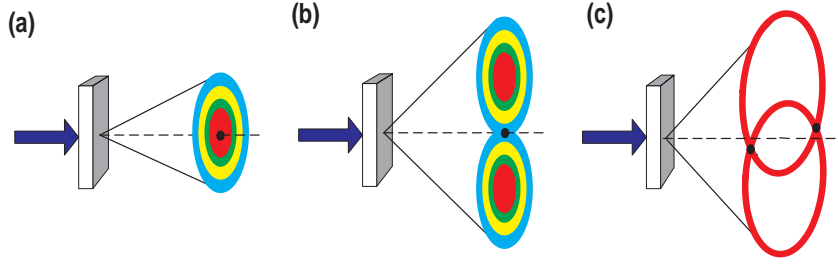}
     \parbox{14cm}{ \vspace{5mm}\caption{Three widely used SPDC setups. (a) Type-I SPDC. 
    (b) Collinear degenerate type-II SPDC. Two rings overlap at one region. (c) Non-collinear
degenerate type-II SPDC.  For clarity, only two degenerate rings, one for e-polarization and 
the other for o-polarization, are shown.  Notice, the color rainbows represent the distribution
function of a signal-idler pair.   One signal-idler pair yields the entire rainbow. 
}\label{fig:spdcbasic}}
\end{figure}

The way of achieving phase matching, i.e., the delta
functions, in Eq.~(\ref{state10}) basically determines how the
signal-idler pair ``looks".  For example, in a negative uniaxial
crystal, one can use a linearly polarized pump laser beam as an
extraordinary ray of the crystal to generate a signal-idler pair
both polarized as the ordinary rays of the crystal, which is
defined as type-I phase matching.  One can alternatively generate
a signal-idler pair with one ordinary polarized and another
extraordinary polarized, which is defined as type II phase
matching. Fig.~\ref{fig:spdcbasic} shows three examples of an SPDC
two-photon source.  All three schemes have been widely used for 
different experimental purposes.  Technical details can be found in
text books and research references in nonlinear optics.   

The two-photon state in the forms of Eq.~(\ref{state1}) or  Eq.~(\ref{state10}) 
is a pure state, which mathematically describes the behavior of a signal-idler 
photon pair.  The surprise comes from the coherent superposition of the 
two-photon modes:  

\vspace{3mm}
\begin{center}
\parbox{14cm}{
Does the signal or the idler photon in the EPR state of Eq.~(\ref{state1}) or 
Eq.~(\ref{state10}) have a  defined energy and momentum
regardless of whether we measure it or not?  Quantum mechanics answers: No!  
However, if one of the subsystems is measured with a certain energy and momentum, 
the other one is determined with certainty, despite the distance between them. } 
\end{center}
\vspace{3mm}

It is indeed a mystery from a classical point of view.  
There has been, nevertheless, classical models to avoid the surprises.   One of
the classical realistic models insists that the state of Eq.~(\ref{state1}) or Eq.~(\ref{state10}) 
only describes the behavior of an ensemble of photon pairs.  In this model, the energy and 
momentum of the signal photon and the idler photon in each individual pair are defined 
with certain values and the resulting state is a statistical mixture. 
Mathematically, it is incorrect to use a pure state to characterize a statistical mixture.
The concerned statistical ensemble should be characterized 
by the following density operator
\begin{eqnarray}\label{statemix}
\hat{\rho} = \int d {\bf k}_{s} \, d {\bf k}_{i} \,
 \delta(\omega_{p}-\omega_{s}-\omega_{i}) \, \delta({\bf k}_{p}-{\bf 
k}_{s}-{\bf k}_{i}) \ a^{\dagger}_{s}({\bf k}_{s}) \, 
a^{\dagger}_{i}({\bf k}_{i})\, | \, 0 \, \rangle \langle \, 0\, | \, 
a_{s}({\bf k}_{s}) \, a_{i}({\bf k}_{i}) 
\end{eqnarray}
which is very different from the pure state of SPDC.  We will show later that a statistical 
mixture of Eq.~(\ref{statemix}) can never have delta-function-like two-photon temporal 
and/or spatial correlation that is shown by the measurement of SPDC.  

\vspace{5mm}

For finite dimensions of the nonlinear interaction region, the entangled two-photon state of 
SPDC may have to be estimated in a more general format.   
Following the earlier discussions, we write the state of the signal-idler photon
pair as
\begin{equation}\label{spdc-00}
| \, \Psi \, \rangle=\int d \mathbf{k_s} \, d \mathbf{k_i} \,
F({\bf k}_s, {\bf k}_i) \, a_i^{\dag}({\bf k}_{s}) \, a_s^{\dag}({\bf k}_{i}) |\, 0 \, \rangle
\end{equation}
where 
\begin{eqnarray} \label{spdc_F}
F({\bf k}_s,{\bf k}_i) \hspace*{-2mm} & = & \hspace*{-1mm} \epsilon \, 
\delta(\omega_p-\omega_s-\omega_i) \, f(\Delta_{z}L) \, h_{tr}(\vec{\kappa}_1 
+ \vec{\kappa}_2) \nonumber \\
f(\Delta_{z}L) \hspace*{-2mm} & = & \hspace*{-2mm} \int_{L} dz \, 
e^{-i(k_{p}-k_{sz}-k_{iz})z} \nonumber \\ 
h_{tr}(\vec{\kappa}_1 + \vec{\kappa}_2) \hspace*{-2mm} & = & \hspace*{-2mm} \int_{A} d \vec{\rho} \, 
\tilde{h}_{tr}(\vec{\rho}) \, e^{-i(\vec{\kappa}_s + \vec{\kappa}_i) \cdot \vec{\rho}} \\
\Delta_{z} \hspace*{-2mm} & = & \hspace*{-1mm} k_{p}-k_{sz}-k_{iz}  \nonumber
\end{eqnarray}
where $\epsilon$ is named as the parametric gain index. $\epsilon$ is proportional 
to the second order electric susceptibility $\chi^{(2)}$ and is usually treated as a constant,
$L$ is the length of the nonlinear interaction, the integral in $\vec{\kappa}$ is evaluated over 
the cross section $A$ of the nonlinear material illuminated by the pump, $\vec{\rho}$ is 
the transverse coordinate vector,  $\vec{\kappa}_j$ (with $j=s,i$) is the transverse 
wavevector of the signal and idler, and $f(|\,\vec{\rho}\,|)$ is the transverse profile of the pump, 
which can be treated as a Gaussion in most of the experimental conditions.  The functions 
$f(\Delta_{z}L)$ and  $h_{tr}(\vec{\kappa}_1 + \vec{\kappa}_2)$ turn to $\delta$-functions for 
an infinitely long 
($L \sim \infty$) and wide ($A \sim \infty$) nonlinear interaction region.  The reason we have 
chosen the form of Eq.~(\ref{spdc_F}) is to separate the ``longitudinal" and the ``transverse" 
correlations.  We will show that $\delta(\omega_p-\omega_s-\omega_i)$ and 
$f(\Delta_{z}L)$ together can be rewritten as a function of $\omega_{s}-\omega_{i}$.  
To simplify the mathematics, we assume near co-linearly SPDC.   
In this situation, $|\, \vec{\kappa}_{s,i} \,| \ll |\, \mathbf{k}_{s,i} \,|$.  

\vspace{5mm} 

Basically, the function $f(\Delta_{z}L)$ determines the ``longitudinal" space-time correlation.
Finding the solution of the integral is straightforward:
\begin{eqnarray}
f(\Delta_{z}L) = \int_{0}^{L} dz \, e^{-i(k_{p}-k_{sz}-k_{iz})z} 
= e^{-i\Delta_{z}L/2} \, sinc(\Delta_{z}L/2). 
\end{eqnarray}

Now, we consider $f(\Delta_{z}L)$ with $ \delta(\omega_p-\omega_s-\omega_i)$ together,
and taking advantage of the $\delta$-function in frequencies by introducing a detuning 
frequency $\Omega$ to evaluate function $f(\Delta_{z}L)$:
\begin{eqnarray}
	\omega_{s} & = & \omega_{s}^{0} + \Omega \nonumber  \\
	\omega_{i} & = & \omega_{i}^{0} - \Omega \label{nu}  \\
	\omega_{p} & = & \omega_s + \omega_i =  \omega_{s}^{0} +  \omega_{i}^{0}.  \nonumber  \\
	 \Omega & = & (\omega_{s} - \omega_{i})/2. \nonumber
\end{eqnarray}
The dispersion relation $k(\omega)$ allows us to express the wave numbers 
through the frequency detuning $\Omega$:
\begin{eqnarray}
k_{s} & \approx & k(\omega_{s}^{0} ) + \Omega \, \frac{dk}{d\omega} \, \Big\vert_{\omega_{s}^{0} }
	= k(\omega_{s}^{0} ) + \frac{\Omega}{u_{s}},\nonumber\\
k_{i} & \approx & k(\omega_{i}^{0} ) - \Omega \, \frac{dk}{d\omega} \, \Big\vert_{\omega_{i}^{0} }
	= k(\omega_{i}^{0} ) - \frac{\Omega}{u_{i}}
	\label{k(nu)}
\end{eqnarray}
where $u_{s}$ and $u_{i}$ are group velocities for the signal and the idler, respectively.
Now, we connect $\Delta_{z}$ with the  detuning frequency $\Omega$:  
\begin{eqnarray}
\Delta_{z}&=& k_{p}-k_{sz}-k_{iz} \nonumber\\
&=&k_{p}-\sqrt{(k_{s})^{2}-(\vec{\kappa}_{s})^{2}}-\sqrt{(k_{i})^{2}-(\vec{\kappa}_{i})^{2}}
\nonumber\\
&\cong &k_{p}-k_{s}-k_{i}+\frac{(\vec{\kappa}_{s})^{2}}{2k_{s}}+\frac{(\vec{\kappa}_{i})^{2}}{2k_{i}} 
\label{Delta1} \\
&\cong &k_{p}-k(\omega_{s}^{0} )-k(\omega_{i}^{0} )+\frac{\Omega}{u_{s}}-\frac{\Omega}{u_{i}}+
\frac{(\vec{\kappa}_{s})^{2}}{2k_{s}}+\frac{(\vec{\kappa}_{i})^{2}}{2k_{i}} 
\nonumber\\
&\cong&D\Omega \nonumber
\end{eqnarray}
where $D\equiv 1/u_{s} - 1/u_{i}$. We have also applied 
$k_{p}-k(\omega_{s}^{0} )-k(\omega_{i}^{0})=0$ and 
$|\, \vec{\kappa}_{s,i}\,| \ll |\, \mathbf{k}_{s,i} \,|$.  The ``longitudinal"
wavevector correlation function is rewritten as a function of the detuning frequency
$\Omega = (\omega_{s} - \omega_{i})/2$: $f(\Delta_{z}L) \cong f(\Omega DL)$.  
In addition to the above approximations, we have inexplicitly 
assumed the angular independence of the wavevector $k=n(\theta)\omega/c$.  
For type II SPDC, the refraction index of the extraordinary-ray depends on the angle 
between the wavevector and the optical axis and an additional term appears in the expansion.  
Making the approximation valid, we have restricted our calculation to a near-collinear process.  
Thus, for a good approximation, in the near-collinear experimental setup
\begin{equation}
\Delta_{z}L \cong \Omega DL =  (\omega_{s} - \omega_{i})DL/2.	
	\label{Deltacol}
\end{equation}

Type-I degenerate SPDC is a special case.  Due to the fact that $u_{s}=u_{i}$, and hence, 
$D=0$, the expansion of $k(\omega)$ should be carried out up to the 
second order.  Instead of (\ref{Deltacol}), we have
\begin{equation}
\Delta_{z}L \cong -\Omega^{2} D'L = -(\omega_{s} - \omega_{i})^{2} D'L/4
	\label{Deltacol1}
\end{equation}
where 
$$
D' \equiv\frac{d}{d\omega}(\frac{1}{u}) \, \Big\vert_{\omega^{0} }.
$$

\vspace{3mm}

\hspace{-6.5mm}The two-photon state of the signal-idler pair is then approximated as
\begin{equation}\label{spdc-22}
| \, \Psi \, \rangle=\int d \Omega \, d\vec{\kappa}_s \, d\vec{\kappa}_i \,
f(\Omega) \, h_{tr}(\vec{\kappa}_s + \vec{\kappa}_i) \, 
a_s^{\dag}(\omega^{0}_{s}+\Omega, \vec{\kappa}_{s}) \, 
a_i^{\dag}(\omega^{0}_{i}-\Omega, \vec{\kappa}_{i}) |\, 0 \, \rangle
\end{equation}
where the normalization constant has been absorbed into $f(\Omega)$.

\section{Correlation measurement of entangled state}

\hspace{6.5mm} EPR state is a pure state which characterizes the behavior of a pair 
of entangled particles.  In principle, one EPR pair contains 
all information of the correlation.  A question naturally arises: 
Can we then observe the EPR correlation from the measurement of 
one EPR pair?   The answer is no.  Generally speaking, we may never learn any meaningful 
physics from the measurement of one particle or one pair of particles.  To learn the 
correlation, an ensemble measurement of a large number of \textit{identical} pairs are 
necessary, where ``identical" means that all pairs which are involved in the ensemble 
measurement must be prepared in the same state, except for an overall phase factor.   
This is a basic requirement of quantum measurement theory.   

Correlation measurements are typically statistical and involve a large 
number of measurements of individual quanta.  Quantum mechanics does not 
predict a precise outcome for a measurement. Rather, quantum mechanics 
predicts the probabilities for certain 
outcomes.  In photon counting measurements, the outcome of a measurement is either 
a ``\textit{yes}" (a count or a ``click") or a ``\textit{no}" (no count).  In a joint measurement 
of two photon counting detectors, the outcome of ``\textit{yes}" means a ``\textit{yes}-\textit{yes}" 
or a ``click-click" joint registration.  If the outcome of a joint measurement shows $100\%$ 
``\textit{yes}" for a certain set of values of a physical observable or a certain relationship between 
physical variables, the measured quantum system is correlated in that observable. 
As a good example, EPR's \textit{gedankenexperiment} suggested to us a 
system of quanta with perfect correlation $\delta(x_1 -x_2 + x_0)$ in position.   To examine 
the EPR correlation, we need to have a $100\%$ ``\textit{yes}" when the 
positions of the two distant detectors satisfy $x_1 -x_2 =x_0$, and $100\%$ ``\textit{no}" 
otherwise, when $x_1 -x_2 \neq x_0$.    
To show this experimentally, a realistic 
approach is to measure the correlation function of $| f(x_1-x_2) |^{2}$ by 
observing the joint detection counting rates of $R_{1,2} \propto | f(x_1-x_2) |^{2}$ while 
scanning all possible values of $x_1- x_2$.  In quantum optics, this means the
measurement of the second-order correlation function, or 
$G^{(2)}(\mathbf{r}_1, t_1; \mathbf{r}_2, t_2)$, in the form of longitudinal correlation
$G^{(2)}(\tau_1-\tau_2)$ and/or transverse correlation $G^{(2)}(\vec{\rho}_1- \vec{\rho}_2)$, 
where $\tau_j=t_j-z_j/c$, $j=1,2$, and $\vec{\rho}_j$ is the transverse coordinate of the 
$jth$ point-like photon counting detector.  

\vspace{6mm}

Now, we study the two-photon correlation of the entangled photon pair of SPDC.   
The probability of jointly detecting the signal and idler at space-time points
$({\bf r}_{1}, t_{1})$ and $({\bf r}_{2}, t_{2})$ is given by the Glauber theory \cite{Glauber}:
\begin{equation}
G^{(2)}({\bf r}_{1}, t_{1};  {\bf r}_{2}, t_{2}) = \langle \, E^{(-)}({\bf r}_{1}, t_{1}) 
E^{(-)}({\bf r}_{2}, t_{2}) E^{(+)}({\bf r}_{2}, t_{2}) E^{(+)}({\bf r}_{1}, t_{1}) \, \rangle
\end{equation}
where $E^{(-)}$ and $E^{(+)}$ are the negative-frequency and the
positive-frequency field operators of the detection events at
space-time points $({\bf r}_{1}, t_{1})$ and $({\bf r}_{2}, t_{2})$.  
The expectation value of the joint detection operator is calculated 
by averaging over the quantum states of the signal-idler photon pair.  
For the two-photon state of SPDC, 
\begin{eqnarray}
G^{(2)}({\bf r}_{1}, t_{1};  {\bf r}_{2}, t_{2}) =
|\, \langle \, 0 \,|\, E^{(+)}({\bf r}_{2}, t_{2}) E^{(+)}({\bf r}_{1}, t_{1})\,|\, \Psi \, \rangle\,|^{2}
=  | \, \psi({\bf r}_{1}, t_{1};  {\bf r}_{2}, t_{2}) \, |^{2}
\end{eqnarray}
where $|\, \Psi \, \rangle$ is the two-photon state, and $\Psi({\bf r}_{1}, t_{1};  {\bf r}_{2}, t_{2})$ 
is named the effective two-photon wavefunction.  To evaluate 
$G^{(2)}({\bf r}_{1}, t_{1};  {\bf r}_{2}, t_{2})$ and $\psi({\bf r}_{1}, t_{1};  {\bf r}_{2}, t_{2})$,
we need to propagate the field operators from the two-photon source to space-time points
$({\bf r}_{1}, t_{1})$ and  $({\bf r}_{2}, t_{2})$.

In general, the field operator $E^{(+)}({\bf r}, t)$ at space-time point $({\bf r}, t)$ can be written in
terms of the Green's function, which propagates a quantized mode from
space-time point $({\bf r}_0, t_0)$ to $({\bf r}, t)$ \cite{Green}\cite{goodman}:
\begin{equation}\label{Green-0}
E^{(+)}({\bf r}, t) = \sum_{\bf{k}} \, g({\bf k}, {\bf r}-{\bf r}_0, t-t_0) \,
E^{(+)}({\bf k}, {\bf r}_0, t_0).
\end{equation}
where $g({\bf k}, {\bf r}-{\bf r}_0, t-t_0)$ is the Green's function, which is also named 
the optical transfer function.  For a different experimental setup, 
$g({\bf k}, {\bf r}-{\bf r}_0, t-t_0)$ can be quite different. To simplify the
notation, we have assumed one polarization.  

\begin{figure}[hbt]
    \centering\vspace{-5mm}
    \includegraphics[width=100mm]{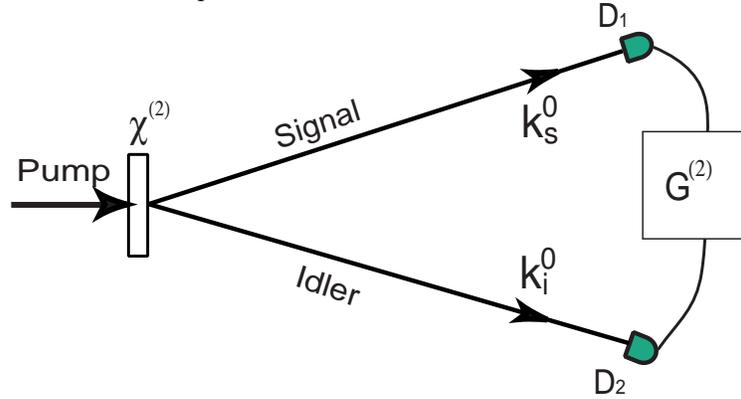} 
    \parbox{14cm}{ \vspace*{3mm}\caption{Collinear propagated signal-idler photon pair,
    either degenerate or non-degenerate, 
    are received by two distant point photo-detectors $D_{1}$ and $D_{2}$, 
    respectively, for longitudinal $G^{(2)}(\tau_1-\tau_2)$ and transverse 
     $G^{(2)}(\vec{\rho}_1-\vec{\rho}_2)$ correlation measurements.   To simplify the 
     mathematics, we assume paraxial approximation is applicable to the signal-idler fields.
     The $z_1$ and $z_2$ are chosen along the central wavevector $\mathbf{k}^{0}_s$ 
     and $\mathbf{k}^{0}_i$.} 
    \label{Quantum}}
\end{figure}

Considering an idealized simple experimental setup, shown in Fig.~\ref{Quantum}, in which 
collinear propagated signal and idler pairs are received by two point photon counting detectors 
$D_1$ and $D_2$, respectively, for longitudinal $G^{(2)}(\tau_1-\tau_2)$ and transverse 
$G^{(2)}(\vec{\rho}_1-\vec{\rho}_2)$ correlation measurements.  To simplify the mathematics,
we further assume paraxial experimental condition.  It is convenient, in the
discussion of longitudinal and transverse correlation measurements, to write the field 
$E^{(+)}({\bf r}_{j}, t_{j})$ in terms of its longitudinal and transversal space-time variables 
under the Fresnel paraxial approximation:
\begin{eqnarray}\label{gg-1}
&& E^{(+)}(\vec{\rho}_j, z_j, t_j) \\ \nonumber 
&\cong&  \int d\omega \, d\vec{\kappa}
\,\, g(\vec{\kappa}, \omega; \vec{\rho}_j, z_j) \, 
e^{-i\omega t_{j}}a(\omega, \vec{\kappa}) 
\cong \int d\omega \, d\vec{\kappa} \, 
\gamma(\vec{\kappa}, \omega; \vec{\rho}_j, z_j)\,
e^{-i\omega \tau_{j}}a(\omega, \vec{\kappa}) 
\end{eqnarray}
where $g(\vec{\kappa}, \omega; \vec{\rho}_j, z_j) = 
\gamma(\vec{\kappa}, \omega; \vec{\rho}_j, z_j) e^{i \omega z_j/c}$ 
is the spatial part of the Green's function, 
$\vec{\rho}_j$ and $z_j$ are the transverse and longitudinal coordinates of the $jth$ 
photo-detector and $\vec{\kappa}$ is the transverse wavevector.  We have chosen 
$z_0 =0$ and $t_0 = 0$ at the output plane of the SPDC.  For convenience, 
all constants associated with the field are absorbed into 
$g(\vec{\kappa}, \omega; \vec{\rho}_j, z_j)$.  

The two-photon effective wavefunction  
$\Psi(\vec{\rho}_{1}, z_1, t_1;  \vec{\rho}_{2}, z_2, t_2)$ is thus calculated as follows
\begin{eqnarray}\label{Wavefunction-L-T}
& & \Psi(\vec{\rho}_{1}, z_1, t_1;  \vec{\rho}_{2}, z_2, t_2) \nonumber \\
&=& \langle \, 0 \,| \int d\omega^{,} \, d\vec{\kappa}^{,}
\,g(\vec{\kappa}^{,}, \omega^{,}; \vec{\rho}_2, z_2) \, e^{-i\omega^{,} t_{2}} \, a(\omega^{,}, 
\vec{\kappa}^{,}) \nonumber \\ && \times \int d\omega^{,,} \, d\vec{\kappa}^{,,}
\,g(\vec{\kappa}^{,,}, \omega^{,,}; \vec{\rho}_1, z_1) \, e^{-i\omega^{,,} t_{1}} \, 
a(\omega^{,,}, \vec{\kappa}^{,,}) \nonumber \\
& & \times \int d \Omega \, d\vec{\kappa}_s \, d\vec{\kappa}_i \,
f(\Omega) \, h_{tr}(\vec{\kappa}_s +  \vec{\kappa}_i) \, 
a_s^{\dag}(\omega^{0}_{s}+\Omega,  \vec{\kappa}_{s}) \, 
a_i^{\dag}(\omega^{0}_{i}-\Omega,  \vec{\kappa}_{i}) |\, 0 \, \rangle \nonumber \\
&=&  \Psi_0 \, e^{-i (\omega^{0}_{s} \tau_{1}+\omega^{0}_{i} \tau_{2})} \nonumber \\
&& \times \int d\Omega \, d\vec{\kappa}_s \, d\vec{\kappa}_i \, f(\Omega) \, 
h_{tr}(\vec{\kappa}_s +  \vec{\kappa}_i) 
\, e^{-i \Omega (\tau_{1} -\tau_{2})}  
\gamma(\vec{\kappa}_s, \Omega; \vec{\rho}_1, z_1) \, 
\gamma(\vec{\kappa}_i, -\Omega; \vec{\rho}_2, z_2). 
\end{eqnarray}
Although Eq.~(\ref{Wavefunction-L-T}) cannot be factorized into a trivial product of
longitudinal and transverse integrals, it is not difficult to measure the temporal correlation
and the transverse correlation separately by choosing suitable experimental conditions.    

Experiments may be designed for measuring either temporal (longitudinal) 
or spatial (transverse) correlation only.  Thus, based on different experimental setups, 
we may simplify the calculation to either the temporal (longitudinal) part:  
\begin{eqnarray}\label{Wavefunction-LL}
\Psi(\tau_{1};  \tau_{2}) = \Psi_0 \,
e^{-i (\omega^{0}_{s} \tau_{1}+\omega^{0}_{i} \tau_{2})} \int d \Omega \, f(\Omega) \, 
e^{-i \Omega (\tau_{1} -\tau_{2})} = \Psi_0 \, 
e^{-i (\omega^{0}_{s} \tau_{1}+\omega^{0}_{i} \tau_{2})} 
\mathcal{F}_{\tau_{1} -\tau_{2}}\big\{ f(\Omega) \big\}
\end{eqnarray}
or the spatial part:
\begin{eqnarray}\label{Wavefunction-TT}
\Psi(\vec{\rho}_{1}, z_1; \vec{\rho}_{2}, z_2) = \Psi_0  
\int d\vec{\kappa}_s \, d\vec{\kappa}_i \, h_{tr}(\vec{\kappa}_s +  
\vec{\kappa}_i) \, g(\vec{\kappa}_s, \omega_s; \vec{\rho}_1, z_1) \, 
g(\vec{\kappa}_i, \omega_i; \vec{\rho}_2, z_2).
\end{eqnarray}
In Eq.~(\ref{Wavefunction-LL}), ${\mathcal F}_{\tau_{1}-\tau_{2}} \, \big\{ f(\Omega)\big\}$
is the Fourier transform of the spectrum amplitude function $f(\Omega)$.
In Eq.~(\ref{Wavefunction-TT}), we may treat $h_{tr}(\vec{\kappa}_s + \vec{\kappa}_i) \sim
\delta(\vec{\kappa}_s + \vec{\kappa}_i)$ by assuming certain experimental conditions.

\subsection*{Two-photon temporal correlation}

\begin{figure}[htb]
\centering 
     \vspace*{6mm}
    \includegraphics[width=70mm]{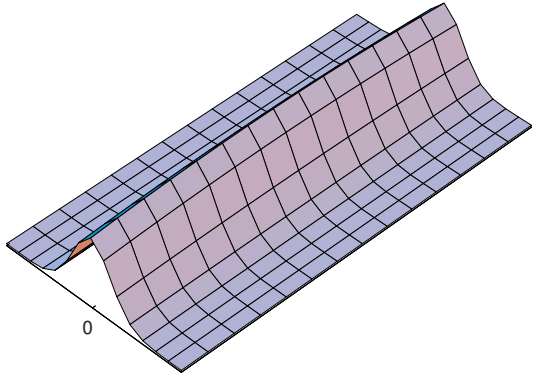}
    \parbox{14cm}{\caption{A schematic envelope of a two-photon 
    wavepacket with a Gaussian shape along $\tau_{1}-\tau_{2}$ corresponding to
    a Gaussian function of $ f(\Omega)$.   In the case of SPDC, the envelope is
    close to a $\delta$-function in $\tau_{1}-\tau_{2}$ corresponding to a broad-band 
    $f(\Omega)=$ constant.  The wavepacket 
    is uniformly distributed along $\tau_{1}+\tau_{2}$ due to the assumption of 
    $\omega_{p}=$ constant. } \label{wavepacket}}
\end{figure}

\hspace*{5.5mm} To measure the two-photon temporal correlation of 
SPDC, we select a pair of transverse wavevectors $\vec{\kappa}_{s} = - \vec{\kappa}_{i}$ 
in Eq.~(\ref{Wavefunction-L-T}) by using appropriate optical apertures.  The  effective 
two-photon wavefunction is thus simplified to that of Eq.~(\ref{Wavefunction-LL}) 
\begin{eqnarray}\label{2wavefunction-SPDC}
\Psi (\tau_{1}; \tau_{2})
&\cong& \Psi_0 \, e^{-i (\omega^{0}_{s} \tau_{1}+
\omega^{0}_{i} \tau_{2})} \int d \Omega \, f(\Omega) \, 
e^{-i \Omega (\tau_{1} -\tau_{2})} \\ \nonumber
&=& \big{[}\, \Psi_{0} \, e^{-\frac{i}{2} (\omega^{0}_{s} + \omega^{0}_{i}) (\tau_{1} + \tau_{2})} \, \big{]} \, 
\big{[} \, {\mathcal F}_{\tau_{1}-\tau_{2}} \, \big\{ f(\Omega)\big\} \,
e^{- \frac{i}{2} (\omega_{s}^{0}- \omega_{i}^{0})(\tau_{1} - \tau_{2})} \, \big{]}
\end{eqnarray}
where, again, ${\mathcal F}_{\tau_{1}-\tau_{2}} \, \big\{ f(\Omega)\big\}$
is the Fourier transform of the spectrum amplitude function $f(\Omega)$.
Eq.~(\ref{2wavefunction-SPDC}) indicates a 2-D wavepacket: a 
narrow envelope along the $\tau_{1}-\tau_{2}$ axis with constant amplitude along the 
$\tau_{1}+\tau_{2}$ axis.  In certain experimental conditions, the function 
$f(\Omega)$ of SPDC can be treated
as constant from $-\infty$ to $\infty$ and thus ${\mathcal F}_{\tau_{1}-\tau_{2}} 
\sim \delta(\tau_{1}-\tau_{2})$.   In this case, for fixed positions of $D_1$ and 
$D_2$, the 2-D wavepacket means the following: the signal-idler pair may be 
jointly detected at any time; however, if the signal is registered at a certain time $t_1$, 
the idler must be registered at a unique time of $t_2 \sim t_1 - (z_1 - z_2)/c$.    
In other words, although the joint detection of the pair may happen at any times of 
$t_1$ and $t_2$ with equal probability ($\Delta (t_1+ t_2) \sim \infty$), the
registration time difference of the pair must be a constant $\Delta (t_1- t_2) \sim 0$.  
A schematic of the two-photon wavepacket is shown in Fig.~\ref{wavepacket}.    
It is a non-factorizeable 2-D wavefunction indicating the entangled nature of the
two-photon state.
The longitudinal correlation function $G^{(2)}(\tau_1- \tau_2)$ is thus
$$
G^{(2)}(\tau_1- \tau_2) \propto | \, {\mathcal F}_{\tau_{1}-\tau_{2}} \, 
\big\{ f(\Omega)\big\} \, |^{2},
$$
which is a $\delta$-function-like function in the case of SPDC.  Thus, we have 
shown the entangled signal-idler photon pair of SPDC hold a typical EPR correlation
in energy and time:
\begin{eqnarray*}
& & \Delta(\omega_s+\omega_i)\sim 0 \,\, \,\, \& \,\, 
\Delta (t_1- t_2)\sim 0 \,\, \\ \nonumber
&\text{with} &  \Delta \omega_s \sim \infty, \hspace{2mm} 
\Delta \omega_i \sim \infty, \hspace{2mm}
\Delta t_1 \sim \infty, \hspace{2mm} \Delta t_2 \sim \infty.
\end{eqnarray*}

\vspace{8mm}

Now we examine a statistical model of SPDC for temporal correlation.  
As we have discussed earlier, realistic statistical models have been proposed 
to simulate the EPR two-particle state. 
Recall that for a mixed state in the form of
\begin{equation*}
\hat{\rho}=\sum_{j} P_{j} \, |\, \Psi _{j} \, \rangle \langle \, \Psi _{j} \, |
\end{equation*}     
where $P_{j}$ is the probability for specifying a given set of state 
vectors $|\, \Psi _{j} \, \rangle$,  
the second-order correlation function of fields $E(\mathbf{r}_{1}, t_{1})$ and 
$E(\mathbf{r}_{2}, t_{2})$ is given by
\begin{eqnarray*}
& &G^{(2)}(\mathbf{r}_{1}, t_{1}; \mathbf{r}_{2}, t_{2}) \nonumber \\ 
&=&Tr[ \, \hat{\rho} \, E^{(-)}(\mathbf{r}_{1}, t_{1})\, E^{(-)}(\mathbf{r}_{2}, t_{2})\, 
E^{(+)}(\mathbf{r}_{2}, t_{2}) \, E^{(+)}(\mathbf{r}_{1}, t_{1}) \, ] 
\nonumber \\ &=& \sum_{j} \, P_{j} \, \langle \, \Psi _{j}  \, | \, E^{(-)}(\mathbf{r}_{1}, t_{1})\, 
E^{(-)}(\mathbf{r}_{2}, t_{2})\, E^{(+)}(\mathbf{r}_{2}, t_{2}) \, E^{(+)}(\mathbf{r}_{1}, t_{1}) \, 
|\, \Psi _{j} \, \rangle \nonumber \\ 
&=& \sum_{j} P_{j} \, G^{(2)}_{j}(\mathbf{r}_{1}, t_{1}; 
\mathbf{r}_{2}, t_{2}), 
\end{eqnarray*}
which is a weighted sum over all individual contributions of $G^{(2)}_{j}$.   
Considering the following simplified version of Eq.~(\ref{statemix}) to
simulate the state of SPDC as a mixed state:
\begin{eqnarray}\label{2Mix-SPDC-1D}
\hat{\rho} =\int d\Omega \, | f(\Omega) |^{2} \, 
a^{\dag}(\omega_{s}^{0} +\Omega) \, a^{\dag}(\omega_{i}^{0}-\Omega) 
| \, 0 \, \rangle \langle \, 0 \, | a(\omega_{i}^{0} -\Omega) \, a(\omega_{s}^{0}+\Omega),
\end{eqnarray}
with 
\begin{equation}
|\, \Psi _{\Omega} \, \rangle = a^{\dag}(\omega_{s}^{0} +\Omega) \, 
a^{\dag}(\omega_{i}^{0}-\Omega) | \, 0 \, \rangle, \hspace{3mm} 
P_{j} = d\Omega \, | f(\Omega) |^{2}.
\end{equation}
It is easy to find $G_\Omega^{(2)}(\tau_{1}-\tau_{2})=$ 
constant, and thus $G^{(2)}(\tau_{1}-\tau_{2})=$ constant.   
This means that the uncertainty of the measurement on $t_{1}-t_{2}$ for the mixed state of 
Eq.~(\ref{2Mix-SPDC-1D}) is infinite: $\Delta(t_{1}-t_{2}) \sim \infty$. 
Although the energy (frequency) or momentum (wavevector) for each photon may be defined 
with constant values pair by pair,  the corresponding temporal correlation 
measurement of the ensemble can never achieve a $\delta$-function-like relationship.  
In fact, the correlation is undefined, i.e., taking an infinite uncertainty.  
Thus, the statistical model of SPDC cannot satisfy the EPR inequalities of 
Eq.~(\ref{QMineq}).

\subsection*{Two-photon spatial correlation}

\hspace*{5.5mm} Similar to that of the two-photon temporal correlation, as an example, 
we analyze the effective two-photon wavefunction of the signal-idler pair of SPDC. 
To emphasize the spatial part of the two-photon correlation, we choose a pair of 
frequencies $\omega_s$ and $\omega_i$ with $\omega_s + \omega_i = \omega_p$.  
In this case, the effective two-photon wavefunction of Eq.~(\ref{Wavefunction-L-T}) 
is simplified to that of Eq.~(\ref{Wavefunction-TT})
$$
\Psi(\vec{\rho}_{1}, z_1; \vec{\rho}_{2}, z_2) = \Psi_0 
\int d\vec{\kappa}_s \, d\vec{\kappa}_i \, \delta(\vec{\kappa}_s +  
\vec{\kappa}_i) \, g(\vec{\kappa}_{s}, \omega_s, \vec{\rho}_{1}, z_1) \, 
g(\vec{\kappa}_{i}, \omega_i, \vec{\rho}_{2}, z_2)
$$
where we have assumed $h_{tr}(\vec{\kappa}_s +  \vec{\kappa}_i) \sim 
\delta(\vec{\kappa}_s + \vec{\kappa}_i)$, which is reasonable by assuming
a large enough transverse cross-session laser beam of pump.

We now design a simple joint detection measurement between two point photon 
counting detectors $D_1$ and $D_2$ located at $(\vec{\rho}_{1}, z_1)$ and 
$(\vec{\rho}_{2}, z_2)$, respectively, for the detection of the signal 
and idler photons.   We have assumed that the two-photon source has 
a finite but large transverse dimension.  Under this simple experimental 
setup, the Green's function, or the optical transfer function describing 
arm-$j$, $j=1,2$, in which the signal and the idler freely propagate to 
photo-detector $D_1$ and $D_2$, respectively, is given by Eq.~(\ref{gg-final})
of the Appendix.
Substitute the $g_{j}(\omega, \vec{\kappa}; z_j, \vec{\rho}_j)$, $j=1,2$, into
Eq.~(\ref{Wavefunction-TT}), the effective wavefunction is then given by
\begin{eqnarray}\label{WF-10}
&& \Psi(\vec{\rho}_{1}, z_1; \vec{\rho}_{2}, z_2) \\ \nonumber 
&=& \Psi_0 \int d\vec{\kappa}_s \, d\vec{\kappa}_i \, \delta(\vec{\kappa}_s +  
\vec{\kappa}_i) 
\big{(} \frac{-i \omega_s}{ 2 \pi c z_1} \, e^{i \frac{\omega_s}{c} z_1} \big{)} 
\, \big{(} \frac{-i \omega_i}{ 2 \pi c z_2} \, e^{i \frac{\omega_i}{c} z_2} \big{)}
\\ \nonumber 
& & \times \int_{A} \, d\vec{\rho}_s \,  d\vec{\rho}_i \, 
G(|\vec{\rho}_1 - \vec{\rho}_s|, \frac{\omega_s}{c z_1}) \, 
e^{i \vec{\kappa}_s \cdot \vec{\rho}_s} \,
G(|\vec{\rho}_2 - \vec{\rho}_i|, \frac{\omega_i}{c z_2}) \,
e^{i \vec{\kappa}_i \cdot \vec{\rho}_i}
\end{eqnarray}
where $\vec{\rho}_s$ ($\vec{\kappa}_s$) and $\vec{\rho}_i$ ($\vec{\kappa}_i$) 
are the transverse coordinates (wavevectors) for the signal and the idler fields,
respectively, defined on the output plane of the two-photon source.   The integral
of $d\vec{\rho}_s$ and $ d\vec{\rho}_i$ is over area $A$, which is determined by 
the transverse dimension of the nonlinear interaction.   The Gaussian function
$G(|\vec{\alpha}|, \beta) = e^{i (\beta/2) |\vec{\alpha}|^2}$ represents the Fresnel 
phase factor that is defined in the Appendix. 
The integral of $d\vec{\kappa}_s$ and $d\vec{\kappa}_i$ can be evaluated easily 
with the help of the EPR type two-phonon transverse wavevector distribution function 
$\delta(\vec{\kappa}_s + \vec{\kappa}_i)$:
\begin{eqnarray}\label{WF-11}
\int d\vec{\kappa}_s \, d\vec{\kappa}_i \, \delta(\vec{\kappa}_s +  
\vec{\kappa}_i) \, e^{i \vec{\kappa}_s \cdot \vec{\rho}_s} \,
e^{i \vec{\kappa}_i \cdot \vec{\rho}_i} \sim \delta(\vec{\rho}_s -  
\vec{\rho}_i).
\end{eqnarray}

Thus, we have shown that the entangled signal-idler photon pair of SPDC 
holds a typical EPR correlation in transverse momentum and position
while the correlation measurement is on the output plane of the two-photon source, 
which is very close to the original proposal of EPR:
\begin{eqnarray*}
& & \Delta(\vec{\kappa}_s+\vec{\kappa}_i)\sim 0 \,\,  \,\, \& \,\, 
\Delta (\vec{\rho}_s- \vec{\rho}_i)\sim 0 \,\,  \\ \nonumber
&\text{with} &  \Delta \vec{\kappa}_s \sim \infty, \hspace{2mm} 
\Delta \vec{\kappa}_i \sim \infty, \hspace{2mm}
\Delta \vec{\rho}_s \sim \infty, \hspace{2mm} \Delta \vec{\rho}_i \sim \infty.
\end{eqnarray*}
In EPR's language, we may never know where the signal photon and the idler 
photon are emitted from the output plane of the source.  However, 
if the signal (idler) is found at a certain position, the idler (signal) must be 
observed at a corresponding unique position.   The signal and the idler 
may have also any transverse momentum. However, if the transverse momentum 
of the signal (idler) is measured at a certain value in a certain direction, the 
idler (signal) must be of equal value but pointed to a certain opposite direction.  
In \emph{collinear} SPDC,
the signal-idler pair is always emitted from the same point in the output plane 
of the two-photon source, $\vec{\rho}_s=\vec{\rho}_i$, and if one of them 
propagates slightly off from the collinear axes, the other one must propagate 
to the opposite direction with $\vec{\kappa}_s=-\vec{\kappa}_i$.    

The interaction of spontaneous parametric down-conversion is nevertheless a local 
phenomenon.  The nonlinear interaction coherently creates mode-pairs 
that satisfy the phase matching conditions of Eq.~(\ref{eq:phsmtch}) which are
also named as energy and momentum conservation.  The signal-idler 
photon pair can be excited to any of these coupled modes or in all of these coupled 
modes simultaneously, resulting in a particular two-photon superposition.  
It is this superposition among those particular ``selected" two-photon states which 
allows the signal-idler
pair to come out from the same point of the source and propagate to opposite directions
with $\vec{\kappa}_s=-\vec{\kappa}_i$.

The two-photon superposition becomes more interesting when the signal-idler is 
separated and propagated to a large distance, either by free propagation or guided by optical 
components such as a lens.  A classical picture would consider the signal photon and 
the idler photon independent whenever the pair is released from 
the two-photon source because there is no interaction 
between the distant photons in free space.  Therefore, the signal photon and the idler photon 
should have independent and random distributions in terms of their transverse position 
$\vec{\rho}_{1}$ and $\vec{\rho}_{2}$.  This classical picture, however, is incorrect.  It is
found that the signal-idler two-photon system would not lose its entangled nature 
in the transverse position. This interesting behavior has been experimentally observed in 
quantum imaging by means of an EPR type correlation in transverse position.
The sub-diffraction limit spatial resolution observed in the ``quantum lithography"
experiment and the nonlocal correlation observed in the ``ghost imaging" experiment
are both the results of this peculiar superposition among those ``selected" two-photon 
amplitudes, namely that of two-photon superposition, corresponding to 
different yet indistinguishable alternative ways of triggering a joint photo-electron 
event at a distance. Two-photon superposition 
does occur in a distant joint detection event of a signal-idler photon pair.
There is no surprise that one has difficulties facing this phenomenon. 
The two-photon superposition is a nonlocal concept in this case. 
There is no counterpart for such a concept in classical theory and it may never be 
understood classically.    

Now we consider propagating the signal-idler pair away from the source to 
$(\vec{\rho}_{1}, z_1)$ and $(\vec{\rho}_{2}, z_2)$, respectively, and 
taking the result of Eq.~(\ref{WF-11}), i.e., $\vec{\rho}_s = \vec{\rho}_i = \vec{\rho}_0$ 
on the output plane of the SPDC source, 
the effective two-photon wavefunction becomes
\begin{eqnarray}\label{WF-20}
&& \Psi(\vec{\rho}_{1}, z_1; \vec{\rho}_{2}, z_2) \\ \nonumber 
&=& - \frac{\omega_s \, \omega_i}{ (2 \pi c)^2 z_1 z_2} \ 
e^{i (\frac{\omega_s}{c} z_1 + \frac{\omega_i}{c} z_2)}  
\int_{A} \, d\vec{\rho}_0 \, 
G(|\vec{\rho}_1 - \vec{\rho}_0|, \frac{\omega_s}{c z_1}) \, 
G(|\vec{\rho}_2 - \vec{\rho}_0|, \frac{\omega_i}{c z_2})
\end{eqnarray}
where $\vec{\rho}_0$ is defined on the output plane of the two-photon source.
Eq.~(\ref{WF-20}) indicates that the propagation-diffraction of the signal and the 
idler cannot be considered as independent.  The signal-idler photon pair are 
created and diffracted together in a peculiar entangled manner.  This point turns out to be
both interesting and useful when the two photodetectors coincided, or are replaced by 
a two-photon sensitive material. Taking $z_1 = z_2$ and $\vec{\rho}_1 = \vec{\rho}_2$,
Eq.~(\ref{WF-20}) becomes 
\begin{eqnarray}\label{WF-22}
\Psi(\vec{\rho}, z; \vec{\rho}, z) =
 - \frac{\omega_s \, \omega_i}{ (2 \pi c z)^2} \ 
e^{i (\frac{\omega_p}{c} z)}  \int_{A} \, d\vec{\rho}_0 \, 
G(|\vec{\rho} - \vec{\rho}_0|, \frac{\omega_p}{c z})
\end{eqnarray}
where $\omega_p$ is the pump frequency, which means that the signal-idler
pair is diffracted as if they have twice the frequency or half the wavelength.  
This effect is named as ``two-photon diffraction".   
This effect is useful for enhancing the spatial resolution of imaging.

\section{Quantum imaging}

\hspace{5.5mm} Although questions regarding fundamental issues of quantum
theory still exist, quantum entanglement has started to play important
roles in practical engineering applications.  Quantum imaging is
one of these exciting areas \cite{IEEE07}. Taking advantage of entangled states, 
Quantum imaging has so far demonstrated two peculiar features: 
(1) enhancing the spatial resolution of imaging beyond the diffraction limit, and
(2) reproducing ghost images in a ``nonlocal" manner. 
Both the apparent ``violation" of the uncertainty principle and the ``nonlocal" 
behavior of the momentnm-momentum position-position  correlation are due 
to the two-photon coherent effect of entangled 
states, which involves the superposition of two-photon amplitudes, a nonclassical 
entity corresponding to different yet indistinguishable alternative ways of triggering 
a joint-detection event in the quantum theory of photodetection.  
In this section, we will focus our discussion on the physics of imaging resolution 
enhancement.  The nonlocal phenomenon of ghost imaging will be discussed in 
the following section.  

\begin{figure}[htb]
\centering
    \includegraphics[width=82mm]{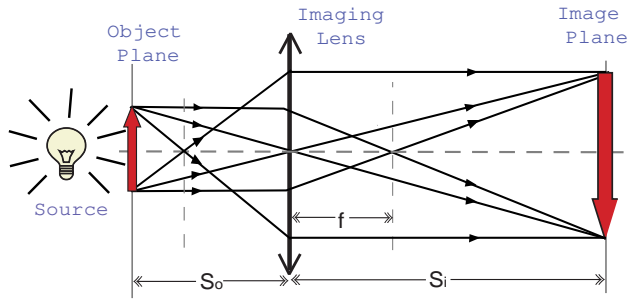}
     \parbox{14cm}{\vspace{2mm}\caption{A lens produces an
\textit{image} of an object in the plane defined by the Gaussian
thin lens equation $1/s_i+1/s_o=1/f$. The concept of an image is
based on the existence of a point-to-point relationship between
the object plane and the image plane.}\label{fig:Projection-1}}
\end{figure}

The concept of imaging is well defined in classical optics. 
Fig.~\ref{fig:Projection-1} schematically illustrates a
standard imaging setup.   
A lens of finite size is used to image
the object onto an image plane which is defined by 
the ``Gaussian thin lens equation"
\begin{equation}\label{Lens-Eq}
\frac{1}{s_i}+\frac{1}{s_o}=\frac{1}{f}
\end{equation}
where $s_o$ is the distance between 
object and lens, $f$ is the
focal length of the lens, and $s_i$ is the distance between the lens
and image plane.  If light always follows the laws
of geometrical optics, the image plane and the object plane would
have a perfect point-to-point correspondence, which means a
perfect image of the object, either magnified or demagnified. 
Mathematically, a perfect image is the result of a convolution 
of the object distribution function $f(\vec{\rho}_{o})$ and a 
$\delta$-function.  The $\delta$-function characterizes the perfect point-to-point 
relationship between the object plane and the image plane: 
\begin{eqnarray}\label{Image-Eq}
F(\vec{\rho}_i) =
\int_{obj} d\vec{\rho}_{o} \, f(\vec{\rho}_{o}) \,
\delta(\vec{\rho}_{o} +  \frac{\vec{\rho}_{i}}{m}) = f(\vec{\rho}_{o}) \otimes 
\delta(\vec{\rho}_{o} + \frac{\vec{\rho}_{i}}{m})
\end{eqnarray}
where $\vec{\rho}_{o}$ and $\vec{\rho}_{i}$ are 2-D vectors of the 
transverse coordinate in the object plane and the image plane,
respectively, and $m$ is the magnification factor.  The symbol $\otimes$
means convolution.

Unfortunately, light behaves like a wave. The diffraction effect turns the
point-to-point correspondence into a point-to-``spot"
relationship.  The $\delta$-function in the convolution of 
Eq.~(\ref{Image-Eq}) will be replaced by a point-spread function.  
\begin{eqnarray}\label{Image-Eq-2}
F(\vec{\rho}_i) = \int_{obj} d\vec{\rho}_{o} \, f(\vec{\rho}_{o}) \,
somb\big{[} \frac{R}{s_o}\,\frac{\omega}{c} \big{|}\vec{\rho}_{o} + 
\frac{\vec{\rho}_{i}}{m} \big{|} \big{]} 
= f(\vec{\rho}_{o}) \otimes
somb\big{[} \frac{R}{s_o}\,\frac{\omega}{c} \big{|}\vec{\rho}_{o} + 
\frac{\vec{\rho}_{i}}{m} \big{|} \big{]}
\end{eqnarray}
where 
$$
somb(x) = \frac{2J_1(x)}{x},
$$
and $J_1(x)$ is the first-order Bessel function, $R$ is the radius of the imaging lens. 
$R/s_o$ is named as the numerical aperture of the imaging system. 
The finite size of the spot, which is defined by 
the point-spread function, determines the spatial resolution of
the imaging setup, and thus, limits the ability of making
demagnified images.  It is clear from Eq.~(\ref{Image-Eq-2}), the use of 
a larger imaging lens and shorter wavelength light of source 
will result in a narrower point-spead function.  To improve the spatial resolution, 
one of the efforts in the lithography industry is the use of shorter
wavelengths.  This effort is, however, limited to a certain level
because of the inability of lenses to effectively work beyond a certain
``cutoff" wavelength.  

\begin{figure}[htb]
\centering
    \includegraphics[width=85mm]{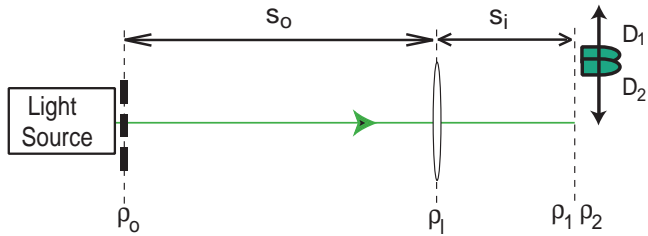}
     \parbox{14cm}{\vspace{2mm}\caption{Typical imaging setup. 
     A lens of finite size is used to produce a 
demagnified image of a object with limited spatial resolution. Replacing 
classical light with an entangled N-photon system, the spatial resolution can be 
improved by a factor of N, despite the Rayleigh diffraction limit.}
\label{fig:lithography-1}}
\end{figure}

Eq.~(\ref{Image-Eq-2}) imposes a diffraction
limited spatial resolution on an imaging system while the aperture size
of the imaging system and the wavelength of the light source 
are both fixed.  This limit is fundamental in both classical
optics and in quantum mechanics.  Any violation would be considered 
as a violation of the uncertainty principle.  

Surprisingly, the use of quantum entangled states gives a different result: 
by replacing classical light sources in
Fig.~\ref{fig:lithography-1} with entangled N-photon states, the
spatial resolution of the image can be improved by a factor of N, 
despite the Rayleigh diffraction limit.  Is this a violation of the uncertainty
principle?  The answer is no!  The uncertainty relation for an entangled N-particle
system is radically different from that of N independent particles.  In terms of the 
terminology of imaging, what we have found is that the $somb(x)$ in the 
convolution of Eq.~(\ref{Image-Eq-2}) has a different
form in the case of an entangled state.  For example, an entangled two-photon 
system has 
$$
x = \frac{R}{s_o}\,\frac{2 \omega}{c} \big{|}\vec{\rho}_{o} + \frac{\vec{\rho}_{i}}{m} \big{|}.
$$
Comparing with Eq.~(\ref{Image-Eq-2}), the factor of $2\omega$ yields a  
point-spread function half the width of that from Eq.~(\ref{Image-Eq-2}) and 
results in a doubling spatial resolution for imaging.    

\begin{figure}[htb]
\centering
    \includegraphics[width=72mm]{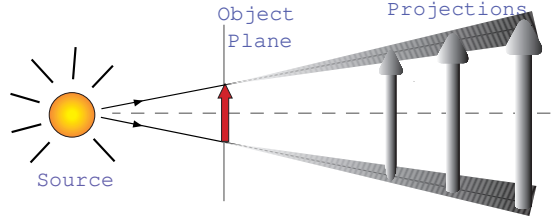}
     \parbox{14cm}{\vspace{2mm}\caption{Projection: a light source
illuminates an object and no image forming system is present, no
image plane is defined, and only projections, or shadows, of the
object can be observed.}\label{fig:Projection-2}}
\end{figure}
It should be further emphasized that one must not confuse a ``projection" 
with an image.  A projection is the shadow of an object, which is obviously 
different from the image of an object.  Fig.~\ref{fig:Projection-2} distinguishes
a projection shadow from an image.  In a projection, the object-shadow 
correspondence is essentially a ``momentum" correspondence, which is 
defined only by the propagation direction of the light rays.  

\vspace{6mm} 

We now analyze classical imaging.
The analysis starts with the propagation of the field from
the object plane to the image plane. In classical optics, such
propagation is described by an optical transfer function
$h(\mathbf{r}-\mathbf{r}_0, t-t_0)$, which accounts for 
the propagation of all modes of the field. 
To be consistent with quantum optics calculations, we prefer to 
work with the single-mode propagator 
$g(\mathbf{k}, \mathbf{r}-\mathbf{r}_0, t-t_0)$,  and to
write the field $E(\mathbf{r}, t)$ in terms of its longitudinal ($z$)
and transverse ($\vec{\rho}$) coordinates under the Fresnel
paraxial approximation:
\begin{eqnarray}\label{e-g}
E(\vec{\rho}, z, t) =  \int d\omega \, d\vec{\kappa} \,\, \tilde{E}(\vec{\kappa}, \omega) \,
g(\vec{\kappa}, \omega; \vec{\rho}, z) \, e^{-i\omega t}
\end{eqnarray}
where $\tilde{E}(\omega, \vec{\kappa})$ is the complex amplitude
of frequency $\omega$ and
transverse wave-vector $\vec{\kappa}$.  In Eq.~(\ref{e-g}) we have
taken $z_0=0$ and $t_0 = 0$ at the object plane as usual.
To simplify the notation, we have assumed one polarization.

Based on the experimental setup of Fig.~\ref{fig:lithography-1},
$g(\vec{\kappa}, \omega; \vec{\rho}, z)$ is found to be
\begin{eqnarray}\label{g-1}
& & g(\vec{\kappa}, \omega; \vec{\rho}_i, s_o+s_i) \nonumber \\
&=&  \int_{obj} d\vec{\rho}_{o} \int_{lens} d\vec{\rho}_l \, \Big\{ A(\vec{\rho}_{o}) \,
e^{i \vec{\kappa} \cdot \vec{\rho}_{o}} \Big\} \, \Big\{\frac{-i \omega}{2 \pi c} \, 
\frac{e^{i \frac{\omega}{c} s_o} }{s_o}\, G(|\vec{\rho}_{l}-\vec{\rho}_{o}|, \frac{\omega}{c s_o}) 
\Big\} \nonumber \\ & &  \times \, 
\,\, \Big\{ G(| \vec{\rho}_l |, -\frac{\omega}{c f}) \Big\}\,
\Big\{ \frac{-i \omega}{2\pi c } \, 
\frac{e^{i \frac{\omega}{c} s_i}}{s_i}\, G(|\vec{\rho}_i-\vec{\rho}_l |, \frac{\omega}{2 c s_i}) \Big\}
\end{eqnarray}
where $\vec{\rho}_{o}$, $\vec{\rho}_l$, and $\vec{\rho}_i$ are
two-dimensional vectors defined, respectively, on the object, the lens, and the image planes.
The first curly bracket includes the object-aperture function $A(\vec{\rho}_{o})$  
and the phase factor $e^{i \vec{\kappa} \cdot \vec{\rho}_{o}}$ contributed 
to the object plane by each transverse mode $\vec{\kappa}$.   Here we have
assumed a far-field finite size source. Thus, a phase factor 
$e^{i \vec{\kappa} \cdot \vec{\rho}_{o}}$ appears on the object plane of
$z=0$.  If a collimated laser beam is used, this phase factor
turns out to be a constant.  The terms in the second and the fourth curly brackets describe 
free-space Fresnel propagation-diffraction from the source/object plane to the 
imaging lens, and from the imaging lens to the detection plane, respectively.
The Fresnel propagator includes a spherical wave
function $e^{i \frac{\omega}{c} (z_j -z_k)} / (z_j -z_k)$ and a Fresnel phase factor 
$G(|\vec{\alpha}|, \beta) = e^{i (\beta/2) |\vec{\alpha}|^2} =
e^{i \omega |\vec{\rho}_{j}-\vec{\rho}_{k}|^2 / {2 c (z_j -z_k)} }$.  
The third curly bracket adds the phase factor, 
$G(| \vec{\rho}_l |, -\frac{\omega}{c f}) = e^{-i \frac{\omega}{2 c f}}$,
which is introduced by the imaging lens.

Applying the properties of the Gaussian function, Eq.~(\ref{g-1}) can be simplified
into the following form
\begin{eqnarray}\label{g-2}
& & g(\vec{\kappa}, \omega; \vec{\rho}_i, z =s_o+s_i) \nonumber \\
&=&  \frac{- \omega^2}{(2 \pi c)^2 s_o s_i} \, e^{i \frac{\omega}{c} (s_o+s_i)} 
\, G(|\vec{\rho}_i|, \frac{\omega}{c s_i})
 \int_{obj} d\vec{\rho}_{o} \, A(\vec{\rho}_{o}) \,
G(|\vec{\rho}_{o}|, \frac{\omega}{c s_o})\, e^{i \vec{\kappa} \cdot \vec{\rho}_{o}}
\nonumber \\ & & \times  \int_{lens} d\vec{\rho}_l \, G(|\vec{\rho}_l|, \frac{\omega}{c}
[\frac{1}{s_o} + \frac{1}{s_i} - \frac{1}{f}]) \, e^{-i \frac{\omega}{c}
(\frac{\vec{\rho}_{o}}{s_o} + \frac{\vec{\rho}_i}{s_i}) \cdot \vec{\rho}_l}.
\end{eqnarray}
The image plane is defined by the Gaussian thin-lens equation
of Eq.~(\ref{Lens-Eq}). Hence, the second integral in
Eq.~(\ref{g-2}) simplifies and gives, for a finite sized lens of
radius $R$, the so called point-spread
function of the imaging system: $somb(x) = 2J_1(x) / x$, where $x=[\frac{R}{s_o}\,
\frac{\omega}{c} |\vec{\rho}_{o} + \rho_{i} / m|]$,
 $J_1(x)$ is the first-order Bessel function and $m=s_i/s_o$
is the magnification of the imaging system.

Substituting the result of Eqs.~(\ref{g-2}) into Eq.~(\ref{e-g}) 
enables one to obtain
the classical self-correlation of the field, or, equivalently, the intensity
on the image plane
\begin{equation}\label{i-0}
I(\vec{\rho}_i, z_i, t_i) = \langle \, E^*(\vec{\rho}_i, z_i, t_i) \, E(\vec{\rho}_i, z_i, t_i) \, \rangle
\end{equation}
where $\langle ... \rangle$ denotes an ensemble average.  
We assume monochromatic light for classical imaging 
as usual. \footnote{Even if assuming a perfect lens without
chromatic aberration, Fresnel diffraction is wavelength dependent.
Hence, large broadband ($\Delta \omega \sim \infty$) would result in 
blurred images in classical imaging.  Surprisingly, the situation is
different in quantum imaging: no aberration blurring.}     

\vspace{2mm}
Case (I): \textit{incoherent imaging.}
The ensemble average of
$\langle \, \tilde{E}^*(\vec{\kappa}, \omega) \,\tilde{E}(\vec{\kappa'}, \omega) \, \rangle$
yields zeros except when $\vec{\kappa}=\vec{\kappa'}$. The image is thus
\begin{eqnarray}\label{i-2}
I(\vec{\rho}_i) &\propto&
 \int d\vec{\rho}_{o} \, \big{|}A(\vec{\rho}_{o})\big{|}^2 \,
\big{|}somb[\frac{R}{s_o}\, \frac{\omega}{c} |\vec{\rho}_{o} + \frac{\vec{\rho}_{i}}{m}|] \big{|}^2.
\end{eqnarray}
An incoherent image, magnified by a factor of $m$, is
thus given by the convolution between the squared moduli of the
object aperture function and the point-spread function. The
spatial resolution of the image is thus determined by the finite
width of the $|somb|^2$-function.

\vspace{2mm}
Case (II): \textit{coherent imaging.} The coherent superposition of
the $\vec{\kappa}$ modes in both $E^*(\vec{\rho}_i, \tau)$ and
$E(\vec{\rho}_i, \tau)$ results in a wavepacket. The image, or the
intensity distribution on the image plane, is  thus
\begin{eqnarray}\label{i-1}
I(\vec{\rho}_i) \propto
\Big{|} \int_{obj} d\vec{\rho}_{o} \, A(\vec{\rho}_{o}) \,
e^{i \frac{\omega}{2 c s_o} |\vec{\rho}_{o}|^2}
somb[\frac{R}{s_o} \frac{\omega}{c} |\vec{\rho}_{o} + \frac{\vec{\rho}_{i}}{m}|] \Big{|}^2. 
\end{eqnarray}
A coherent image, magnified by a factor of $m$, is thus
given by the squared modulus of the convolution between the object
aperture function (multiplied by a Fresnel phase  factor) and
the point-spread function. 

For $s_i<s_o$ and $s_o>f$, both Eqs.~(\ref{i-2}) and (\ref{i-1})
describe a real demagnified inverted image. In both cases, a
narrower $somb$-function yields a higher spatial resolution.
Thus, the use of shorter wavelengths allows for improvement of 
the spatial resolution of an imaging system.

\vspace{6mm}

To demonstrate the working principle of quantum imaging, we
replace classical light with an entangled two-photon source such 
as spontaneous parametric down-conversion (SPDC) and
replace the ordinary film with a two-photon absorber, which is
sensitive to two-photon transition only, on the image plane. We
will show that, in the same experimental setup of
Fig.~\ref{fig:lithography-1}, an entangled two-photon system gives
rise, on a two-photon absorber, to a point-spread function half the
width of the one obtained in classical imaging at the same
wavelength. Then, without employing shorter wavelengths, entangled
two-photon states improve the spatial resolution of a
\emph{two-photon image} by a factor of 2 \cite{Dowling}\cite{Milena}. 
We will also show that 
the entangled two-photon system yields a peculiar Fourier transform 
function as if it is produced by a light source with $\lambda/2$.  

In order to cover two different measurements, one on the 
image plane and one on the Fourier transform plane, we generalize the Green's
function of Eq.~(\ref{g-1}) from the image plane of $z=s_o + s_i$ to an arbitrary 
plane of $z=s_o + d$, where $d$ may take any values for different experimental 
setups:
\begin{eqnarray}\label{g}
&& g(\vec{\kappa}_j, \omega_j; \vec{\rho}_k, z=s_o+d) \nonumber \\
&= & \int_{obj} d\vec{\rho}_{o} \int_{lens} d\vec{\rho}_l \: A(\vec{\rho}_{o}) \, 
\{\frac{-i \omega_j}{2 \pi c s_o} \, e^{i \vec{\kappa}_j \cdot \vec{\rho}_{o}} \,
e^{i \frac{\omega_j}{c} s_o} \, G(|\vec{\rho}_{o}-\vec{\rho}_l|, \frac{\omega_j}{c s_o})\} 
\nonumber \\ & &\times \ G(|\vec{\rho}_l|, -\frac{\omega_j}{c f}) \, 
\{ \frac{-i \omega_j}{2\pi c d} \, \,
e^{i \frac{\omega_j}{c} d} \, G(|\vec{\rho}_l-\vec{\rho}_k|, \frac{\omega_j}{c d})\},
\end{eqnarray}
where $\vec{\rho}_{o}$, $\vec{\rho}_l$, and $\vec{\rho}_j$ are two-dimensional
vectors defined, respectively, on the (transverse) output plane of the source 
(which coincide with the object plane), on the transverse plane of the imaging 
lens and on the detection plane; 
and $j = s ,i$, labels the signal and the idler; $k = 1, 2$, labels the
photodetector $D_1$ and $D_2$.
The function $A(\vec{\rho}_{o})$ is the object-aperture 
function, while the terms in the first and second curly brackets of
Eq.~(\ref{g}) describe, respectively, free propagation from the output
plane of the source/object to the imaging lens, and from the imaging lens to the detection plane.   

Similar to the earlier calculation, by employing the second and third expressions given in 
Eq.~(\ref{Gaussian-10}), Eq.~(\ref{g}) simplifies to
\begin{eqnarray} \label{g_fin}
&& g(\vec{\kappa}_j, \omega_j; \vec{\rho}_k, z =s_o+d) \nonumber \\
&= & \frac{- \omega_j^2}{(2 \pi c)^2 s_o d} \, e^{i \frac{\omega_j}{c} (s_o+d)} \,
G(|\vec{\rho}_k|, \frac{\omega_j}{c d})  \int_{obj} d\vec{\rho}_{o} \, A(\vec{\rho}_{o}) \,
G(|\vec{\rho}_{o}|, \frac{\omega_j}{c s_o})\, e^{i \vec{\kappa}_j \cdot \vec{\rho}_{o}}  
\nonumber \\ & & \times  \int_{lens} d\vec{\rho}_l \, G(|\vec{\rho}_l|, \frac{\omega_j}{c} 
[\frac{1}{s_o} + \frac{1}{d} - \frac{1}{f}]) \, e^{-i \frac{\omega_j}{c} 
(\frac{\vec{\rho}_{o}}{s_o} + \frac{\vec{\rho}_k}{d}) \cdot \vec{\rho}_l}.
\end{eqnarray}

Substituting the Green's functions into Eq.~(\ref{Wavefunction-L-T}),    
the effective two-photon wavefunction 
$\Psi(\vec{\rho}_1, z; \vec{\rho}_2, z)$ is thus 
\begin{eqnarray}\label{Wavefunction-L-T-2}
\Psi(\vec{\rho}_1, z; \vec{\rho}_2, z) &=& \Psi_0\; \int d\Omega \, f(\Omega) \, 
G(|\vec{\rho}_1|, \frac{\omega_s}{c d}) \, G(|\vec{\rho}_2|, \frac{\omega_i}{c d} )  
\nonumber \\ & \times & \int_{obj} d\vec{\rho}_{o} \; 
A(\vec{\rho}_{o}) \; G(|\vec{\rho}_{o}|, \frac{\omega_s}{c s_o})  \int_{obj} d\vec{\rho'}_{o} \; 
A(\vec{\rho'}_{o}) \; G(|\vec{\rho'}_{o}|, \frac{\omega_i}{c s_o}) \nonumber \\ & \times &   
\int_{lens} d\vec{\rho}_l \; G(|\vec{\rho}_l|, \frac{\omega_s}{c} [\frac{1}{s_o} + \frac{1}{d} 
- \frac{1}{f}]) \; e^{-i \frac{\omega_s}{c} (\frac{\vec{\rho}_{o}}{s_o} + \frac{\vec{\rho}_1}{d}) 
\cdot \vec{\rho}_l}   \nonumber \\ & \times &   \int_{lens} d\vec{\rho'}_l \; 
G(|\vec{\rho'}_l|, [\frac{\omega_i}{c} [\frac{1}{s_o} + \frac{1}{d} - \frac{1}{f}]) \; 
e^{-i \frac{\omega_i}{c} (\frac{\vec{\rho'}_{o}}{s_o} + \frac{\vec{\rho}_2}{d}) \cdot \vec{\rho'}_l} 
\nonumber \\ & \times & \int d\vec{\kappa}_s \, d\vec{\kappa}_i \; \delta(\vec{\kappa}_s + \vec{\kappa}_i) \,
e^{i (\vec{\kappa}_s \cdot \vec{\rho}_{o} + \vec{\kappa}_i \cdot \vec{\rho'}_{o})}
\end{eqnarray}
where we have absorbed all constants into $\Psi_0$, including the phase
$$
e^{i \frac{\omega_s}{c} (s_o+d)} \, e^{i \frac{\omega_i}{c} (s_o+d)}  = e^{i \frac{\omega_p}{c} (s_o+d)}.
$$
The double integral of $d\vec{\kappa}_s$ and 
$d\vec{\kappa}_i$ yields a $\delta$-function of $\delta(\vec{\rho}_{o} - \vec{\rho'}_{o})$, 
and Eq.~(\ref{Wavefunction-L-T-2}) is simplified as:
\begin{eqnarray}\label{biphoton}
&& \Psi(\vec{\rho}_1, z; \vec{\rho}_2, z) \nonumber \\
&=& \Psi_0 \int d\Omega \, f(\Omega) \, 
G(|\vec{\rho}_1|, \frac{\omega_s}{c d}) \, G(|\vec{\rho}_2|, \frac{\omega_i}{c d} ) 
 \int_{obj} d\vec{\rho}_{o} \; A^2(\vec{\rho}_{o}) \; 
G(|\vec{\rho}_{o}|, \frac{\omega_p}{c s_o})   \nonumber \\ & & \times \   
\int_{lens} d\vec{\rho}_l \; G(|\vec{\rho}_l|, \frac{\omega_s}{c} [\frac{1}{s_o} + \frac{1}{d} 
- \frac{1}{f}]) \; e^{-i \frac{\omega_s}{c} (\frac{\vec{\rho}_{o}}{s_o} + \frac{\vec{\rho}_1}{d}) 
\cdot \vec{\rho}_l}   \nonumber \\ & & \times \   \int_{lens} d\vec{\rho'}_l \; 
G(|\vec{\rho'}_l|, [\frac{\omega_i}{c} [\frac{1}{s_o} + \frac{1}{d} - \frac{1}{f}]) \; 
e^{-i \frac{\omega_i}{c} (\frac{\vec{\rho}_{o}}{s_o} + \frac{\vec{\rho}_2}{d}) \cdot \vec{\rho'}_l}. 
\end{eqnarray}

\vspace{3mm}
\hspace{-6.5mm}We consider the following two cases:

\vspace{3mm}
\noindent  Case (I) on the imaging plane and $\vec{\rho}_1 = \vec{\rho}_2 = \vec{\rho}$.
\vspace{1.5mm}

In this case, Eq.~(\ref{biphoton}) is simplified as
\begin{eqnarray}\label{calc}
\Psi(\vec{\rho}, z; \vec{\rho}, z)  &\propto&
 \int_{obj} d\vec{\rho}_{o} \; A^2(\vec{\rho}_{o})
G(|\vec{\rho}_{o}|, \frac{\omega_p}{c s_o})   
\int d\vec{\rho}_l \, e^{-i \frac{\omega_p}{2
c} (\frac{\vec{\rho}_{o}}{s_o} + \frac{\vec{\rho}}{s_{i}}) \cdot
\vec{\rho}_l}
\int  d\vec{\rho'}_l \; e^{-i \frac{\omega_p}{2 c} (\frac{\vec{\rho}_{o}}{s_o} + 
\frac{\vec{\rho}}{s_{i}}) \cdot \vec{\rho'}_l} \nonumber \\
&& \times \  \Big{\{} \int d \Omega \, f(\Omega) \, e^{-i \Omega [(\frac{\vec{\rho}_{o}}{c s_o} + 
\frac{\vec{\rho}}{c s_{i}}) \cdot (\vec{\rho}_l - \vec{\rho'}_l)]} \Big{\}} 
\end{eqnarray}
where we have used $\omega_s = \omega_{p}/2 + \Omega$ and 
$\omega_s = \omega_{p}/2 - \Omega$ following
$\omega_s + \omega_i=\omega_p$.
The integral of $d \Omega$ gives a 
$\delta$-function of $\delta[(\frac{\vec{\rho}_{o}}{c s_o} + 
\frac{\vec{\rho}}{c s_{i}}) (\vec{\rho}_{l} - \vec{\rho'}_{l})]$ while taking 
the integral to infinity with a constant $f(\Omega)$.  This result
indicates again that the propagation-diffraction of the signal and the idler are not 
independent.  The ``two-photon diffraction" couples the 
two integrals in $\vec{\rho}_{o}$ and $\vec{\rho'}_{o}$
as well as the two integrals in $\vec{\rho}_{l}$ and $\vec{\rho'}_{l}$ 
and thus gives the $G^{(2)}$ function
\begin{eqnarray}\label{eq_image2}
G^{(2)}(\vec{\rho}, \vec{\rho})  
\propto \Big{|}\int_{obj} d\vec{\rho}_{o} \;
A^2(\vec{\rho}_{o}) \, e^{i \frac{\omega_p}{2 c s_o} |\vec{\rho}_{o}|^2} 
\frac{2 J_{1}\Big{(}\frac{R}{s_o}
\frac{\omega_p}{c} \big{|}\vec{\rho}_{o} +
\frac{\vec{\rho}}{m}\big{|}\Big{)}}{\Big{(}\frac{R}{s_o}
\frac{\omega_p}{c} \big{|}\vec{\rho}_{o} +
\frac{\vec{\rho}}{m}\big{|}\Big{)}^{2}} \Big{|}^2
\end{eqnarray}
which indicates that a coherent image (see Eq. (\ref{i-1})) magnified by a factor of 
$m=s_i/s_o$ is reproduced on the image plane by joint-detection or by 
two-photon absorption.

In Eq.~(\ref{eq_image2}), the point-spread function is
characterized by the pump wavelength $\lambda_p = \lambda_{s,i}/2$; 
hence, the point-spread function is half the width of
the (first order) classical case (Eqs.~(\ref{i-1}) and
(\ref{i-2})). An entangled two-photon state thus gives an image in joint-detection
with double spatial resolution when compared to the image
obtained in classical imaging. Moreover, the spatial resolution of
the two-photon image obtained by perfect SPDC radiation is further
improved because it is determined by the function $2 J_1(x)/ x^2$, which is much
narrower than the $somb(x)$.  

It is interesting to see that, different from the classical case,
the frequency integral over $\Delta \omega_s \sim \infty$ does not
give any blurring, but rather enhances the spatial
resolution of the two-photon image.  

\vspace{3mm}
\noindent  Case (II): on the Fourier transform plane and $\vec{\rho}_1 = \vec{\rho}_2 = \vec{\rho}$.
\vspace{1.5mm}

The detectors are now placed in the focal plane, i.e., $d=f$. In this case, 
the spatial effective two-photon wavefunction $\Psi(\vec{\rho}, z; \vec{\rho}, z)$ becomes:
\begin{eqnarray}\label{calc-2}
\Psi(\vec{\rho}, z; \vec{\rho}, z) &\propto& \int d\Omega \, f(\Omega) \,  
 \int_{obj} d\vec{\rho}_{o} \; A^2(\vec{\rho}_{o}) \; 
G(|\vec{\rho}_{o}|, \frac{\omega_p}{c s_o}) 
\int_{lens} d\vec{\rho}_l \; G(|\vec{\rho}_l|, \frac{\omega_s}{c s_o}) \; 
e^{-i \frac{\omega_s}{c} (\frac{\vec{\rho}_{o}}{s_o} + \frac{\vec{\rho}}{f}) 
\cdot \vec{\rho}_l}   \nonumber \\ & & \times \, \int_{lens} d\vec{\rho'}_l \; 
G(|\vec{\rho'}_l|, \frac{\omega_i}{c s_o}) \; 
e^{-i \frac{\omega_i}{c} (\frac{\vec{\rho}_{o}}{s_o} + \frac{\vec{\rho}}{f}) 
\cdot \vec{\rho'}_l}. 
\end{eqnarray}
We will first evaluate the two integrals over the lens.  To simplify the mathematics we 
approximate the integral to infinity.  Differing from the calculation for imaging
resolution, the purpose of this evaluation is to determine the Fourier transform.  Thus,
the approximation of an infinite lens is appropriate.  By applying Eq.~(\ref{Gaussian-10}), 
the two integrals over the lens contribute the following function of $\vec{\rho}_o$ to the 
integral of $d\vec{\rho}_{o}$ in Eq.~(\ref{calc-2}):
\begin{eqnarray*}
C \, G(|\vec{\rho}_o|, -\frac{\omega_p}{c s_o}) \, 
e^{-i \frac{\omega_p}{c f} \vec{\rho}_o \cdot \vec{\rho}}
\end{eqnarray*}
where $C$ absorbs all constants including a phase factor 
$G(|\vec{\rho}|, -\frac{\omega_p}{c f^2/s_o})$.  
Replacing the two integrals of $d\vec{\rho}_l$ and $d\vec{\rho'}_l$ 
in Eq.~(\ref{calc-2}) with this result, we obtain:
\begin{eqnarray}\label{biphoton_int}
\Psi(\vec{\rho}, z; \vec{\rho}, z)  
&\propto& \int d\Omega \, f(\Omega) \,  
\int_{obj} d\vec{\rho}_{o} \; A^2(\vec{\rho}_{o}) \; e^{-i \frac{\omega_p}{c f} \vec{\rho} 
\cdot \vec{\rho}_o} \propto
\mathcal{F}_{[\frac{\omega_p}{c f} \vec{\rho}\,]} \,
\big\{ A^2(\vec{\rho}_{o})\big\},
\end{eqnarray}
which is the Fourier transform of the object-aperture function.  When the two 
photodetectors scan together (i.e., $\vec{\rho}_1=\vec{\rho}_2=\vec{\rho}$), 
the second-order transverse correlation $G^{(2)}(\vec{\rho}, z; \vec{\rho}, z)$, where 
$z = s_o + f$, is reduced to:
\begin{equation}\label{g2-biphoton}
G^{(2)}(\vec{\rho}, z; \vec{\rho}, z) \propto \big{|}\, \mathcal{F}_{[\frac{\omega_p}{c f} \vec{\rho} \,]} 
\big\{ A^2(\vec{\rho}_{o})\big\} \big{|}^{2}.
\end{equation}
Thus, by replacing classical light with entangled two-photon sources, in the double-slit setup of 
Fig.~\ref{fig:lithography-1}, a Young's double-slit interference/diffraction 
pattern with twice the interference modulation and half the pattern width, compared to 
that of classical light at wavelength $\lambda_{s,i}= 2 \lambda_p$, is observed in the 
joint detection.  This effect has also been examined in a recent  ``quantum lithography" 
experiment \cite{Milena}.

\begin{figure}[hbt]
    \centering
    \includegraphics[width=80mm]{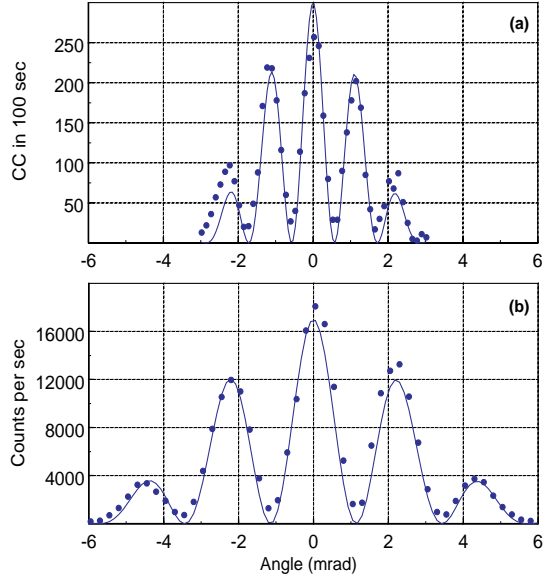}
    \parbox{14cm}{\caption{(a) Two-photon Fourier transform of a double-slit.  
    The light source was a collinear degenerate SPDC of $\lambda_{s,i}=916nm$.
    (b) Classical Fourier transform of the same double-slit.  A classical light source 
    of $\lambda=916nm$ was used.} 
    \label{fig:lithography-2}}
\end{figure}

Due to the lack of two-photon sensitive material, the first experimental demonstration 
of quantum lithography was measured on the Fourier transform plane, instead of the 
image plane.   Two point-like photon counting detectors were scanned jointly, 
similar to the setup illustrated in Fig.~\ref{fig:lithography-1}, for the observation 
of the interference/diffraction pattern of Eq.~(\ref{g2-biphoton}).  The published 
experimental result is shown in Fig.~\ref{fig:lithography-2} \cite{Milena}.  It is clear
that the two-photon Young's double-slit interference-diffraction pattern has half the width 
with twice the interference modulation compared to that of the classical case although 
the wavelengths are both $916nm$.  

Following linear Fourier optics, it is not difficult to see that, with the help of another
lens (equivalently building a microscope), one can transform the Fourier transform 
function of the double-slit back onto its image plane to observe its image with twice 
the spatial resolution.   

\begin{figure}[hbt]
    \centering
    \includegraphics[width=85mm]{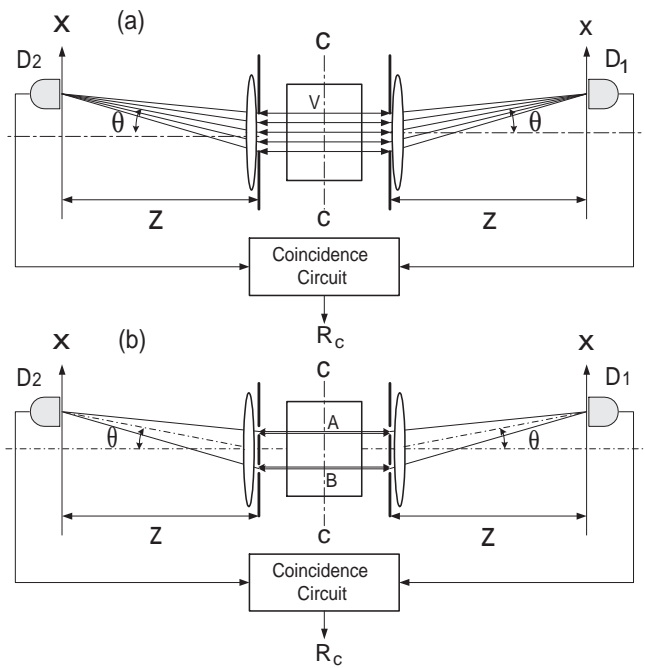}
    \parbox{14cm}{\caption{Unfolded experimental setup.  The joint measurement is 
    on the Fourier transform plane.  Each point of the object is ``illuminated" by the 
    signal-idler pair ``together", resulting in twice narrower interference-diffraction pattern 
    width in the Fourier transform plane through the joint detection of the signal-idler pair,
    equivalent to the use of classical light of $\lambda/2$.}   
    \label{fig:lithography-3}}
\end{figure}

The key to understanding the physics of this experiment is again through entangled nature of the
signal-idler two-photon system.  As we have discussed earlier, the pair is always emitted 
from the same point on the output plane of the source, thus always passing the same slit 
together if the double-slit is placed close to the surface of the nonlinear crystal.  There is no 
chance for the signal-idler pair to pass different slits in this setup.  In other words, each point
of the object is ``illuminated" by the pair ``together" and the pair ``stops" on the image
plane ``together".  The point-``spot" correspondence between the object and image
planes are based on the physics of two-photon diffraction, resulting in a twice narrower 
Fourier transform function in the Fourier transform plane and twice the image resolution in 
the image plane.  The unfolded schematic setup, which is shown in Fig.~\ref{fig:lithography-3},
may be helpful for understanding the physics.  
It is not difficult to calculate the interference-diffraction function under the experimental
condition indicated in Fig.~\ref{fig:lithography-3}.  The non-classical observation is due
to the superposition of the two-photon amplitudes, 
which are indicated by the straight lines connecting $D_1$ and $D_2$.  The two-photon
diffraction, which restricts the spatial resolution of a two-photon image, is very different from
that of classical light.  Thus, there should be no surprise in having an improved spatial 
resolution even beyond the classical limit.  

\vspace{6mm}

It is worthwhile to emphasize the following important aspects of physics in this 
simplified illustration: 

\vspace{3mm}
\hspace{-5mm} (1) The goal of lithography is the reproduction of demagnified 
images of complicated patterns.  The sub-wavelength interference feature does not 
necessarily translate into an improvement of the lithographic performance.  
In fact, the Fourier transform argument works for \emph{imaging setups} 
only; sub-wavelength interference in a Mach-Zehnder type interferometer, 
for instance, does not necessarily lead to an image.

\vspace{3mm}
\hspace{-5mm} (2) In the imaging setup, it is the peculiar nature of the entangled 
N-photon system that allows one to generate an image with N-times the spatial
resolution: the entangled photons come out from one point of the
object plane, undergo N-photon diffraction, and stop in the image plane within 
a N-times narrower spot than that of classical imaging.   The historical experiment 
by D'Angelo \emph{et al}, in which the working principle of quantum lithography was first 
demonstrated,  has taken advantage of the entangled two-photon state of 
SPDC: the signal-idler photon pair comes out from either the upper slit or the lower slit
that is in the object plane, undergoes two-photon diffraction, and 
stops in the image plane within a twice narrower image than that of
the classical one.   It is easy to show that a second Fourier transform, 
by means of the use of a second lens to set up a simple microscope, will produce 
an image on the image plane with double spatial resolution.  

\vspace{3mm}
\hspace{-5mm} (3) Certain 
``clever" tricks allow the production of doubly modulated interference patterns by
using classical light in joint photo-detection.  These tricks, however, may never 
be helpful for imaging.  Thus, they may never be useful for lithography. 

\section{Ghost imaging}

\hspace{5.5mm} The \emph{nonlocal} position-position and momentum-momentum 
EPR correlation of the entangled two-photon state of SPDC was successfully 
demonstrated in 1995 \cite{Pittman} inspired by the theory of Klyshko \cite{KlyshkoImg}
The experiment was immediately named as ``ghost imaging"  
in the physics community due to its surprising nonlocal nature.   
The important physics demonstrated in the experiment, 
however, may not  be the so called ``ghost".  Indeed, the original purpose of the experiment 
was to study the EPR correlation in position and in momentum and to test the EPR 
inequality of Eq.~(\ref{QMineq}) for the entangled signal-idler photon pair of SPDC 
\cite{IEEE07}\cite{Milena2}.   The experiments of ``ghost imaging" \cite{Pittman}
and ``ghost interference" \cite{GhostInt} together stimulated the 
foundation of quantum imaging in terms of geometrical and physical optics.  

\begin{figure}[hbt]
    \centering
    \includegraphics[width=110mm]{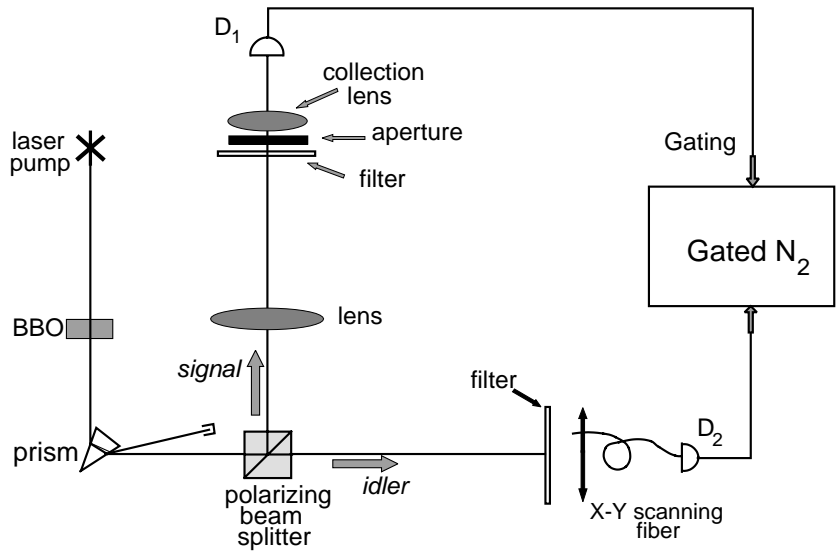}
    \parbox{14cm}{\caption{Schematic set-up of the ``ghost'' image experiment.}
    \label{Imageset}}
\end{figure}  

The schematic setup of the  ``ghost" imaging experiment is shown in 
Fig. \ref{Imageset}. A CW laser is used to pump a nonlinear
crystal, which is cut for degenerate type-II phase matching to produce a pair 
of orthogonally polarized signal (e-ray of the crystal) and idler (o-ray of the crystal) 
photons. The pair emerges from the
crystal as collinear, with $\omega _{s}\cong \omega _{i}\cong \omega
_{p}/2$. The pump is then separated from the signal-idler
pair by a dispersion prism, and the remaining
signal and idler beams are sent in different directions by a polarization
beam splitting Thompson prism. The signal beam passes through a convex lens
with a $400mm$ focal length and illuminates a chosen aperture (mask). As an
example, one of the demonstrations used the letters ``UMBC'' for the object mask. 
Behind the aperture is the ``bucket'' detector package $D_{1}$, which consists of a short focal
length collection lens in whose focal spot is an avalanche photodiode.  
$D_{1}$ is mounted in a fixed position during the experiment.  The idler beam is met by 
detector package $D_{2} $, which consists of an optical fiber whose output is mated 
with another avalanche photodiode. The input tip of the fiber is scanned in the transverse 
plane by two step motors. The output pulses of each detector, which are operating in 
photon counting mode, are sent to a coincidence counting circuit for the 
signal-idler joint detection.  
\begin{figure}[hbt]
    \centering
    \includegraphics[width=11cm]{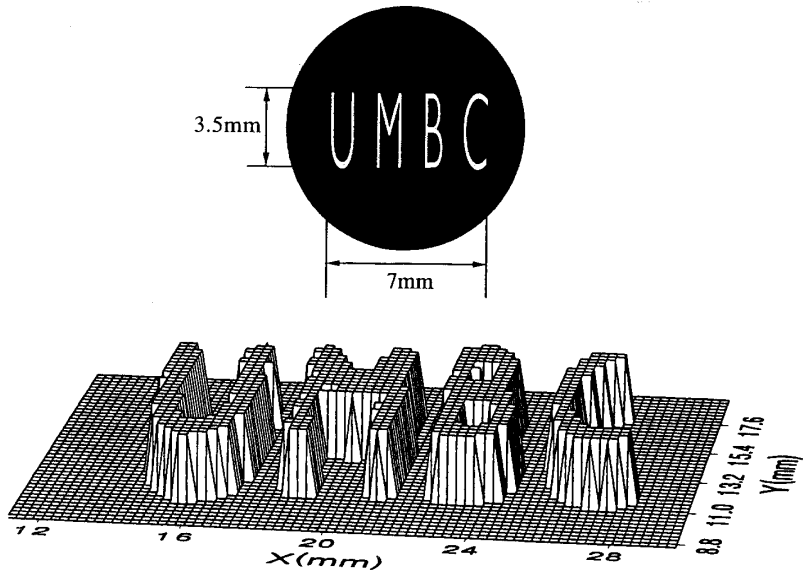}
    \parbox{14cm}{\caption{(a) A reproduction of the
actual aperature ``UMBC'' placed in the signal beam. (b) The image
of ``UMBC'': coincidence counts as a function of the fiber tip's
transverse plane coordinates. The step size is 0.25mm. The image
shown is a ``slice'' at the half maximum value.}
    \label{UMBC}}
\end{figure}

By recording the coincidence counts as a function of the fiber tip's
transverse plane coordinates, the image of the chosen aperture (for
example, ``UMBC'') is observed, as reported in Fig. \ref{UMBC}. It is interesting to
note that while the size of the ``UMBC'' aperture inserted in the signal beam is only about 
$3.5mm\times 7mm$, the observed image measures $7mm\times14mm$.  The image is 
therefore magnified by a factor of 2.  The observation also confirms that the focal length 
of the imaging lens, $f$, the aperture's optical distance from the lens, $S_{o}$, and the 
image's optical distance from the lens, $S_{i}$ (which is from the imaging lens going 
backward along the signal photon path to the two-photon source of the SPDC crystal then 
going forward along the path of idler photon to the image), satisfy the Gaussian thin lens 
equation.  In this experiment, $S_{o}$ was chosen to be $S_{o}=600mm$, and the twice 
magnified clear image was found when the fiber tip was on the plane of $S_{i}=1200mm$.
While $D_2$ was scanned on other transverse planes not defined by the Gaussian thin lens
equation, the images blurred out.

The measurement of the signal and the idler subsystem themselves 
are very different.  The single 
photon counting rate of $D_{2}$ was recorded during the scanning of the image and was 
found fairly constant in the entire region of the image.   This means that the transverse 
coordinate uncertainty of either signal or idler is considerably large compared to that of 
the transverse correlation of the entangled signal-idler photon pair: 
 $\Delta x_{1}$ ($\Delta y_{1}$) and $\Delta x_{2}$ ($\Delta y_{2}$) are much 
greater than $\Delta (x_{1} - x_{2})$ ($\Delta (y_{1} - y_{2})$).

\begin{figure}[hbt]
    \centering
    \includegraphics[width=90mm]{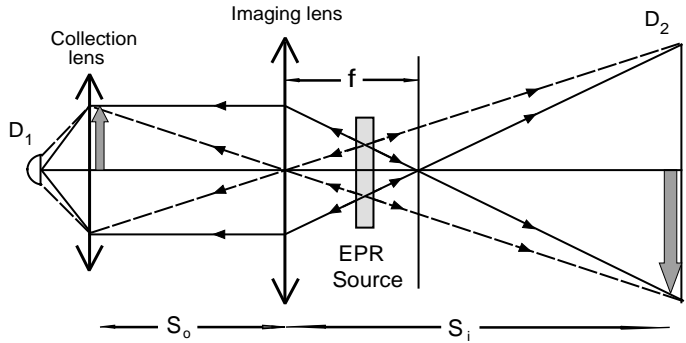}
    \parbox{14cm}{\vspace{5mm} \caption{An unfolded setup of the ``ghost'' imaging experiment, 
    which is helpful for understanding the physics.  Since the two-photon ``light" propagates along 
    ``straight-lines", it is not difficult to find that any geometrical light point on the subject plane 
    corresponds to an unique geometrical light point on the image plane.  Thus, a ``ghost" image 
    of the subject is made nonlocally in the image plane.  Although the placement of the lens, the 
    object, and detector $D_{2}$ obeys the Gaussian thin lens equation, it is important to 
    remember that the geometric rays in the figure actually represent the two-photon amplitudes 
    of an entangled photon pair.   The point 
    to point correspondence is the result of the superposition of these two-photon amplitudes.} 
    \label{fig:imaging-unfold}}
\end{figure}

\vspace{6mm}
The EPR $\delta$-functions, $\delta(\vec{\rho}_s - \vec{\rho}_i)$ and
$\delta(\vec{\kappa}_s + \vec{\kappa}_i)$ in transverse dimension, are the key to 
understanding this interesting phenomenon.  In degenerate SPDC, although the 
signal-idler photon pair has equal
probability to be emitted from any point on the output surface of the nonlinear crystal, the 
transverse position $\delta$-function indicates that if one of them is observed at one position,
the other one must be found at the same position.  In other words, the pair is always emitted
from the same point on the output plane of the two-photon source.  
The transverse momentum $\delta$-function,
defines the angular correlation of the signal-idler pair: the transverse momenta of a 
signal-idler amplitude are equal but pointed in opposite directions:
$\vec{\kappa}_s =-\vec{\kappa}_i$.  In other words, 
the two-photon amplitudes are always existing at roughly equal 
yet opposite angles relative to the pump.  This then allows for a simple
explanation of the experiment in terms of ``usual'' geometrical optics in
the following manner: we envision the nonlinear crystal as a ``hinge point'' and
``unfold'' the schematic of Fig. \ref{Imageset} into that shown in 
Fig. \ref{fig:imaging-unfold}.  The signal-idler two-photon amplitudes can then be 
represented by straight lines (but keep in mind the different propagation directions) 
and therefore, the image is well produced in coincidences when the aperture, lens, 
and fiber tip are located according to the Gaussian thin lens equation of 
Eq.(\ref{Gaussian}).  The image is exactly the same as one
would observe on a screen placed at the fiber tip if detector $D_{1}$ were
replaced by a point-like light source and the nonlinear crystal by a reflecting
mirror. 

Following a similar analysis in geometric optics,  it is not difficult to find that any 
geometrical ``light spot" on the subject plane, which is the intersection point of all 
possible two-photon amplitudes coming from the two-photon light source, 
corresponds to a unique geometrical ``light spot" on the image 
plane, which is another intersection point of all the possible two-photon amplitudes.  
This point to point correspondence made the ``ghost" image of the subject-aperture possible.  
Despite the completely different physics from classical geometrical optics, the remarkable 
feature is that the relationship between the focal length of the lens, $f$, the aperture's 
optical distance from the lens, $S_{o}$, and the image's optical distance from the lens, 
$S_{i}$, satisfy the Gaussian thin lens equation:
\begin{equation*}
\frac{1}{s_{o}}+\frac{1}{s_{i}}=\frac{1}{f}.  \label{Gaussian}
\end{equation*}
Although the placement of the lens, the object, and the detector $D_{2}$ obeys the Gaussian 
thin lens equation, it is important to remember that the geometric rays in the figure actually 
represent the two-photon amplitudes of a signal-idler photon pair and the point to point 
correspondence is the result of the superposition of these two-photon amplitudes.  
The ``ghost" image is a realization of the 1935 EPR {\em gedankenexperiment}.  

\vspace{6mm}

Now we calculate $G^{(2)}(\vec{\rho}_o, \vec{\rho}_i)$ for the ``ghost" imaging 
experiment, where $\vec{\rho}_o$ and $ \vec{\rho}_i$ are the transverse coordinates 
on the object plane and the image plane.  We will show that there exists a $\delta$-function 
like point-to-point relationship between the object plane and the image plane, i.e., if one 
measures the signal photon at a position of $\vec{\rho}_o$ on the object plane the idler 
photon can be found only at a certain unique position of $\vec{\rho}_i$ on the image plane 
satisfying $\delta(m\vec{\rho}_o - \vec{\rho}_i)$, where $m=-(s_i/s_o)$ is the image-object 
magnification factor.  After demonstrating the $\delta$-function, we show how the object-aperture 
function of $A(\vec{\rho}_o)$ is transfered to the image plane as a magnified image 
$A(\vec{\rho}_i/m)$.  Before showing the calculation, it is worthwhile to emphasize again 
that the ``straight lines" in Fig.~\ref{fig:imaging-unfold} schematically represent 
the two-photon amplitudes belonging to a pair of signal-idler photon.  
A ``click-click" joint measurement at ($\mathbf{r}_{1}, t_{1}$), which is on the object
plane, and ($\mathbf{r}_{2}, t_{2}$), which is on the image plane, in the form of an
EPR $\delta$-function, is the result of the coherent superposition of all these
two-photon amplitudes.  

\begin{figure}[hbt]
    \centering
    \includegraphics[width=100mm]{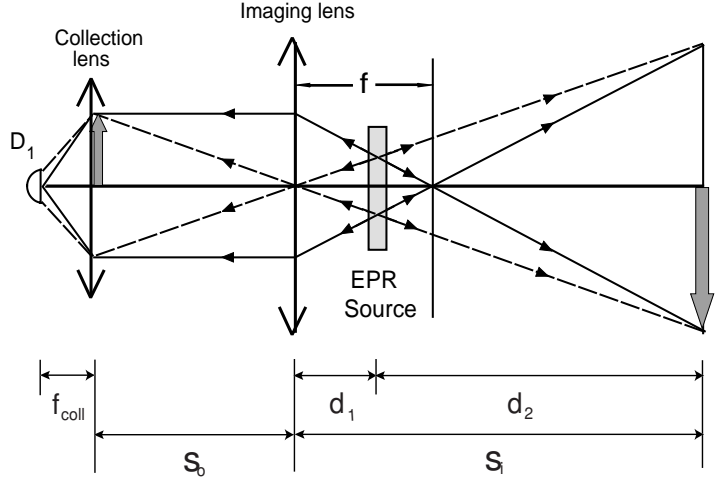}
    \parbox{14cm}{\caption{In arm-$1$, the signal propagates freely over a distance $d_1$ 
    from the output plane of the source to the imaging lens, then passes an object aperture 
    at distance $s_o$, and then is focused onto photon counting detector $D_1$ by a 
    collection lens.  In arm-$2$, the idler propagates freely over a distance $d_2$ from the 
    output plane of the source to a point-like photon counting detector $D_2$.} 
    \label{fig:imaging-unfold-2}}
\end{figure}

We follow the unfolded experimental setup shown in Fig.~\ref{fig:imaging-unfold-2} to 
establish the Green's functions $g(\vec{\kappa}_s, \omega_s, \vec{\rho}_o, z_o)$ and 
$g(\vec{\kappa}_i, \omega_i, \vec{\rho}_2, z_2)$.  In arm-$1$, the signal propagates freely 
over a distance $d_1$ from the output plane of the source to the imaging lens, then 
passes an object aperture at distance $s_o$, and then is focused onto photon 
counting detector $D_1$ by a collection lens.  We will evaluate 
$g(\vec{\kappa}_s, \omega_s, \vec{\rho}_o, z_o)$ by propagating the field from 
the output plane of the two-photon source to the object plane.
In arm-$2$, the idler propagates freely over a distance $d_2$ 
from the output plane of the two-photon source to a point-like detector $D_2$.     
$g(\vec{\kappa}_i, \omega_i, \vec{\rho}_2, z_2)$ is thus a free propagator.

\vspace{3mm}
\hspace{-6.5mm}(I) Arm-$1$ (source to object):
\vspace{3mm}

The optical transfer function or Green's function in arm-$1$, which propagates the field 
from the source plane to the object plane, is given by:
\begin{eqnarray}\label{Arm-1}
&& g(\vec{\kappa}_s, \omega_s; \vec{\rho}_o, z_o=d_{1}+s_{o}) \nonumber \\
&=& e^{i \frac{\omega_s}{c} z_o}
\int_{lens} d\vec{\rho}_l \, \int_{A} d\vec{\rho}_S \,
\Big{\{} \, \frac{-i \omega_s}{2 \pi c d_1} e^{i \vec{\kappa_s} \cdot \vec{\rho}_S}  
G(|\, \vec{\rho}_S-\vec{\rho}_l\, |, \frac{\omega_s}{c d_1}) \Big{\}} \nonumber \\ 
& & \times \,\, \Big{\{}  \, G(| \vec{\rho}_l |, \frac{\omega_s}{c f}) \, \Big{\}} \, 
\Big{\{} \, \frac{-i \omega_s}{2 \pi c s_{o}} 
G(|\,\vec{\rho}_l-\vec{\rho}_o\,|, \frac{\omega_s}{c s_o}) \Big{\}},
\end{eqnarray}
where $\vec{\rho}_S$ and $\vec{\rho}_l$ are the transverse vectors defined, respectively, 
on the output plane of the source and on the plane of the imaging lens. 
The terms in the first and third curly brackets in Eq.~(\ref{Arm-1}) describe free space 
propagation from the output plane of the source to the imaging lens and from the imaging 
lens to the object plane, respectively.  The function 
$G(| \vec{\rho}_l |, \frac{\omega}{c f})$ in the second curly brackets is the transformation function 
of the imaging lens. Here, we treat it as a thin-lens: $G(| \vec{\rho}_l |, \frac{\omega}{c f}) 
\cong e^{-i \frac{\omega}{2 c f} \, | \vec{\rho}_l |^2}$.  

\vspace{3mm}
\hspace{-6.5mm}(II) Arm-$2$ (from source to image):
\vspace{3mm}

In arm-$2$, the idler propagates freely from the source to the plane of $D_2$, which
is also the plane of the image. The Green's function is thus:
\begin{eqnarray}\label{Arm-2} 
g(\vec{\kappa}_i, \omega_i; \vec{\rho}_2, z_2=d_2) 
= \frac{-i \omega_i}{2 \pi c d_2} \, e^{i \frac{\omega_i}{c} d_2} \int_{A} d\vec{\rho'}_S  \,
G(|\, \vec{\rho'}_S-\vec{\rho}_2\, |, \frac{\omega_i}{c d_2}) \, 
e^{i \vec{\kappa}_i \cdot \vec{\rho^{,}_S} }
\end{eqnarray}
where $\vec{\rho'}_S$ and $\vec{\rho}_2$ are the transverse vectors defined, respectively, 
on the output plane of the source, and on the plane of the photo-dector $D_2$.  

\vspace{3mm}
\hspace{-6.5mm}(III) $\Psi(\vec{\rho}_o, \vec{\rho}_i)$ (object plane - image plane):
\vspace{3mm}

To simplify the calculation and to focus on the transverse correlation, 
in the following calculation we assume degenerate 
($\omega_s = \omega_i = \omega$) and collinear SPDC.  
The transverse two-photon effective wavefunction 
$\Psi(\vec{\rho}_o, \vec{\rho}_2)$ 
is then evaluated by substituting the Green's functions 
$g(\vec{\kappa}_s, \omega; \vec{\rho}_o, z_o) $ and 
$g(\vec{\kappa}_i, \omega; \vec{\rho}_2, z_2)$
into the expression given in Eq.~(\ref{Wavefunction-TT}):
\begin{eqnarray}\label{biphoton_x}
&& \Psi(\vec{\rho}_o,\vec{\rho}_2) \nonumber \\
&\propto&  \int d\vec{\kappa}_s \, d\vec{\kappa}_i \, 
\delta(\vec{\kappa}_s +  \vec{\kappa}_i) \, g(\vec{\kappa}_s, \omega; \vec{\rho}_o, z_o)  \, 
g(\vec{\kappa}_i, \omega; \vec{\rho}_2, z_2) \nonumber \\
&\propto& e^{i \frac{\omega}{c} (s_0+s_i)}  \int d\vec{\kappa}_s \, d\vec{\kappa}_i \, 
\delta(\vec{\kappa}_s +  \vec{\kappa}_i) 
\int_{lens} d\vec{\rho}_l \, \int_{A} d\vec{\rho}_S \,
 \, e^{i \vec{\kappa_s} \cdot \vec{\rho}_S}  
G(|\, \vec{\rho}_S-\vec{\rho}_l\, |, \frac{\omega}{c d_1}) \nonumber \\ 
& & \times \ G(| \vec{\rho}_l |, \frac{\omega}{c f}) \,  \, 
G(|\,\vec{\rho}_l-\vec{\rho}_o\,|, \frac{\omega}{c s_o}) \int_{A} d\vec{\rho^{,}_S} \,\,
e^{i \vec{\kappa}_i \cdot \vec{\rho^{,}_S} } \,
G(|\, \vec{\rho^{,}_S}-\vec{\rho}_2\, |, \frac{\omega}{c d_2})
\end{eqnarray}
where we have ignored all the proportional constants.
Completing the double integral of $d\vec{\kappa}_s$ and $d\vec{\kappa}_s$
\begin{eqnarray}\label{delta-source}
\int d\vec{\kappa}_s \, d\vec{\kappa}_i \, \delta(\vec{\kappa}_s +  \vec{\kappa}_i)\,
e^{i \vec{\kappa_s} \cdot \vec{\rho}_S} \, e^{i \vec{\kappa}_i \cdot \vec{\rho^{,}_S} } 
\sim \, \delta(\vec{\rho}_S - \vec{\rho^{,}_S}),
\end{eqnarray}  
Eq.~(\ref{biphoton_x}) becomes:
\begin{eqnarray*}
&& \Psi(\vec{\rho}_o,\vec{\rho}_2) \\ \nonumber
&\propto& 
\int_{lens} d\vec{\rho}_l \, \int_{A} d\vec{\rho}_S \,
G(|\, \vec{\rho}_2-\vec{\rho}_S\, |, \frac{\omega}{c d_2}) \,
G(|\, \vec{\rho}_S-\vec{\rho}_l\, |, \frac{\omega}{c d_1})
\, G(| \vec{\rho}_l |, \frac{\omega}{c f}) \,
G(|\,\vec{\rho}_l-\vec{\rho}_o\,|, \frac{\omega}{c s_o}).
\end{eqnarray*}
We then apply the properties of the Gaussian functions of Eq.~(\ref{Gaussian-10})
and complete the integral on $d\vec{\rho}_S$ by assuming the transverse size of 
the source is large enough to be treated as infinity.  
\begin{eqnarray}\label{biphoton_z}
\Psi(\vec{\rho}_o,\vec{\rho}_2) 
\propto  
\int_{lens} d\vec{\rho}_l \, 
G(|\, \vec{\rho}_2-\vec{\rho}_l\, |, \frac{\omega}{c s_i}) \,
G(| \vec{\rho}_l |, \frac{\omega}{c f}) \,
G(|\,\vec{\rho}_l-\vec{\rho}_o\,|, \frac{\omega}{c s_o}). 
\end{eqnarray}
Although the signal and idler propagate to different directions along two optical arms, 
Interestingly, the Green function in Eq.~(\ref{biphoton_z}) is equivalent to that of a 
classical imaging setup, if we imagine the fields start propagating from a point $\vec{\rho}_o$ 
on the object plane to the lens and then stop at point $\vec{\rho}_2$ 
on the imaging plane.  The mathematics is consistent with our previous qualitative analysis 
of the experiment.  

The integral on $d\vec{\rho}_l$ yields a point-to-point relationship between
the object plane and the image plane that is defined by the Gaussian thin-lens 
equation:   
\begin{eqnarray}\label{biphoton_zz}
\int_{lens} d\vec{\rho}_l \, G(|\, \vec{\rho}_l|, 
\frac{\omega}{c}[\frac{1}{s_o} + \frac{1}{s_i} - \frac{1}{f}]) \,
e^{-i\frac{\omega}{c} (\frac{\vec{\rho}_o}{s_o} + \frac{\vec{\rho}_i}{s_i})\cdot \vec{\rho}_l} 
\propto \delta(\vec{\rho}_o + \frac{\vec{\rho}_i}{m})
\end{eqnarray}
where the integral is approximated to infinity and the Gaussian thin-lens equation 
of $1/s_o +1/s_i =1/f$ is applied.  We have also defined $m=s_i/s_o$ as the magnification 
factor of the imaging system.  The function $\delta(\vec{\rho}_o + \vec{\rho}_i/m)$
indicates that a point $\vec{\rho}_o$ on the object plane corresponds to a unique point 
$\vec{\rho}_i$ on the image plane.  The two vectors point in opposite directions and 
the magnitudes of the two vectors hold a ratio of $m=|\vec{\rho}_i|/|\vec{\rho}_o|$.  

If the finite size of the imaging lens has to be taken into 
account (finite diameter $D$), the integral yields a point-spread function of $somb(x)$:
\begin{eqnarray}\label{somb}
\int_{lens} d\vec{\rho}_l \,
e^{-i\frac{\omega}{c} (\frac{\vec{\rho}_o}{s_o} + \frac{\vec{\rho}_i}{s_i})\cdot \vec{\rho}_l} 
\propto somb\Big{(} \frac{R}{s_o}\, \frac{\omega}{c} [\vec{\rho}_{o} + \frac{\vec{\rho}_{i}}{m}] \Big{)}
\end{eqnarray}
where $somb(x) = 2J_1(x)/x$, $J_1(x)$ is the first-order Bessel function and 
$R/s_o$ is named as the numerical aperture.  The point-spread function turns 
the point-to-point correspondence between the object plane and the image plane into 
a point-to-``spot" relationship and thus limits the spatial resolution.  This point has been 
discussed in detail in the last section.   

Therefore, by imposing the condition of the Gaussian thin-lens equation, the transverse 
two-photon effective wavefunction is approximated as a $\delta$ function 
\begin{eqnarray}\label{biphoton_x_fin}
\Psi(\vec{\rho}_o,\vec{\rho}_i) \propto
\delta(\vec{\rho}_o + \frac{\vec{\rho}_i}{m})
\end{eqnarray}
where $\vec{\rho}_o$ and $\vec{\rho}_i$, again, are the transverse coordinates on the 
object plane and the image plane, respectively, defined by the Gaussian thin-lens 
equation.  Thus, the second-order spatial correlation function 
$G^{(2)}(\vec{\rho}_o,\vec{\rho}_i)$ turns out to be:
\begin{eqnarray}\label{biphoton_x_fin-2}
G^{(2)}(\vec{\rho}_o,\vec{\rho}_i) = |\, \Psi(\vec{\rho}_o,\vec{\rho}_i) \,|^{2}
\propto |\, \delta(\vec{\rho}_o + \frac{\vec{\rho}_i}{m}) \,|^{2}.
\end{eqnarray}
Eq.~(\ref{biphoton_x_fin-2}) indicates a point to point EPR 
correlation between the object plane and the image plane, i.e., if one 
observes the signal photon at a position $\vec{\rho}_o$ on the object plane, the idler 
photon can only be found at a certain unique position $\vec{\rho}_i$ on the image plane 
satisfying $\delta(\vec{\rho}_o + \vec{\rho}_i/m)$ with $m=s_i/s_o$. 

\vspace{6mm}

We now include an object-aperture function, a collection lens and a photon counting
detector $D_1$ into the optical transfer function of arm-$1$ as shown in Fig.~\ref{Imageset}.   

We will first treat the collection-lens-$D_1$ package  as a ``bucket" detector.  The ``bucket" 
detector integrates all $\Psi(\vec{\rho}_o, \vec{\rho}_2)$ which passes 
the object aperture $A(\vec{\rho}_o)$ as a joint photo-detection event.  This process 
is equivalent to the following convolution :
\begin{eqnarray}\label{biphoton_final-2}
R_{1, 2} \propto  \int_{obj} d\vec{\rho}_o \, \big{|}  A(\vec{\rho}_o)  \big{|}^2 \,
\big{|}  \Psi(\vec{\rho}_o, \vec{\rho}_i)  \big{|}^2  
\simeq \big{|} A(\frac{-\vec{\rho}_i}{m}) \big{|}^{2}
\end{eqnarray}
where, again, $D_2$ is scanning in the image plane, $\vec{\rho}_2 = \vec{\rho}_i$. 
Eq.~(\ref{biphoton_final-2}) indicates a magnified (or demagnified) image of the 
object-aperture function by means of the joint-detection events between distant
photodetectors $D_1$ and $D_2$. The ``-" sign in $A(-\vec\rho_i/m)$
indicates opposite orientation of the image.  The model of the ``bucket" detector is a 
good and realistic approximation. 

Now we consider a detailed evaluation by including the object-aperture function, 
the collection lens and the photon counting detector $D_1$ into arm-$1$.   
The Green's function of Eq.~(\ref{Arm-1}) becomes:
\begin{eqnarray}\label{ga-collect-lens}
& & g(\vec{\kappa}_s, \omega_s; \vec{\rho}_1, z_1=d_{1}+s_{o} + f_{coll}) \nonumber \\
&=& e^{i \frac{\omega_s}{c} z_1} \int_{obj} d\vec{\rho}_o
\int_{lens} d\vec{\rho}_l \int_{A} d\vec{\rho}_S
\Big{\{} \, \frac{-i \omega_s}{2 \pi c d_1} e^{i \vec{\kappa_s} \cdot \vec{\rho}_S}  
G(|\, \vec{\rho}_S-\vec{\rho}_l\, |, \frac{\omega_s}{c d_1}) \Big{\}} \nonumber \\ 
& & \times \,\, G(| \vec{\rho}_l |, \frac{\omega_s}{c f}) \, \Big{\{} \, \frac{-i \omega_s}{2 \pi c s_{o}} 
G(|\,\vec{\rho}_l-\vec{\rho}_o\,|, \frac{\omega_s}{c s_o}) \Big{\}} \, A(\vec{\rho}_o) \nonumber \\
& & \times \,\, G(|\vec{\rho}_o|, \frac{\omega_s}{c f_{coll}}) \Big{\{} \, \frac{-i \omega_s}{2 \pi c f_{coll}} \,
G(|\vec{\rho}_o-\vec{\rho}_1|, \frac{\omega_s}{c f_{coll}})\Big{\}}
\end{eqnarray}
where $f_{coll}$ is the focal-length of the collection lens and $D_1$ is placed on the 
focal point of the collection lens.  Repeating the
previous calculation, we obtain the transverse two-photon effective wavefunction:
\begin{eqnarray}\label{biphoton_final}
\Psi(\vec{\rho}_1,\vec{\rho}_2) \propto 
\int_{obj} d\vec{\rho}_o \, A(\vec{\rho}_o) \, 
\delta(\vec{\rho}_o + \frac{\vec{\rho}_2}{m}) = 
A(\vec{\rho}_o)\otimes \delta(\vec{\rho}_o + \frac{\vec{\rho}_2}{m})
\end{eqnarray}
where $\otimes$ means convolution.  Notice, in Eq.~(\ref{biphoton_final}) 
we have ignored the phase factors which have no contribution to the formation 
of the image.
The joint detection counting rate, $R_{1,2}$, between photon counting detectors 
$D_1$ and $D_2$ is thus:
\begin{equation}\label{rate-collect-lens}
R_{1, 2} \propto G^{(2)}(\vec{\rho}_1, \vec{\rho}_2) 
\propto \big{|}\,A(\vec{\rho}_o)\otimes \delta(\vec{\rho}_o + \frac{\vec{\rho}_2}{m})\,\big{|}^{2} 
= \big{|}\, A(\frac{-\vec{\rho}_2}{m})\,\big{|}^{2}
\end{equation} 
where, again, $\vec{\rho}_2 = \vec{\rho}_i$. 

As we have discussed earlier, the point-to-point EPR correlation is the result 
of the coherent superposition of a special selected set of two-photon states.  In principle, 
one signal-idler pair contains all the necessary two-photon amplitudes 
that generate the ghost image - a nonclassical characteristic which we name as a 
\textit{two-photon coherent} image.

\section{Popper's experiment}

\hspace{5.5mm}In quantum mechanics, one can never expect to measure both 
the precise position and
momentum of a particle simultaneously. It is prohibited. We say that the
quantum observable ``position'' and ``momentum'' are ``complementary'' because
the precise knowledge of the position (momentum) implies that all possible
outcomes of measuring the momentum (position) are equally probable.

Karl Popper, being a ``metaphysical realist", however, took a different point of
view. In his opinion, the quantum formalism {\em could} and {\em should} be
interpreted realistically: a particle must have a precise position and momentum \cite{Popper0}.
This view was shared by Einstein. In this regard, he invented a thought
experiment in the early 1930's aimed to support his realistic interpretation 
of quantum mechanics \cite{Popper}. What Popper intended to show in his thought experiment
is that a particle can have both precise position and momentum simultaneously
through the correlation measurement of an entangled two-particle system.  

\begin{figure}[hbt]
    \centering
    \includegraphics[width=85mm]{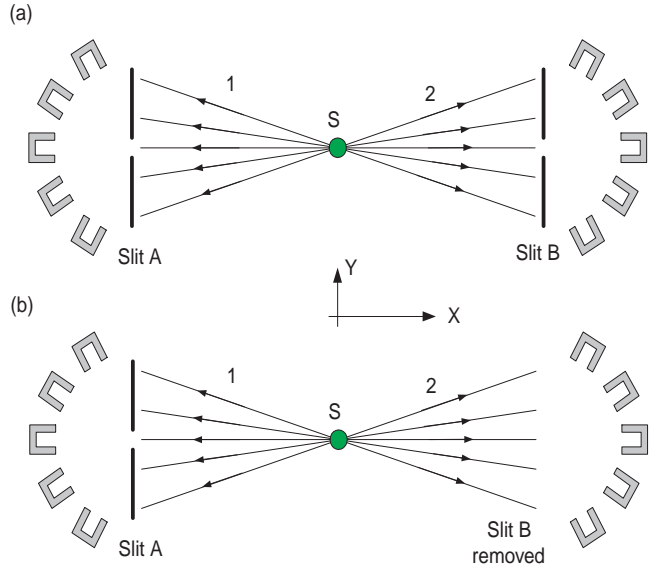}
    \parbox{14cm}{\caption{Popper's thought experiment. An entangled pair of particles are
emitted from a point source with momentum conservation. A narrow slit on screen
A is placed in the path of particle 1 to provide the precise knowledge of its
position on the $y$-axis and this also determines the precise $y$-position of
its twin, particle 2, on screen B. (a) Slits A and B are both adjusted very
narrowly. (b) Slit A is kept very narrow and slit B is left wide open.} 
    \label{Popperidea}}
\end{figure}

Similar to EPR's {\em gedankenexperiment}, Popper's thought experiment is 
also based on the feature of {\em two-particle entanglement}: if the position or
momentum of particle 1 is known, the corresponding observable of its twin,
particle 2, is then 100\% determined. Popper's original thought
experiment is schematically shown in Fig. \ref{Popperidea}. A point source S,
positronium as Popper suggested, is placed at the center of the experimental
arrangement from which entangled pairs of particles 1 and 2 are emitted in
opposite directions along the respective positive and negative $x$-axes towards
two screens A and B. There are slits on both screens parallel to the $y$-axis
and the slits may be adjusted by varying their widths $\Delta y$. Beyond the
slits on each side stand an array of Geiger counters for the joint
measurement of the particle pair as shown in the figure. The entangled pair
could be emitted to any direction in $4\pi$ solid angles from the point source.
However, if particle 1 is detected in a certain direction, particle 2 is then
known to be in the opposite direction due to the momentum conservation of the
pair.

First, let us imagine the case in which slits A and B are both adjusted very
narrowly. In this circumstance, particle 1 and particle 2 experience diffraction 
at slit A and slit B, respectively, and exhibit greater $\Delta p_{y} $ for smaller 
$\Delta y$ of the slits. There seems to be no disagreement in this situation between 
Copenhagen and Popper.

Next, suppose we keep slit A very narrow and leave slit B wide
open. The main purpose of the narrow slit A is to provide the precise knowledge
of the position $y$ of particle 1 and this subsequently determines the precise
position of its twin (particle 2) on side B through quantum entanglement. Now,
Popper asks, in the absence of the physical interaction with an actual slit,
does particle 2 experience a greater uncertainty in $\Delta p_{y}$ due to the
precise knowledge of its position?  Based on his beliefs,
Popper provides a straightforward prediction: {\em particle 2 must not
experience a greater $\Delta p_{y}$ unless a real physical narrow slit B is
applied}. However, if Popper's conjecture is correct, this would imply the
product of $\Delta y$ and $\Delta p_{y}$ of particle 2 could be smaller than
$h$ ($\Delta y \, \Delta p_{y}< h $). This may pose a serious difficulty for
Copenhagen and perhaps for many of us.  On the other hand, if particle 2
going to the right does scatter like its twin, which has passed though slit A,
while slit B is wide open, we are then confronted with an apparent {\it
action-at-a-distance}!

\begin{figure}[hbt]
    \centering
    \includegraphics[width=95mm]{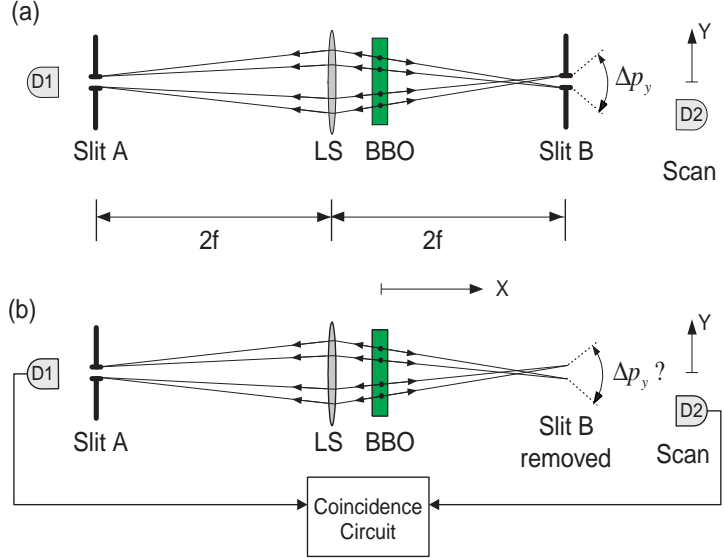}
    \parbox{14cm}{\caption{Modified version of Popper's experiment.  An 
    entangled photon pair is generated by SPDC. A lens and a narrow slit A 
    are placed in the path of photon 1 to provide the precise knowledge of its 
    position on the $y$-axis and also to determine the precise $y$-position of
     its twin, photon 2, on screen B by means of two-photon ``ghost" imaging. 
     Photon counting detectors $D_{1}$ and $D_{2}$ are used to scan in
     $y$-directions for joint detections. (a) Slits A and B are both adjusted
     very narrowly. (b) Slit A is kept very narrow and slit B is left wide open.} 
    \label{PopperSPDC}}
\end{figure}

The use of a ``point source" in Popper's proposal has been criticized
historically as the fundamental mistake Popper made \cite{Point}.   It is true
that a point source can never produce a pair of entangled particles 
which preserves the EPR correlation in momentum as Popper expected.
However, notice that a ``point source" is {\it not} a necessary requirement for
Popper's experiment.  What is required is a precise position-position 
EPR correlation: if the position of particle 1 is precisely known, the
position of particle 2 is 100\% determined.  As we have shown
in the last section, ``ghost" imaging is a perfect tool to achieve this.  

In 1998, Popper's experiment was realized with the help of  two-photon 
``ghost" imaging \cite{Kim}.  Fig.~\ref{PopperSPDC} is a schematic diagram 
that is useful for comparison with
the original Popper's thought experiment.   It is easy to see that this is a typical
``ghost" imaging experimental setup.  An entangled photon pair is used 
to image slit A onto the distant image plane of ``screen" B.  In the setup, $s_o$ 
is chosen to be twice the focal length of the imaging lens $LS$, $s_o = 2f$. 
According to the Gaussian thin lens equation, an equal size ``ghost" image of 
slit A appears on the two-photon image plane at $s_i = 2f$.  The use of
slit A provides a precise knowledge of the position of photon 1 
on the $y$-axis and also determines the precise $y$-position of its twin, 
photon 2, on screen B by means of the two-photon ``ghost" imaging. 
The experimental condition specified in Popper's experiment is then achieved.
When slit A is adjusted to a certain narrow width and slit B is wide open, slit
A provides precise knowledge about the position of photon 1 on the $y$-axis up to
an accuracy $\Delta y$ which equals the width of slit A, and the corresponding
``ghost image" of pinhole A at screen B determines the precise position $y$
of photon 2 to within the same accuracy $\Delta y$. $ \Delta p_{y}$ of photon
2 can be independently studied by measuring the width of its ``diffraction
pattern'' at a certain distance from ``screen" B. This is obtained by recording
coincidences between detectors $D_{1}$ and $ D_{2}$ while scanning detector
$D_{2}$ along its $y$-axis, which is behind screen B at a certain distance.

\begin{figure}
    \centering
    \includegraphics[width=95mm]{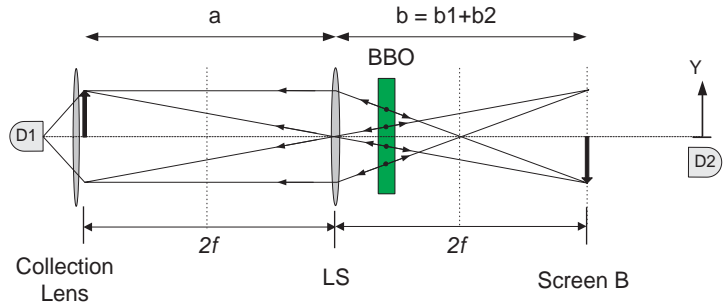}
    \parbox{14cm}{\caption{An unfolded schematic of ghost imaging.   We assume 
    the entangled signal-idler photon pair holds a perfect momentum correlation 
    $\delta({\bf k}_{s}+{\bf k}_{i})\sim 0$. The locations of the slit A, the imaging lens LS, 
    and the ``ghost" image must be governed by the Gaussian thin lens equation.  In 
    this experiment, we have chosen $s_o = s_i =2f$. Thus, the ``ghost" image of slit A
    is expected to be the same size as that of slit A.} 
    \label{Popperunfold}}
\end{figure}

Figure \ref{Popperunfold} is a conceptual diagram to connect the modified 
Popper's experiment with two-photon  ``ghost" imaging.  In this unfolded 
``ghost" imaging setup, we assume the entangled signal-idler photon pair 
holds a perfect transverse momentum correlation with $\vec{k}_{s}+ \vec{k}_{i} \sim 0$,
which can be easily realized in SPDC.   In this experiment, we have chosen
$s_o = s_i =2f$. Thus, an equal size ``ghost" image of slit A is expected to appear
on the image plane of screen B.  

\begin{figure}[hbt]
    \centering
    \includegraphics[width=95mm]{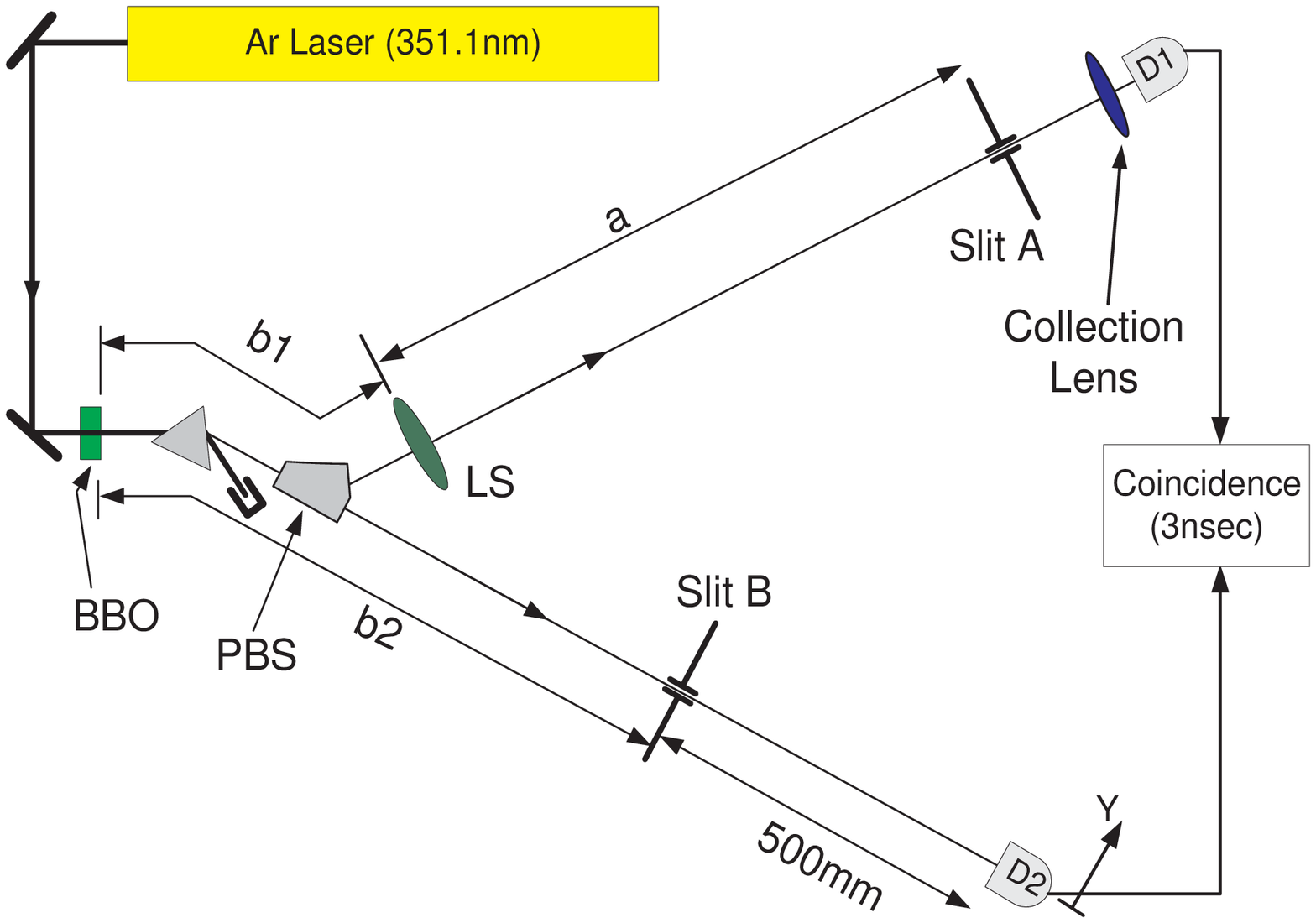}
    \parbox{14cm}{\caption{Schematic of
the experimental setup. The laser beam is about $3mm$ in diameter. The
``phase-matching condition" is well reinforced. Slit A ($ 0.16mm$) is placed
$1000mm=2f$ behind the converging lens, LS ($f=500mm$). The one-to-one ``ghost
image" ($0.16mm$) of slit A is located at B. The optical distance from LS in
the signal beam taken as back through PBS to the SPDC crystal ($b_1=255mm$) and
then along the idler beam to ``screen B" ($b_2=745mm$) is $1000mm=2f$
($b=b_1+b_2$). } 
    \label{Poppersetup}}
\end{figure}

The detailed experimental setup is shown in Fig.\ref{Poppersetup} with
indications of the various distances.  A CW Argon ion laser line of
$\lambda _{p}=351.1nm$ is used to pump a $3mm$ long beta barium borate (BBO)
crystal for type-II SPDC to generate an orthogonally polarized
signal-idler photon pair. The laser beam is about $3mm$ in diameter with a
diffraction limited divergence. It is important to keep the pump beam a large size so
that the transverse phase-matching condition, $\vec{k}_{s}+\vec{k}_{i} \sim 0$ ($\vec{k}_{p} = 0$), 
is well reinforced in the SPDC process, where $\vec{k}_{j}$
$(j=s,i)$ is the transverse wavevector of the signal (s) and idler (i),
respectively. The collinear signal-idler beams, with $\lambda _{s}=\lambda
_{i}=702.2nm=2\lambda _{p}$ are separated from the pump beam by a fused quartz
dispersion prism, and then split by a polarization beam splitter PBS. The
signal beam (photon 1) passes through the converging lens LS with a $500mm$
focal length and a $25mm$ diameter. A $ 0.16mm$ slit is placed at location A
which is $1000mm$ $(=2f)$ behind the lens LS.  A short
focal length lens is used with $D_{1}$ for focusing the signal beam that
passes through slit A. The point-like photon counting detector $D_{2}$ is located 
$500mm$ behind ``screen B".  ``Screen B" is the image plane defined by the Gaussian
thin lens equation.   Slit B, either adjusted as the same size as that of slit A or opened 
completely, is placed to coincide with the ``ghost" image.  The output pulses from the
detectors are sent to a coincidence circuit.  During the measurements, detector
$D_{1}$ is fixed behind slit A while detector $D_{2}$ is scanned on the
$y$-axis by a step motor.

\begin{figure}[hbt]
    \centering
    \includegraphics[width=100mm]{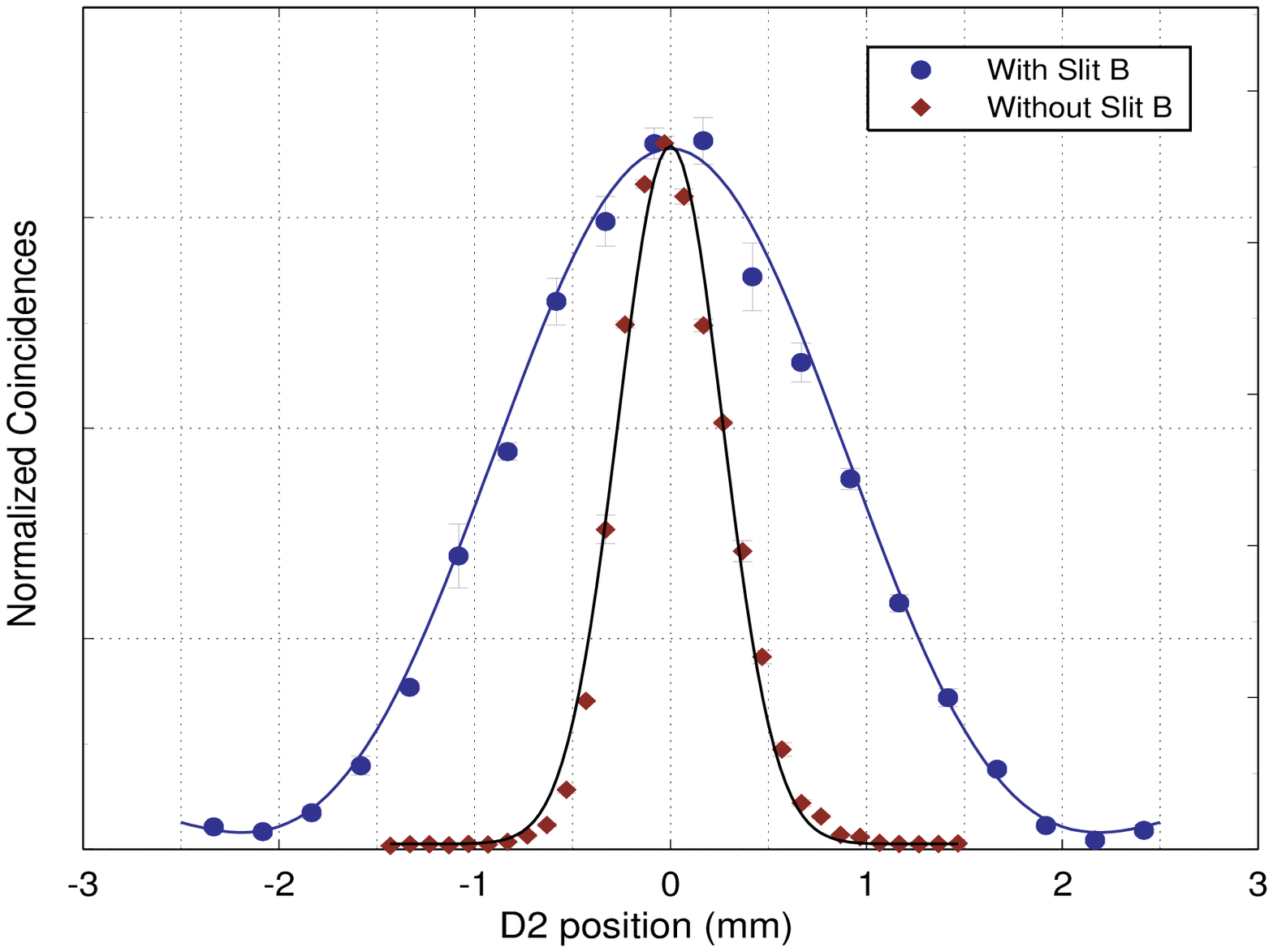}
    \parbox{14cm}{\caption{The
observed coincidence patterns. The $y$-coordinate of $D_{1}$ was chosen to be
$0$ (center) while $D_{2}$ was allowed to scan along its $y$-axis. Circled dot
points: {\em Slit A = Slit B = $0.16mm$}. Diamond dot points: {\em Slit A =
$0.16mm$, Slit B wide open}. The width of the $sinc$ function curve fitted by
the circled dot points is a measure of the minimum $ \Delta p_{y}$ diffracted
by a $0.16mm$ slit. } 
    \label{Popperdata}}
\end{figure}

\vspace{5mm}

{\bf Measurement 1}: Measurement 1 studied the case in which both slits A and B were
adjusted to be $0.16mm$. The $y$-coordinate of $D_{1}$ was chosen to be $0$
(center) while $D_{2}$ was allowed to scan along its $y$-axis. The circled dot
data points in Fig. \ref{Popperdata} show the {\em coincidence} counting rates
against the $y$-coordinates of $D_{2}$. It is a typical single-slit diffraction
pattern with $\Delta y \, \Delta p_{y}=h $. Nothing is special in this measurement
except that we have learned the width of the diffraction pattern for the $0.16mm$
slit and this represents the minimum uncertainty of $ \Delta p_{y}$. 
We should emphasize at this point that the {\em single}
detector counting rate of $D_{2}$ as a function of its position $y$ is basically the same 
as that of the coincidence counts except for a higher counting rate.

\vspace{5mm}

{\bf Measurement 2}: The same experimental conditions were maintained except
that slit B was left wide open. This measurement is a test of Popper's
prediction. The $y$-coordinate of $D_{1}$ was chosen to be $0$ (center) while
$D_{2}$ was allowed to scan along its $y$-axis. Due to the entangled nature of the
signal-idler photon pair and the use of a coincidence measurement circuit, only those twins
which have passed through slit A and the ``ghost image" of slit A at screen
B with an uncertainty of $\Delta y=0.16mm$ (which is the same width as the real
slit B we have used in measurement 1) would contribute to the coincidence
counts through the joint detection of $D_{1}$ and $D_{2}$. The diamond
dot data points in Fig. \ref{Popperdata} report the measured coincidence
counting rates against the $y$ coordinates of $D_{2}$. The measured width of
the pattern is narrower than that of the diffraction pattern shown in
measurement 1.  It is
also interesting to notice that the single detector counting rate of $D_{2}$ keeps
constant in the entire scanning range, which is very different from that in
measurement 1. The experimental data has provided a clear indication of $\Delta
y \, \Delta p_{y}< h$ in the joint measurements of the entangled photon pairs.

\vspace{6mm}

Given that $\Delta y \, \Delta p_{y} < h$, is this a violation of the
uncertainty principle?  Does quantum mechanics agree with
this peculiar experimental result?  If quantum
mechanics does provide a solution with $\Delta y \, \Delta p_{y}< h $
for photon 2.  We would indeed be forced to face a paradox as
EPR had pointed out in 1935.   

Quantum mechanics does provide a solution that agrees with the 
experimental result.  However, the solution is for a joint measurement 
of an entangled photon pair that involves both photon 1 and photon 2,
but not just for photon 2 itself . 

We now examine the experimental results with the quantum mechanical 
calculation by adopting the formalisms from the ghost image experiment 
with two modifications: 

\vspace{3mm}\hspace{-6.5mm}Case (I): slits A $= 0.16mm$,
slit B $ = 0.16mm$. \vspace{2mm}

This is the experimental condition for measurement one: slit B is 
adjusted to be the same as slit A.   There is nothing surprising about this 
measurement.   The measurement simply provides us with the knowledge 
for $\Delta p_y$ of photon 2 caused by the diffraction of slit B 
($\Delta y = 0.16mm$).   The experimental data shown in 
Fig.~\ref{Popperdata} agrees with the calculation.
Notice that slit B is about $745mm$ away
from the $3mm$ two-photon source, the angular size of the light
source is roughly the same as $\lambda / \Delta y$, 
$\Delta\theta \sim \lambda / \Delta y$, where $\lambda = 702nm$ 
is the wavelength and $\Delta y = 0.16mm$ is the width 
of the slit.   The calculated diffraction pattern is very close to that of  the
``far-field" Fraunhofer diffraction of a $0.16mm$ single-slit.  

\vspace{3mm}\hspace{-6.5mm}Case (II): slit A $= 0.16mm$, 
slits B $\sim \infty$ (wide open). \vspace{2mm} 

Now we remove slit B from the ghost image plane.  The calculation of the transverse
effective two-photon wavefunction and the second-order correlation is the same as that 
of the ghost image except the observation plane of $D_2$ is moved behind the image 
plane to a distance of $500mm$.  The two-photon image of slit A is located at a distance 
$s_i = 2f = 1000mm$ ($b_1 + b_2$) from the imaging lens, in this measurement $D_2$
is placed at $d = 1500mm$ from the imaging lens.  The measured pattern is simply a 
``blurred" two-photon image of slit A.  The ``blurred" two-photon image can be calculated
from Eq.~(\ref{Blurred}) which is a slightly modified version of Eq.~(\ref{biphoton_z})
\begin{eqnarray}\label{Blurred}
\Psi(\vec{\rho}_o,\vec{\rho}_2) 
&\propto& \int_{lens} d\vec{\rho}_l \, 
G(|\, \vec{\rho}_2-\vec{\rho}_l\, |, \frac{\omega}{c d}) \,
G(| \vec{\rho}_l |, \frac{\omega}{c f}) \,
G(|\,\vec{\rho}_l-\vec{\rho}_o\,|, \frac{\omega}{c s_o}) \nonumber \\
&\propto& \int_{lens} d\vec{\rho}_l \, G(|\, \vec{\rho}_l|, 
\frac{\omega}{c}[\frac{1}{s_o} + \frac{1}{d} - \frac{1}{f}]) \,
e^{-i\frac{\omega}{c} (\frac{\vec{\rho}_o}{s_o} + \frac{\vec{\rho}_i}{d})\cdot \vec{\rho}_l} 
\end{eqnarray}
where $d$ is the distance between the imaging lens and $D_2$.  In this measurement, 
$D_2$ was placed $500mm$ behind the image plane, i.e., $d = s_i + 500mm$.   
The numerically calculated ``blurred" image, which is narrower then that of the diffraction 
pattern of the $0.16mm$ slit B, agrees with the measured result of Fig.~\ref{Popperdata} 
within experimental error.

\vspace{6mm}

The measurement does show a result of $\Delta y \, \Delta p_{y} < h$.  
The measurement, however, has nothing to do with the uncertainty relation,
which governs the behavior of photon 2 (the idler).     
Popper and EPR were correct in the prediction of the outcomes of their
experiments. Popper and EPR, on the other hand, made the same error by 
applying the results of two-particle physics to the explanation of the behavior 
of an individual subsystem. 

In both the Popper and EPR experiments, the measurements are ``joint detection"
between two detectors applied to entangled states. Quantum mechanically, an
entangled two-particle state only provides {\em the precise knowledge of the
correlations of the pair}. The behavior of ``photon 2'' observed in the joint 
measurement is conditioned upon the measurement of
its twin.  A quantum must obey the uncertainty principle but the ``conditional
behavior" of a quantum in an entangled two-particle system is different in principle.  
We believe paradoxes are unavoidable if one insists the {\em conditional behavior} 
of a particle is the {\em behavior} of the particle. This is the central problem in
the rationale behind both Popper and EPR.  $\Delta y \, \Delta p_{y} \geq h$ is not
applicable to the \emph{conditional behavior} of either ``photon 1" or ``photon 2" in
the cases of Popper and EPR.

The behavior of photon 2 being conditioned upon the measurement of photon 1 is well 
represented by the two-photon amplitudes.  Each of the {\em straight lines} in the 
above discussion corresponds to a two-photon amplitude.  Quantum mechanically, 
the superposition of these two-photon amplitudes are responsible for
a ``click-click" measurement of the entangled pair.  A ``click-click'' joint
measurement of the two-particle entangled state projects out certain
two-particle amplitudes, and only these two-particle amplitudes are featured in the
quantum formalism.  In the above analysis we never consider ``photon 1'' or
``photon 2'' {\em individually}.  Popper's question about the momentum
uncertainty of photon 2 is then inappropriate. 

Once again, the demonstration of Popper's experiment calls
our attention to the important message: the physics of an entangled
two-particle system must be inherently very different from that of individual
particles.  

\section{Subsystem in an entangled two-photon state}

\hspace{5.5mm}The entangled EPR two-particle state is a pure state with zero
entropy.  The precise correlation of the subsystems is completely described by 
the state.  The measurement, however,  is not necessarily always on the two-photon 
system.  It is an experimental choice to study a single subsystem and to 
ignore the other.  What can be learn about a subsystem from these kinds of 
measurements?  Mathematically, it is easy to show that by taking a partial trace 
of a two-particle pure state, the state of each subsystem is in a mixed state 
with entropy greater than zero.  One can only learn statistical properties of the 
subsystems in this kind of measurement.  

In the following, again, we use the signal-idler pair of SPDC as an example to
study the physics of a subsystem. The two-photon state of SPDC is a pure state that satisfies
\begin{equation*}
\hat{\rho}^{2}=\hat{\rho},\quad \hat{\rho}\equiv \left| \Psi \right\rangle
\left\langle \Psi \right|  
\end{equation*}
where $\hat{\rho}$ is the density operator corresponding to the two-photon
state of SPDC. The single photon states of the signal and idler
\begin{equation*}
\hat{\rho}_{s}=tr_{i}\left| \Psi \right\rangle \left\langle \Psi \right|,
\quad \hat{\rho}_{i}=tr_{s}\left| \Psi \right\rangle \left\langle \Psi \right|  
\end{equation*}
are not pure states. To calculate the signal (idler) state from the two-photon state, 
we take a partial trace, as usual, summing over the idler (signal) modes. 

We assume a type II SPDC.  The orthogonally polarized signal and idler are
degenerate in frequency around $\omega^0_{s} = \omega^0_{i} = \omega _{p}/2$.  
To simplify 
the discussion, by assuming appropriate experimental conditions, we trivialize 
the transverse part of the state and write the two-photon state in the following 
simplified form: 
\begin{equation*}
\left| \Psi \right\rangle =\Psi_{0} \int d\Omega \ \Phi ({\rm DL}\Omega)\, a_{s}^{\dagger
}(\omega^0_s +\Omega )\, a_{i}^{\dagger }(\omega^0_i - \Omega )\left| 0\right\rangle
\end{equation*}
where $\Phi({\rm DL}\Omega )$ is a $sinc$-like function:
\[
\Phi ({\rm DL}\Omega)=\frac{1-e^{-i{\rm DL}\Omega}}{i{\rm DL}\Omega}
\]
which is a function of the crystal length L, and the difference of inverse
group velocities of the signal (ordinary) and the idler (extraordinary), 
${\rm D}\equiv 1/u_{o}-1/u_{e}$.  The constant $\Psi_{0}$ is calculated from the
normalization $tr\, \hat{\rho}=\left\langle \Psi \mid \Psi \right\rangle =1$.
It is easy to calculate and to find $\hat{\rho}^{2}=\hat{\rho}$ for the two-photon
state of the signal-idler pair.

Summing over the idler modes, the density matrix of signal is given by
\begin{equation}  \label{densityIs}
\hat{\rho} _s=\Psi_0^2\int d\Omega \ \left| \Phi(\Omega)\right| ^2\, a_s^{\dagger
}(\omega^0_s +\Omega)\left| 0\right\rangle \left\langle 0\right| \, a_s(\omega^0_s +\Omega)
\end{equation}
with
\begin{equation*} 
\left| \Phi (\Omega)\right|^2={\rm sinc}^2 \, \frac{{\rm DL}\Omega}{2}
\end{equation*}
where all constants coming from the integral have been absorbed into $\Psi_0$.
First, we find immediately that $\hat{\rho} _s^2\neq \hat{\rho} _s$. It means
the state of the signal is a mixed state (as is the idler). Second, it
is very interesting to find that the spectrum of the signal
depends on the group velocity of the idler.  This, however, 
should not come as a surprise, because the state of the signal photon is 
calculated from the two-photon state by summing over the idler modes.

\begin{figure}[hbt]
    \centering
    \includegraphics[width=95mm]{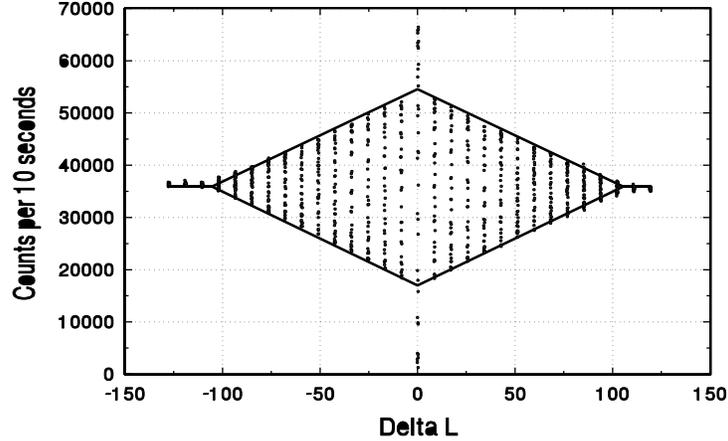}
    \parbox{14cm}{\caption{Experimental data indicated a
``double notch'' envelope. Each of the doted single vertical line
contines many cycles of sinusoidal modulation.}
    \label{fig:notch}}
\end{figure}

The spectrum of the signal and idler has been experimentally verified by 
Strekalov \emph{et al} using a Michelson interferometer in a standard 
Fourier spectroscopy type measurement \cite{Strekalov}.   
The measured interference pattern is shown in Fig.~\ref{fig:notch}. 
The envelope of the sinusoidal modulations (in segments) is fitted 
very well by two ``notch'' functions (upper and lower part of the envelope). 
The experimental data agrees with the theoretical analysis of the experiment.  

The following is a simple calculation to explain the observed ``notch" function.   
We first define the field operators: 
$$
E^{(+)}(t, z_d) = E^{(+)}(t-\frac{z_1}{c}, z_0) + E^{(+)}(t-\frac{z_2}{c}, z_0) 
$$
where $z_d$ is the position of the photo-detector, $z_0$ is the input point of the
interferometer, $t_1 = t-\frac{z_1}{c}$ and $t_2= t-\frac{z_2}{c}$, respectively, are 
the early times before propagating to the photo-detector at time $t$ with time delays 
of $z_1/c$ and $z_2/c$, where $z_1$ and $z_2$ are the optical paths in arm 1 
and arm 2 of the interferometer.  We have defined a very general field operator 
which is a superposition of two early fields propagated individually through 
arm 1 and arm 2 of any type of interferometer.  The counting rate of the photon counting 
detector is thus 
\begin{eqnarray}
R_d &=& tr \, \big{[} \hat{\rho} _s E^{(-)}(t, z_d) E^{(+)}(t, z_d) \big{]} \nonumber \\
&=& \Psi_0^2\int d\Omega \, 
\left| \Phi(\Omega)\right|^2 \big{|}  \langle 0 | E^{(+)}(t, z_d) \, 
a_s^{\dagger}(\omega^{0}_s +\Omega)\left| 0\right\rangle \big{|}^2 \nonumber \\
&=& \Psi_0^2\int d\Omega \, 
\left| \Phi(\Omega)\right|^2 \big{|}  \langle 0 | \big{[} E^{(+)}(t-\frac{z_1}{c}, z_0) +
E^{(+)}(t-\frac{z_2}{c}, z_0) \big{]}\, 
a_s^{\dagger}(\omega^{0}_s +\Omega)\left| 0\right\rangle \big{|}^2 \nonumber \\
&\propto& 1+ Re \, \Big{[} e^{-i\omega^{0} \tau} \int d\Omega \, 
\, {\rm sinc}^2\frac{{\rm DL}\Omega}{2} \,
e^{-i \Omega \tau} \Big{]} 
\end{eqnarray}
where $\tau = (z_1 - z_2)/c$.  The Fourier transform of ${\rm sinc}^2({\rm DL}\Omega/2)$
has a ``notch" shape.  It is noticed that the base of the ``notch" function is determined 
by parameter DL of the SPDC, which is easily confirmed from the experiment.  

\vspace{6mm}

Now we turn to another interesting aspect of physics, namely the physics 
of entropy.  In classical information theory, the concept of entropy, named as 
Von Neuman entropy, is defined by \cite{Entropy}
\begin{equation}\label{entropy }
S = - \, tr\,(\hat{\rho} \, \log \hat{\rho})
\end{equation}
where $\hat{\rho}$ is the density operator.  It is easy to find that the entropy of 
the entangled two-photon pure
state is zero.  The entropy of its subsystems, however, are both greater
than zero.  The value of the Von Neuman entropy can be numerically 
evaluated from the measured spectrum.  Note that the density operator of the 
subsystem is diagonal.  Taking its trace is simply performing an integral over the 
frequency spectrum with the measured spectrum function.  It is straightforward 
to find the entropy of the subsystems $S_s > 0$.  
This is an expected result due to the statistical mixture nature of the
subsystem.  Considering that the entropy of the two-photon system is
zero and the entropy of the subsystems are both greater than zero, 
does this mean that negative entropy is present somewhere
in the entangled two-photon system?  According to classical ``information
theory'', for the entangled two-photon system, $S_{s}+S_{s\mid
i}=0$, where $S_{s\mid i}$ is the conditional entropy. It is this
conditional entropy that must be negative, which means that {\em given the result
of a measurement over one particle, the result of a measurement over the other
must yield negative information} \cite{Entropy2}. This paradoxical statement is similar and,
in fact, closely related to the EPR ``paradox''.  
It comes from the same philosophy as that of the EPR. 

\section*{Summary}

\hspace{5.5mm} The physics of an entangled system is very different from that of either 
classically independent or correlated systems.  We use $2 \neq 1 + 1$ to emphasize the 
nonclassical behavior of an entangled two-particle system.  The entangled system is 
characterized by the properties of an entangled state which does not specify the state 
of an individual system, but rather describes the correlation between the subsystems.
An entangled two particle state is a pure state which involves the superposition of a set 
of ``selected" two-particle states, or two-particle quantum mechanical amplitudes.  
Here, the term ``selection" stems from the physical laws which govern the creation of the 
subsystems in the source, such as energy or momentum conservation.  Interestingly, 
quantum mechanics allows for the superposition of these local two-particle states which 
have been observed in nature.  However, the most surprising physics arises from the  
joint measurement of the two particles when they are released form the source and 
propagated a large distance apart.  The
two well separated interaction-free particles do not lose their entangled properties, i.e., 
they maintain their ``selected" set of two-particle superposition.  In this sense quantum 
mechanics allows for the two-particle superposition of well separated particles which 
has, remarkably, also been observed to exist in nature.  

The two-photon state of SPDC is a good example.  The nonlinear interaction of spontaneous 
parametric down-conversion coherently creates a set of mode in pairs that satisfy the phase 
matching conditions of Eq.~(\ref{eq:phsmtch}) which is also characteristic of energy and 
momentum conservation.  The signal-idler photon pair can be excited to any or all of these 
coupled modes simultaneously, resulting in a superposition of these coupled modes inside 
of the nonlinear crystal.  The physics behind the two-photon superposition becomes even 
more interesting when the signal-idler pair is separated and propagated a large distance 
apart outside the nonlinear crystal, either through free propagation or guided by 
optical components.  Remarkably the entangled pair does not lose its entangled properties 
once the subsystems are interaction free.  As a result the properties of the entangled two-photon 
system, such as the EPR correlation or the EPR inequalities, are still observable in the joint 
detection counting rate of the pair, regardless of the distance between the two photons 
as well as the two individual photo-detection events.   In this situation 
the superposition of the two-photon amplitudes, corresponding to different yet indistinguishable 
alternative ways of triggering a joint photo-electron event at any distance can be regarded as 
nonlocal.  There is no counterpart to such a concept in classical theory and this 
behavior may never be understood in any classical sense.  It is with this intent that we use
$2 \neq 1 + 1$ to emphasize that the physics of a two-photon is not the same as that of 
two photons.

\vspace{3mm}

\section*{A statement from the author}
\hspace{6.5mm}This article was originally prepared as lecture notes for my students a few years ago.  
It was also used in 2006 for a conference.  My colleagues, friends and students have urged me 
to include it in this archive.  They believe that this article is helpful for the general physics 
and engineering community.  Truthfully, I have been hesitant because I
cannot forget my terrible experience in 1996 as a result of Pittman's experiment: 
``Can tow-photon interference be considered the interference of two photons?" \cite{Pittman} 
My email account was bombarded for months.  Of course, I was happy to have scientific 
discussions on the subject, but certain types of messages caused headaches.  For example, 
an individual attempted to force my laboratory to pay a visit for a face-to-face condemnation on 
my guilt for saying $1 + 1 \neq 2$. (I truly believe what I said was $2 \neq 1 + 1$ and anyone 
would be able to see the difference by reading this article).
Another individual expressed their interests in a law suit because we did not acknowledge that 
\emph{they were the first} to show ``Dirac was mistaken".  (I am definitely sure that we have 
nothing to do with their ``discovery". What we said was ``Dirac was correct".)  I decided to 
keep quiet.  I understood that it takes time for people to recognize the truth.  

I have to break my silence now, because we are experiencing the same problem again.  My student 
Scarcelli published a lens-less ghost imaging experiment of chaotic light and raised a reasonable question: 
``Can two-photon correlation of chaotic light be considered as correlation of intensity fluctuations?" 
\cite{Scarcelli}.  The lens-less ghost imaging setup of Scarcelli \emph{et al}. 
is a straightforward modification of the historical Hanbury-Brown and Twiss experiment (HBT) \cite{hbt}. 
Advancing from HBT to the fundamentally interesting and practically useful lens-less ghost imaging, 
what one needs to do is simply move the two HBT photodetectors from far-field to near-field.  
We cannot but stop to ask: 
What has been preventing this simple move for 50 years (1956-2006)?  Some aspect must be terribly 
misleading to give us such misled confidence not to even try the near-field measurement in half a century.  
As we know, unlike the first-order correlation of radiation that is considered as the interference effect 
of the electromagnetic waves, the second-order correlation of light is treated as statistical correlation 
of intensity fluctuations.  Scarcelli \emph{et al}. pointed out that although the theory of statistical correlation 
of intensity fluctuations gives a reasonable explanation to the far-field HBT phenomena, it does not work 
in near-field and consequently does not work for their lens-less near-field ghost imaging experiment 
\cite{Shih-2}.  It was the idea of statistical correlation 
of intensity fluctuation that has prevented this from happening for 50 years.  On the other hand, under the 
framework of Glauber's theory of photodetection, Scarcelli \emph{et al}. proved a successful interpretation 
based on the quantum picture of two-photon interference.   This successes indicates that although the 
concept of multi-photon interference, or the superposition of multi-photon amplitudes, was benefited 
from the research of entangled states, the concept is generally true and applicable to any radiation, 
including ``classical" thermal light.   Unfortunately, this concept has no counterpart in classical 
electromagnetic theory of light.  Now, we are back to 1996.  My student and I have been charged 
with ``guilt" again because we have told the physics community a simple truth of the failure of a classical 
idea and adapted the quantum mechanical concept of two-photon superposition to ``classical" light.  

It was a mistake to keep silence.  I have finally resolved to speak about the subject.  
The concept of multi-photon coherence, or the superposition principle of multi-photon 
amplitudes, is important and worthwhile to do, even if  I might be burned at the stake.  



\vspace*{5mm}

\section*{Appendix: Fresnel propagation-diffraction}
\def\theequation{$A-$\arabic{equation}}
\setcounter{equation}{0}
\def\thefigure{$A-$\arabic{figure}}
\setcounter{figure}{0}

\begin{figure}[hbt]
 \centering
    \includegraphics[width=75mm]{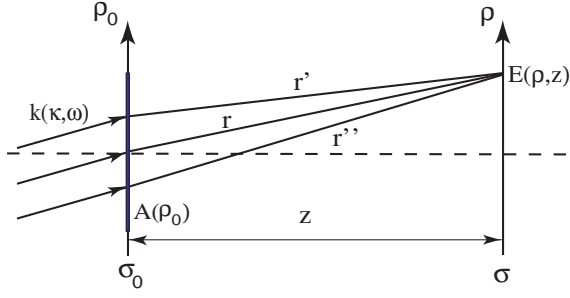}
     \parbox{14cm}{\caption{Schematic of free-space Fresnel propagation.  
    The complex amplitude $\tilde{A}(\vec{\rho}_0)$ 
    is composed by a real function $A(\vec{\rho}_0)$ and a phase 
    $e^{-i \vec{\kappa} \cdot \vec{\rho}_0}$ associated with each of the transverse 
    wavevector $\vec{\kappa}$ on the plane of $\sigma_0$.  Notice: only one mode of 
    wavevector $\mathbf{k}(\vec{\kappa}, \omega)$ is shown in the figure.}
    \label{fig:Fresnel}}
\end{figure}

\hspace{5.5mm}In Fig.~\ref{fig:Fresnel}, the field is freely propagated from the source 
plane $\sigma_0$ to an arbitrary plane $\sigma$.   It is convenient to describe such a 
propagation in the form of Eq.~(\ref{gg-1}). We now evaluate  
$g(\vec{\kappa}, \omega; \vec{\rho}, z)$, namely the Green's function  
for free-space Fresnel propagation-diffraction.  

According to the Huygens-Fresnel principle, the field at a space-time point
$(\vec{\rho}, z, t)$ is the result of a superposition of the spherical 
secondary wavelets originated from each point on the $\sigma_0$ plane,
see Fig.~\ref{fig:Fresnel},
\begin{eqnarray}\label{gg-00}
E^{(+)}(\vec{\rho}, z, t) = \int  d\omega \, d\vec{\kappa} \
a(\omega, \vec{\kappa}) \int_{\sigma_0} \, d\vec{\rho}_0 \, 
\frac{\tilde{A}(\vec{\rho}_0)}{r'} \, e^{-i (\omega t - k r')}
\end{eqnarray}
where $\tilde{A}(\vec{\rho}_0)$ is the complex amplitude, or distribution 
function, in terms of the transverse coordinate $\vec{\rho}_0$, which 
may be a constant, a simple aperture function, or a combination 
of the two.  In Eq.~(\ref{gg-00}), we have taken 
$z_0 = 0$ and $t_0 = 0$ on the source plane of $\sigma_0$ as usual.  

In a paraxial approximation,
we take the first-order expansion of $r'$ in terms of $z$ and $\vec{\rho}$ 
$$
r' = \sqrt{z^2 + |\vec{\rho} - \vec{\rho}_0|^2} \simeq z(1 + 
\frac{|\vec{\rho} - \vec{\rho}_0|^2}{2 z^2}).
$$
$E^{(+)}(\vec{\rho}, z, t)$ is thus approximated as
\begin{eqnarray*}
E^{(+)}(\vec{\rho}, z, t) \simeq  \int d\omega \,  d\vec{\kappa} \
a(\omega, \vec{\kappa}) \int d\vec{\rho}_0 \, \frac{\tilde{A}(\vec{\rho}_0)}{z}
\, e^{i \frac{\omega}{c} z} \, 
e^{i \frac{\omega}{2 c z} |\vec{\rho} - \vec{\rho}_0|^2} e^{-i \omega t}
\end{eqnarray*}
where $e^{i \frac{\omega}{2 c z} |\vec{\rho} - \vec{\rho}_0|^2}$ is named as
the Fresnel phase factor.

Assuming the complex amplitude $\tilde{A}(\vec{\rho}_0)$ is composed of a real 
function $A(\vec{\rho}_0)$ and a phase $e^{-i \vec{\kappa} \cdot \vec{\rho}_0}$, 
associated with the transverse
wavevector and the transverse coordinate on the plane of $\sigma_0$, 
which is reasonable for the setup of Fig.~\ref{fig:Fresnel},
$E(\vec{\rho}, z, t)$ can be written in the following form
\begin{eqnarray*}
E^{(+)}(\vec{\rho}, z, t) =  \int d\omega \, d\vec{\kappa} \ a(\omega, \vec{\kappa}) \,
e^{-i \omega t} \, \frac{e^{i \frac{\omega}{c} z}}{z} 
\int d\vec{\rho}_0 \, A(\vec{\rho}_0) \, e^{i \vec{\kappa} \cdot \vec{\rho}_0} \,  
e^{i \frac{\omega}{2 c z} |\vec{\rho} - \vec{\rho}_0|^2}.
\end{eqnarray*}
The Green's function $g(\vec{\kappa}, \omega; \vec{\rho}, z)$ for free-space
Fresnel propagation is thus 
\begin{equation}\label{gg-10}
g(\vec{\kappa}, \omega; \vec{\rho}, z) = \frac{e^{i \frac{\omega}{c} z}}{z} 
\int_{\sigma_0} \, d\vec{\rho}_0 \, A(\vec{\rho}_0) \, 
e^{i \vec{\kappa} \cdot \vec{\rho}_0} \, 
G(|\vec{\rho} - \vec{\rho}_0|, \frac{\omega}{c z}).
\end{equation}

In Eq.~(\ref{gg-10}) we have defined a Gaussian function 
$G(|\vec{\alpha|}, \beta)= e^{i (\beta/2) |\alpha|^2}$, namely the Fresnel phase factor.  
It is straightforward to find that
the Gaussian function $G(|\vec{\alpha|}, \beta)$ has the following properties:
\begin{eqnarray}\label{Gaussian-10}
G^*(|\vec{\alpha}|, \beta) &=& G(|\vec{\alpha}|, - \beta ), \nonumber \\
G(|\vec{\alpha}|, \beta_1 + \beta_2) &=& G(|\vec{\alpha}|, \beta_1 ) \,
G(|\vec{\alpha}|, \beta_2), \nonumber \\
G(|\vec{\alpha}_1+\vec{\alpha}_2|, \beta)
 &=& G(|\vec{\alpha}_1|, \beta) \, G(|\vec{\alpha}_2|, \beta) \,
e^{i \beta \vec{\alpha}_1 \cdot \vec{\alpha}_2}, \nonumber \\
\int d\vec{\alpha} \,\,
G(|\vec{\alpha}|, \beta ) \, e^{i \vec{\gamma} \cdot \vec{\alpha}} &=& i \frac{2\pi}{\beta} \,
G(|\vec{\gamma}|, -\frac{1}{\beta} ).
\end{eqnarray}
Notice that the last equation in Eq.~(\ref{Gaussian-10}) is the Fourier transform of the 
$G(|\vec{\alpha|}, \beta)$ function. As we shall see in the following, these properties 
are very useful in simplifying the calculations of the Green's functions 
$g(\vec{\kappa}, \omega; \vec{\rho}, z)$.

\vspace{6mm}

Now, we consider inserting an imaginary plane $\sigma'$ between $\sigma_0$ and 
$\sigma$.  This is equivalent having two consecutive Fresnel propagations with a 
diffraction-free $\sigma'$ plane of infinity.   Thus, the calculation of these consecutive
Fresnel propagations should yield the same Green's function as that of the 
above direct Fresnel propagation shown in Eq.~(\ref{gg-10}):
\begin{eqnarray}\label{gg-11}
&& g(\vec{\kappa}, \omega; \vec{\rho}, z) \nonumber \\
&=& C^2 \, \frac{e^{i \frac{\omega}{c}( d_1+ d_2)}}{d_1 d_2} 
\int_{\sigma'}  d\vec{\rho'}  \int_{\sigma_0} d\vec{\rho}_0 \, \tilde{A}(\vec{\rho}_0) \, 
G(|\vec{\rho'} - \vec{\rho}_0|, \frac{\omega}{c d_1}) \,
G(|\vec{\rho} - \vec{\rho'}|, \frac{\omega}{c d_2}) \nonumber \\
&=& C \, \frac{e^{i \frac{\omega}{c} z}}{z} 
\int_{\sigma_0} \, d\vec{\rho}_0 \, \tilde{A}(\vec{\rho}_0) \, 
G(|\vec{\rho} - \vec{\rho}_0|, \frac{\omega}{c z})
\end{eqnarray}
where $C$ is a necessary normalization constant for a valid Eq.~(\ref{gg-11}),
and $z = d_1 + d_2$.  The double integral of $d\vec{\rho}_0 $ and 
$d\vec{\rho'}$ in Eq.~(\ref{gg-11}) can be evaluated as 
\begin{eqnarray*}
&& \int_{\sigma'}  d\vec{\rho'}  \int_{\sigma_0} d\vec{\rho}_0 \, \tilde{A}(\vec{\rho}_0) \, 
G(|\vec{\rho'} - \vec{\rho}_0|, \frac{\omega}{c d_1}) \,
G(|\vec{\rho} - \vec{\rho'}|, \frac{\omega}{c d_2}) \nonumber \\
&=&\int_{\sigma_0} d\vec{\rho}_0 \, \tilde{A}(\vec{\rho}_0) \,  
G(\vec{\rho}_0, \frac{\omega}{c d_1}) \, G(\vec{\rho}, \frac{\omega}{c d_2})
\int_{\sigma'}  d\vec{\rho'} \, G(\vec{\rho'}, \frac{\omega}{c}(\frac{1}{d_1} + 
\frac{1}{d_2})) \,
e^{-i\frac{\omega}{c}(\frac{\vec{\rho}_{0}}{d_1} + 
\frac{\vec{\rho}}{d_2})\cdot \vec{\rho'}} \nonumber \\
&=& \frac{i 2\pi c}{\omega} \frac{d_1 d_2}{d_1 + d_2}
\int_{\sigma_0} d\vec{\rho}_0 \, \tilde{A}(\vec{\rho}_0) \,  
G(\vec{\rho}_0, \frac{\omega}{c d_1}) \, 
G(\vec{\rho}, \frac{\omega}{c d_2}) \,
G(|\frac{\vec{\rho}_{0}}{d_1} + 
\frac{\vec{\rho}}{d_2}|, \frac{\omega}{c}(\frac{d_1 d_2}{d_1 + d_2})) 
\nonumber \\
&=& \frac{i 2\pi c}{\omega} \frac{d_1 d_2}{d_1 + d_2}
\int_{\sigma_0} d\vec{\rho}_0 \, \tilde{A}(\vec{\rho}_0) \,
G(|\vec{\rho} - \vec{\rho}_0|, \frac{\omega}{c (d_1 + d_2)})
\end{eqnarray*}
where we have applied Eq.~(\ref{Gaussian-10}), and the integral of
$d\vec{\rho'}$ has been taken to infinity.  Substituting this result into 
Eq.~(\ref{gg-11}), we thus have
\begin{eqnarray*}
g(\vec{\kappa}, \omega; \vec{\rho}, z) 
&=& C^2 \, \frac{i 2\pi c}{\omega} \frac{e^{i \frac{\omega}{c}( d_1+ d_2)}}{d_1 + d_2} 
\int_{\sigma_0} d\vec{\rho}_0 \, \tilde{A}(\vec{\rho}_0) \,
G(|\vec{\rho} - \vec{\rho}_0|, \frac{\omega}{c (d_1 + d_2)}) \nonumber \\
&=& C \, \frac{e^{i \frac{\omega}{c} z}}{z} 
\int_{\sigma_0} \, d\vec{\rho}_0 \, \tilde{A}(\vec{\rho}_0) \, 
G(|\vec{\rho} - \vec{\rho}_0|, \frac{\omega}{c z}).
\end{eqnarray*}
Therefore, the normalization constant $C$ must take the value of
$
C = - i \omega / 2 \pi c.
$
The normalized Green's function for free-space Fresnel propagation is thus
\begin{eqnarray}\label{gg-final}
g(\vec{\kappa}, \omega; \vec{\rho}, z) 
=  \frac{-i \omega}{ 2 \pi c} \ \frac{e^{i \frac{\omega}{c} z}}{z}
\int_{\sigma_0} \, d\vec{\rho}_0 \, \tilde{A}(\vec{\rho}_0) \, 
G(|\vec{\rho} - \vec{\rho}_0|, \frac{\omega}{c z}).
\end{eqnarray}

\end{document}